\documentclass[12pt,draftclsnofoot, onecolumn]{IEEEtran}
\usepackage{amssymb}
\usepackage{amsmath}
\usepackage{amsthm}
\usepackage[lined,boxed,commentsnumbered,ruled]{algorithm2e}
\usepackage{algorithmic}
\usepackage{mathrsfs}
\usepackage{bm}
\usepackage{tikz}
\usepackage{pgfplots}
\usetikzlibrary{shapes}
\usepackage{subcaption}
\usepackage{graphicx,booktabs,multirow}
\usepackage{enumerate} 
\usepackage{epstopdf}
\usepackage{hyperref}
\usepackage{epsfig}
\usepackage{cite}
\usepackage{float}
\allowdisplaybreaks

\definecolor{color1}{HTML}{D0B22B}
\definecolor{dred}{RGB}{128,0,0}
\definecolor{colorhkust}{RGB}{20,43,140}
\definecolor{colorshanghaitech}{RGB}{162,0,5}
\definecolor{colortsinghua}{RGB}{116,52,129}
\definecolor{colordark}{RGB}{184,134,11}
\theoremstyle{definition}
\newtheorem{lemma}{Lemma}
\newtheorem{theorem}{Theorem}

\newcommand{\transpose}{\mathsf{T}}
\newcommand{\Htranspose}{\mathsf{H}}
\newcommand{\bracket}[1]{\left(#1\right)}
\newcommand{\norm}[1]{\left\|#1\right\|}
\newcommand{\card}[1]{\left|#1\right|}

\begin{document}

\title{Over-the-Air Federated Learning via Second-Order Optimization}
\author{Peng Yang, \textit{Student Member, IEEE},
Yuning Jiang, \textit{Member, IEEE},  \\Ting Wang, \textit{Senior Member, IEEE}, Yong Zhou, \textit{Member, IEEE}, \\Yuanming~Shi, \textit{Senior Member}, \textit{IEEE}, Colin N. Jones, \textit{Member, IEEE} 
\thanks{P. Yang and T. Wang are with the Shanghai Key Lab. of Trustworthy Computing, Software Engineering Institute, East China Normal University, Shanghai 200062, China (e-mail: 51205902030@stu.ecnu.edu.cn, twang@sei.ecnu.edu.cn). Y. Jiang and C. N. Jones are with the Automatic Control Laboratory, EPFL, Laussane 1015, Switzerland (e-mail: yuning.jiang, colin.jones@epfl.ch). Y. Zhou and Y. Shi are with the School of Information Science and Technology, ShanghaiTech University, Shanghai 201210, China (e-mail: zhouyong, shiym@shanghaitech.edu.cn).}
}

\maketitle

\begin{abstract}
Federated learning (FL) is a promising learning paradigm that can tackle the increasingly prominent isolated data islands problem while keeping users’ data locally with privacy and security guarantees. However, FL could result in task-oriented data traffic flows over wireless networks with limited radio resources. 
To design communication-efficient FL, most of the existing studies employ the first-order federated optimization approach that has a slow convergence rate. 
This however results in excessive communication rounds for local model updates between the edge devices and edge server. 
To address this issue, in this paper, we instead propose a novel over-the-air second-order federated optimization algorithm to simultaneously reduce the communication rounds and enable low-latency global model aggregation. 
This is achieved by exploiting the waveform superposition property of a multi-access channel to implement the distributed second-order optimization algorithm over wireless networks. 
The convergence behavior of the proposed algorithm is further characterized, which reveals a linear-quadratic convergence rate with an accumulative error term in each iteration. 
We thus propose a system optimization approach to minimize the accumulated error gap by joint device selection and beamforming design. 
Numerical results demonstrate the system and communication efficiency compared with the state-of-the-art approaches.
\end{abstract}


\section{Introduction}

Artificial intelligence (AI) technologies under rapid development have been widely studied and deployed in various scenarios.
As a data-driven technology, its reliability and accuracy largely depend on the volume and quality of source data. 
However, it is recognized as a big challenge for most enterprises to obtain a dataset with sufficient volume and quality for AI model training. 
In the meantime, data privacy is another crucial issue that needs to be considered among different involved parties~\cite{xu2014information}. 
To this end, 
it is preferred in real-world implementations that data be kept locally, 
forming a variety of isolated data islands. 
This makes it difficult to directly aggregate data in the cloud and centrally train the AI models. 
Therefore, federated learning (FL)~\cite{mcmahan2017communication, bonawitz2019towards, yang2019federated} has emerged as a novel paradigm to address these challenges.
A generic and practical FL framework is essentially a distributed training process, and each iteration of FL includes the following three steps~\cite{yang2019federated}. Firstly, the server broadcasts the current global model parameters to all the involved devices. Next, each device performs local model training based on its local data and then sends the local updates back to the server.
Finally, the server aggregates the local updates and generates new global model parameters for the next iteration of distributed training. In essence, the server and devices aim to collaboratively solve a distributed optimization problem, which is typically referred to as \emph{Federated Optimization}~\cite{konevcny2016federated}. Different from centralized optimization, federated optimization confronts several practical challenges including communication efficiency, data heterogeneity, security, system complexity, etc.~\cite{wang2021field}. 
Among them, communication efficiency is of utmost importance since the communication between the server and devices usually suffers from unreliable network connections, limited resources, and severe latency~\cite{shi2020communication}. 

To deal with the communication issue, a large amount of research has been conducted in federated optimization. On the one hand, reducing the communication volume in each iteration is an effective method.
Specifically, quantization and sparsification techniques are employed to reduce the transmitted bits and remove the redundant updates of parameters, respectively~\cite{bernstein2018signsgd, aji2017sparse}.
These compression techniques have shown remarkable effectiveness for high-dimensional models.
However, their design needs to consider the compatibility for the aggregation operation in FL~\cite{wang2021field}.
On the other hand, minimizing the total communication rounds is another primary method. To this end, 
zeroth-order methods~\cite{chen2019zo, gao2020can} have been investigated for some restrictive circumstances (e.g., black-box adversarial attack, non-smooth objective function) while showing great potential as only the objective function value is required to approximate derivative information~\cite{nesterov2017random}. 
In the situation where gradients are available, first-order methods are widely used. 
By increasing the amount of local computation, various gradient descent based methods have been shown that can significantly decrease the total number of communication rounds~\cite{mcmahan2017communication, woodworth2020local, yuan2020federated, pathak2020fedsplit}.
Nevertheless, these existing approaches, i.e., zeroth-order and first-order approaches, are governed by the linear convergence in the best case.
As a result, the total number of iteration rounds required to achieve the desired accuracy is relatively large~\cite{bischoff2021second}.   
Therefore, the second-order methods (e.g., Newton-type methods) become attractive in such a wireless environment due to their fast local quadratic convergence rate.
Nevertheless, the construction of the canonical Newton update requires both the Hessian and gradient information, where the distributed situation in FL makes gathering Hessian information a severe communication overhead.
To this end, second-order federated optimization algorithms have been investigated to resolve this issue, which can be divided into two categories~\cite{bischoff2021second}.
One is to use second-order information implicitly.
In~\cite{shamir2014communication}, a mirror descent update is carried out on the local function to approximate the Hessian information.
In~\cite{smith2018cocoa}, the dual problems of the objective function are used to serve as the local subproblems.
The other category is to use second-order information explicitly.
In~\cite{wang2018giant}, a globally improved approximate Newton method (GIANT) using local Hessian for aggregation is proposed.
In~\cite{crane2019dingo, crane2020dino}, the optimization of the gradient's norm acts as the surrogate function.
In~\cite{zhang2015disco}, Hessian-vector product computation and conjugate gradient descent are performed on the devices and the server, respectively.
The fast convergence rate with efficient communication makes the application of these second-order algorithms a great benefit to FL.

Despite the potential in the application of second-order algorithms to reduce the total communication rounds and improve the communication efficiency, the transmission of FL model parameters through wireless channels still confronts great challenges as wireless channels are always noisy with limited resources and high latency~\cite{yang2020federatedIOT, letaief2019roadmap, li2021delay}. 
Based on the conventional ``transmit-then-communicate'' principle, the aggregation of FL model parameters can be achieved by digital coded transmission and orthogonal multiple access (OMA) schemes~\cite{amiri2020machine, elgabli2021harnessing, chang2020communication}.
By taking advantage of OMA and error correction techniques, local updates are transmitted separately in the quantized form and then decoded individually at the server.
In this way, the model transmission can be deemed to be reliable and trustworthy. 
However, the increase in the number of devices will inevitably lead to a sharp increase in total communication latency and bandwidth requirement, which is often intolerable.
Therefore, a novel technique called over-the-air computation (AirComp)~\cite{nazer2007computation} has emerged in FL algorithm design to decrease the communication cost based on the ``compute-when-transmit'' principle~\cite{yang2020federated, zhu2019broadband, amiri2020machine, sery2020analog, liu2020privacy, elgabli2021harnessing, liu2020reconfigurable, xiaowen2021optimized, xu2021learning, zhu2020one, wei2022federated, fan2021joint}.
This technique leverages the superposition property of multiple access channels to realize the aggregation operation.
Through the simultaneous transmission of all local updates, which are aggregated over the air, the communication overheads are significantly decreased.
Specifically, the authors in~\cite{yang2020federated} proposed an AirComp-based approach for FL with joint design of device selection and beamforming to improve the statistical learning performance.
In~\cite{sery2020analog}, a novel Gradient-Based Multiple Access (GBMA) algorithm was put forward to perform FL with an energy scaling law for approaching the convergence rate of centralized training. 
In~\cite{xiaowen2021optimized}, the authors investigated the power control optimization for enhancing the learning performance of over-the-air federated learning.
In~\cite{liu2020reconfigurable, wang2020federated}, intelligent reflecting surface (IRS) technology was used to achieve fast yet reliable model aggregation for over-the-air federated learning. 
The authors in~\cite{xu2021learning} proposed the dynamic learning rate design for AirComp-based FL.
Overall, the application of over-the-air computation in FL also improves the communication efficiency a lot.


Based on the above observations, this paper proposes to improve communication efficiency from two aspects, i.e., reducing communication rounds and the communication overhead in each round.
To reduce the communication rounds, we shall utilize second-order information during the training process of FL.
Due to the fast convergence speed, all these existing second-order state-of-the-arts have shown substantial improvement in terms of the total iteration rounds compared with first-order methods.
However, their iterative procedures still have at least two communication rounds per iteration, i.e., the aggregation of gradient and second-order information.
To avoid such two communication rounds, a recently proposed second-order method~\cite{ghosh2020distributed} cuts down the aggregation of gradients and realizes one communication round per iteration.
Motivated by this, we adopt local Newton step aggregation for wireless FL algorithm design.
Specifically, the product of the local Hessian's inversion and the local gradient is used to construct a local Newton step for aggregation. 
By this means, the devices only need to communicate once with the server per iteration, cutting down the transmission of local Hessian matrices and local gradients while keeping the convergence behavior of canonical Newton's method.
Moreover, due to the limited radio resources, we adopt over-the-air computation, which has been widely used in the existing wireless FL schemes, to further reduce the communication overheads in each round.
Based on this efficient local Newton step aggregation and AirComp technique, we propose an over-the-air second-order federated algorithm over wireless networks.
Furthermore, we provide a rigorous theoretical analysis of the convergence behavior of our proposed method. 
The results show that 
the transmission of the above-mentioned product is sufficient to guarantee convergence and our proposed method outperforms first-order algorithms. To be specific, 
the proposed algorithm keeps a linear-quadratic convergence rate, which means it can achieve the optimal point with a quadratic convergence rate and degenerate into the linear convergence rate when it is close enough to the optimal point.
However, as a result of local Newton step aggregation, device selection, and channel noise, there is an error term in each iteration. As the training proceeds, this accumulative error term will deflect the model parameters and affect learning performance. In order to mitigate the impact of this error term, we further propose a joint optimization approach of device selection and receiver beamforming.
Specifically, Gibbs Sampling \cite{geman1984stochastic} is adopted to determine the set of selected devices, and the difference-of-convex-functions (DC) algorithm \cite{tao1997convex} is tailored to optimize the receiver beamforming during the iterative process of Gibbs Sampling.

\subsection{Contributions}
In this paper, we propose a novel over-the-air FL algorithm via the second-order optimization method. 
Then, we theoretically analyze its convergence behavior, which shows that the proposed algorithm keeps a linear-quadratic convergence rate, with an accumulative error term arising during the FL process. 
To minimize the error gap and achieve better performance, we formulate this problem as a combinatorial non-convex problem and propose a system optimization approach to solve it. 
The main contributions of this paper are summarized as follows:
\begin{itemize}
    \item[1)] We design a novel AirComp-based FL algorithm by leveraging the principles of distributed second-order optimization methods and exploiting the waveform superposition property of a wireless multi-access channel for model aggregation.
    This algorithm is fundamentally different from most existing works which only consider gradient descent/SGD in training.
    The utilization of second-order information significantly reduces the total communication rounds in Aircomp-based FL, which further improves the communication efficiency.
    \item[2)] We theoretically analyze the convergence behaviors of our proposed over-the-air second-order federated optimization algorithm with the presence of data heterogeneity (i.e., the different data sizes), device selection, and channel noise. 
    The results show that our algorithm keeps a linear-quadratic convergence rate and outperforms first-order methods;
    \item[3)] We formulate a system optimization problem to minimize the accumulative error gap during the execution of our proposed algorithm. Correspondingly, we propose a system optimization approach. 
    Through the combination of Gibbs Sampling and DC algorithm, we jointly optimize the device selection and receiver beamforming;  
    \item[4)] 
    We conduct extensive experiments to demonstrate that our proposed algorithm and system optimization approach can achieve better performance than other state-of-art approaches.
\end{itemize}

\subsection{Organization and Notations}
The remainder of this paper is organized as follows. 
Section \ref{Federated Edge Learning Model and Algorithm} presents the federated learning model and our FL algorithm.
Section \ref{Theoretical Analysis} provides the convergence analysis of our proposed algorithm. 
Section \ref{System Optimization} analyzes the system optimization problem arising from the error term, and describes our joint optimization method of device selection and beamforming. 
The experimental results are given in Section \ref{Simulation Results}. 
Finally, Section \ref{Conclusion} concludes the whole paper. 

$\|\cdot\|_p$ is the $\ell_p$-norm, $\|\cdot\|_\mathsf{F}$ is the Frobenius norm. 
Italic, boldface lower case and upper case letters represent scalars, vectors and matrices, respectively. 
For a given set $\mathcal{X}$, $|\mathcal{X}|$ denotes the cardinality of $\mathcal{X}$. 
The operators $(\cdot)^\mathsf{T}$, $(\cdot)^\mathsf{H}$, $\text{Tr}(\cdot)$ and $\text{diag}(\cdot)$ denote the transpose, Hermitian transpose, trace, and diagonal matrix, respectively.
$\mathbb{E}[\cdot]$ denotes the statistical expectation.


\section{Federated Learning Model and Algorithm}
\label{Federated Edge Learning Model and Algorithm}
\subsection{Federated Learning System}
\begin{figure}[htbp!]
    \centering
    \includegraphics[width=0.4\linewidth]{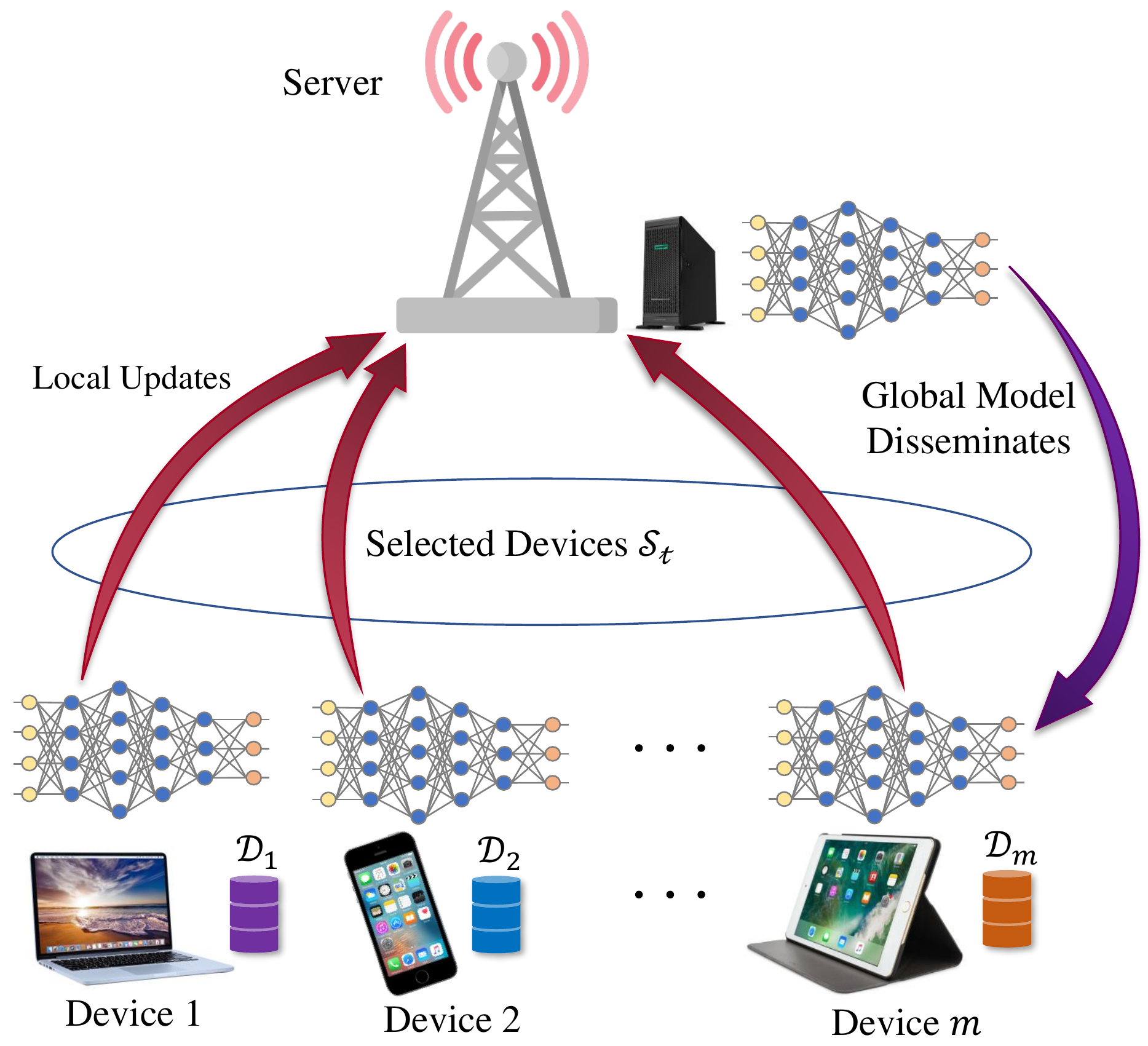}
    \caption{Illustration of wireless FL systems.}
    \label{FL system}
    \vspace{-0.6cm}
\end{figure}
A typical wireless federated learning system consists of a group of distributed devices and one server, where the communication takes place over wireless channels. 
As depicted in Fig.~\ref{FL system}, there are $m$ single-antenna devices and a server equipped with $k$ antennas to collaboratively complete a learning task.
We denote $\mathcal{D}$ as the entire sample set used in the FL task. 
Each device $i \in \mathcal{S}$ stores a sample set $\mathcal{D}_{i} = \left\{\bm{z}_{i, j} := \left(\bm{u}_{i, j}, v_{i, j}\right)\right\}$ and $\mathcal{D} = \bigcup_{i=1}^{m} \mathcal{D}_{i}$ with $\card{\mathcal{D}} = n$, where $\mathcal{S}$ denotes the index set of devices,  $\bm{u}_{i, j}$ is the feature vector and $v_{i, j}$ is the corresponding label.

As an important part of the learning task, the loss function is usually used for model parameter estimation. 
Here, the loss function of the $i$-th device is defined by
\begin{small}
\begin{equation}
    F_{i}\left(\bm{w}\right)=\frac{1}{\left|\mathcal{D}_{i}\right|} \sum_{\bm{z}_{i, j} \in \mathcal{D}_{i}} f\left(\bm{w}, \bm{z}_{i, j}\right)+\frac{\gamma}{2}\|\bm{w}\|_{2}^{2}\;.
    \label{local_loss}
\end{equation}
\end{small}
The first term is the average of $f\left(\bm{w}, \bm{z}_{i, j}\right)$, where $\bm{w} \in \mathbb{R}^{d}$ is the model parameter vector and function $f$ is used to measure the prediction error of $\bm{w}$. The second term is for regularization
with $\gamma$ being the weighting parameter. FL aims to train a suitable model at the server by aggregating the results collected from multiple devices, on which the distributed models are trained based on local datasets. Specifically, the server needs to optimize the following global loss function:
\begin{small}
\begin{equation}
    F \bracket{\bm{w}} = \frac{1}{n}\sum_{i=1}^{m}\card{\mathcal{D}_{i}}F_{i}\bracket{\bm{w}} \;.
    \label{global_loss}
\end{equation}
\end{small}

\subsection{Federated Second-Order Optimization Algorithm}
As typical training algorithms, gradient descent methods (e.g., SGD \cite{bottou2012stochastic}, batch gradient descent) are widely used.
However, the relatively slow convergence rate of gradient descent results in too many communication rounds between the server and devices to complete the learning task. 
Thus, many research works have been done to improve the communication efficiency of gradient descent in FL.
For example, some methods utilize multiple local updates to reduce the number of communication rounds \cite{mcmahan2017communication, woodworth2020local},
while several algorithms employ compression techniques to reduce transmitted bits and save communication costs~\cite{bernstein2018signsgd, aji2017sparse, vogels2019powersgd}. 
Although these schemes have greatly improved the communication efficiency of gradient descent in FL, they are still limited by the linear convergence rate.

To address this issue, this paper considers second-order algorithms with a faster convergence rate such that the communication rounds can be significantly reduced.
The descent direction vector of canonical Newton's method~\cite{nocedal2006numerical} is given by
\begin{small}
\begin{equation}
    \bm{p} = \left(\nabla^{2}F\left(\bm{w}\right)\right)^{-1}\nabla F\left(\bm{w}\right).
    \label{Newton step}
\end{equation}
\end{small}
The canonical Newton's method can achieve a locally quadratic convergence rate so that its total iteration rounds needed to complete the learning task are much fewer than first-order algorithms.
However, in the distributed scenario, the computation of $\nabla^{2}F\left(\bm{w}\right) = \frac{1}{m}\sum_{i=1}^{m} \nabla^{2}F_{i}\left(\bm{w}\right)$ requires the aggregation of the local Hessian $\nabla^{2}F_{i}\left(\bm{w}\right)$. 
The transmission of such $d \times d$ matrices inevitably brings huge communication overheads.
To resolve this issue, numerous second-order distributed machine learning algorithms have been proposed, such as DANE  \cite{shamir2014communication}, DISCO  \cite{zhang2015disco}, GIANT \cite{wang2018giant}, DINGO  \cite{crane2019dingo}, and DINO \cite{crane2020dino}. 
These methods approximate Hessian information in varied forms to avoid the direct transmission of Hessian matrices and approach the performance of canonical Newton's method.
However, at least two communication rounds per iteration are required, including the aggregation of local gradients and second-order descent directions.
Different from these second-order algorithms, which require the aggregation of local gradients $\nabla F_{i}\left(\bm{w}\right)$ to compute the global gradient $\nabla F\left(\bm{w}\right) = \frac{1}{m} \sum_{i=1}^{m} \nabla F_{i}\left(\bm{w}\right)$, a recently proposed COMRADE \cite{ghosh2020distributed} method cuts down this aggregation.
By this means, the number of communication rounds required per iteration is reduced to one, further improving the communication efficiency.
Motivated by this, we leverage the local Newton step aggregation as in~\cite{ghosh2020distributed} to achieve a faster convergence rate with fewer communication rounds. 
The product of the inversion of the local Hessian matrix $\left(\nabla^{2}F_{i}\left(\bm{w}\right)\right)^{-1}$ and the local gradient $\nabla F_{i}\left(\bm{w}\right)$ is used to serve as the local descent direction vector $\bm{p}_{i} = \left(\nabla^{2}F_{i}\left(\bm{w}\right)\right)^{-1}\nabla F_{i}\left(\bm{w}\right)$ for model aggregation.
In this way, with the preserved convergence behavior of Newton's method, only one aggregation of the $d$-dimensional local descent direction vectors will be carried out in each iteration.
To be specific, at $t$-th iteration, the procedure of our proposed method is summarized as follows:
\begin{itemize}
\item[1)] \textbf{Device Selection}: The server decides the set of devices, denoted as $\mathcal{S}_{t}$, to participate in this iteration.

\item[2)] \textbf{Global Model Broadcast}: The server disseminates the current global model parameter vector $\bm{w}_{t}$ to the selected devices through the wireless channel.

\item[3)] \textbf{Local Model Update}: After the $i$-th device receives global model parameter vector $\bm{w}_{t}$,
it first computes the local gradient based on local data samples:
\begin{small}
\begin{equation}
    \bm{g}_{t,i} = \nabla F_{i}\left(\bm{w}_{t}\right)=\frac{1}{\left|\mathcal{D}_{i}\right|} \sum_{\bm{z}_{i, j} \in \mathcal{D}_{i}} \nabla f\left(\bm{w}_{t} ,\bm{z}_{i, j}\right)+\gamma \bm{w}_{t}\;,
    \label{local_gradient}
\end{equation}
\end{small}
where the derivatives are taken with respect to the first argument. 
Afterwards, the $i$-th device calculates the local Hessian matrix according to local gradient and local data samples:
\begin{small}
\begin{equation}
    \bm{H}_{t, i}=\nabla^2F_{i}\left(\bm{w}_{t}\right)=\frac{1}{\left|\mathcal{D}_{i}\right|} \sum_{\bm{z}_{i, j} \in \mathcal{D}_{i}} \nabla^{2}f\left(\bm{w}_{t}, \bm{z}_{i, j}\right)+\gamma \bm{I}_{d}\;.
    \label{local_hessian}
\end{equation}
\end{small}
The $i$-th device then gets a local Newton descent direction vector from previous results:
\begin{small}
\begin{equation}
    \bm{p}_{t,i} = \bm{H}_{t, i}^{-1} \bm{g}_{t,i} = \left(\nabla^{2}F_{i}\left(\bm{w}_{t}\right)\right)^{-1}\nabla F_{i}\left(\bm{w}_{t}\right).
    \label{approx_newton_direction}
\end{equation}
\end{small}
In practice, this step involves the computation of Hessian matrix and its inverse operation.
To reduce the computational complexity, we adopt the conjugate gradient method~\cite{nocedal2006numerical} to obtain an approximate local Newton descent direction vector.
According to the analysis in~\cite{wang2018giant}, this approximate solution will not have a significant impact on the convergence behavior.

\item[4)] \textbf{Model Aggregation}: The devices participating in the $t$-th iteration transmit local Newton descent direction vectors $\{\bm{p}_{t,i}\}$ to the server through the wireless channel, 
and the server aggregates them to obtain the global descent direction vector for this iteration:
\begin{small}
\begin{equation}
    \tilde{\bm{p}}_{t} = \frac{1}{\underset{i \in \mathcal{S}_{t}}{\sum}\card{\mathcal{D}_{i}}}\underset{i \in \mathcal{S}_{t}}{\sum}\card{\mathcal{D}_{i}} \bm{p}_{t,i} \;.
    \label{averaged_local_descent_direction}
\end{equation}
\end{small}
\item[5)] \textbf{Global Model Update}: Finally, the server updates the model parameter vector $\bm{w}_{t}$ through global descent direction vector $\tilde{\bm{p}}_{t}$ and learning rate $\alpha$.
\end{itemize}
\begin{small}
\begin{equation}
    \bm{w}_{t+1} = \bm{w}_{t} - \alpha \tilde{\bm{p}}_{t}\;.
    \label{global_update}
\end{equation}
\end{small}
Notably, the Newton's method has a faster convergence rate than the gradient descent methods because it makes full use of the curvature information of the loss function, but the aggregation of the $d \times d$ local Hessian matrices for Newton descent direction in~\eqref{Newton step} aggravates the communication overheads in another way. 
As implied in Step 3) of our proposed FL scheme, it does not need to compute the global gradient $\nabla F\left(\bm{w}\right)$ and Hessian $\nabla^{2}F\left(\bm{w}\right)$ to get a precise Newton descent direction by aggregating the local $\nabla F_i\left(\bm{w}\right)$ and $\nabla^{2}F_i\left(\bm{w}\right)$.
Note that this approximation also brings a controllable error gap with the exact descent direction vector, and its impact on the convergence rate will be analyzed in Section~\ref{Theoretical Analysis}.

\subsection{Communication Model}

To further reduce the communication overheads, this subsection focuses on the design of the communication model between the server and devices. 
Specifically, there are two communication-related steps in each iteration of our FL algorithm. 
One is global model broadcasting in the downlink. Since only one global parameter vector needs to be broadcasted, the total communication cost of this step is negligible~\cite{sery2020analog, chang2020communication, amiri2020machine, amiri2020federated}.
The other is model aggregation in the uplink, which involves the transmission of $|\mathcal{S}_{t}|$ local descent direction vectors. Accordingly, the uploading process of this step brings the primary communication overhead in FL, which is also the focus of our communication model design.

In this paper, we consider a block fading channel.
Each block is divided into $d$ time slots, ensuring the transmission of one local descent direction vector. 
Suppose the traditional orthogonal multiple access channel is used to perform the model aggregation procedure. 
Each device will use one coherent block to transmit its local descent direction vector.
Consequently, the time consumed for transmission in this step will increase linearly with the number of participating devices $|\mathcal{S}_{t}|$. 
Unfortunately, the number of devices $|\mathcal{S}_{t}|$ is usually very large, which inevitably leads to unacceptable communication overheads.
In order to eliminate this issue, we adopt a state-of-the-art technique named over-the-air computation (AirComp)~\cite{nazer2007computation}, which is shown to be effective in assisting the analog aggregation in FL studies~\cite{yang2020federated, zhu2019broadband, amiri2020machine, sery2020analog, liu2020privacy, elgabli2021harnessing, liu2020reconfigurable, xiaowen2021optimized}. 
This technique captures the nomographic function form of averaging the local descent direction vectors and implements the summation operation by the superposition property of the wireless channel.
In this way, the server can receive the summation by letting all devices transmit their local descent direction vectors simultaneously in each block. 
Therefore, the entire process of model aggregation can be completed over the air in a single coherent block, and the communication overheads can be significantly reduced.  
More specifically, in the $t$-th iteration, the over-the-air computation can be represented as the nomographic function form~\cite{zhu2018mimo} : $ \hat{\bm{p}}_{t}=\psi\left(\sum_{i \in \mathcal{S}_t} \varphi_{i}\left(\bm{p}_{t,i}\right)\right).$
To reduce the transmission power, the pre-processing function $\phi_{i}$ and post-processing function $\psi$ can be designed to normalize and de-normalize the local descent direction vector $\bm{p}_{t,i}$~\cite{zhu2019broadband}. 
However, due to the variety of $\bm{p}_{t,i}$ among devices, the stationary of the information-bearing symbols obtained by such normalization methods can not be guaranteed, which further leads to the inapplicability of the uniform-forcing transceiver design in the following.
Therefore, to guarantee the stationary of the information-bearing symbols, we adopt the data-and-CSI-aware design as in~\cite{guo2020analog}. 
Before transmission, $\bm{p}_{t,i}$ is first pre-processed and encoded as $\bm{s}_{t,i} \in \mathbb{R}^{d}$ at the $i$-th device:
\begin{small}
\begin{equation}
    \bm{s}_{t,i} = \phi_{i}\bracket{\bm{p}_{t,i}} = \frac{\card{\mathcal{D}_{i}}\bm {p}_{t,i}}{\bar{p}_{t,i}} \;,
    \label{preprocess}
\end{equation}
\end{small}
where $\bar{p}_{t,i} = \card{\mathcal{D}_{i}}\norm{\bm{p}_{t,i}}$ is the product of the size of local dataset and the magnitude of $\bm{p}_{t,i}$. 
In this way, the stationary of the information-bearing symbols $\{\bm{s}_{t,i}\}$ can be guaranteed.
Hence, we have $\norm{\bm{s}_{t,i}}^{2} = 1$ and $\mathbb{E}\bracket{|\bm{s}_{t,i}[j]|^{2}} = \frac{1}{d}, \;\forall j\in d$, where $\bm{s}_{t,i}[j]$ denotes the $j$-th entry of $\bm{s}_{t,i}$.
Thereafter, each entry of the transmitted signal sent by the $i$-th device is given by:
\begin{small}
\begin{equation}
    \bm{x}_{t,i}[j] = b_{t,i}\bm{s}_{t,i}[j] \;,
    \label{transmitted_signal}
\end{equation}
\end{small}
where $\bm{x}_{t,i}[j] \in \mathbb{R}$ and $\bm{s}_{t,i}[j] \in \mathbb{R}$ denote two representative entries of $\bm{x}_{t,i}$ and $\bm{s}_{t,i}$, respectively. $b_{t,i} \in \mathbb{R}$ is the transmitted power control factor, and the power constraint for each device in the whole process is given by:
\begin{small}
\begin{equation}
    \mathbb{E}\bracket{|b_{t,i}\bm{s}_{t,i}[j]|^{2}} = b_{t,i}^{2} / d \leq P_{0}, \; \forall t, i\;,
    \label{power_constraint}
\end{equation}
\end{small}
where $P_{0}$ denotes the maximum transmitted power of each device.

Let $\bm{h}_{t,i} \in \mathbb{C}^{k}$ be the channel coefficient vector between the $i$-th device and the server in the $t$-th block, which remains unchanged in each block but differs among blocks.
In addition, we assume that perfect channel state information (CSI) is available at all devices to adjust their transmitted signals based on channel coefficients~\cite{sery2020analog, amiri2020machine, liu2020privacy, zhu2019broadband, yang2020federated, wang2020federated, seif2020wireless, wang2020wireless, fang2021over}. 
Then the received signal $\bm{y}_{t} \in \mathbb{C}^{k}$ at the server can be represented as follows:
\begin{small}
\begin{equation}
    \bm{y}_{t} = \sum_{i\in\mathcal{S}_{t}}\bm{h}_{t,i}\bm{x}_{t,i}[j] + \bm{e}_{t} = \sum_{i\in\mathcal{S}_{t}}\tilde{\bm{h}}_{t,i}b_{t,i}\card{\mathcal{D}_{i}}\bm{p}_{t,i}[j] + \bm{e}_{t}\;,
    \label{received_signal}
\end{equation}
\end{small}
where $\tilde{\bm{h}}_{t,i} = \frac{\bm{h}_{t,i}}{\bar{p}_{t,i}}$ is the effective channel coefficient introduced in~\cite{guo2020analog}, $\bm{e}_{t} \in \mathbb{C}^{k}$ denotes the additive white Gaussian noise vector with the power of $\sigma^{2}$.
We define the signal-to-noise ratio (SNR) as $P_{0} / \sigma^{2}$.

After the server receives $\bm{y}_{t}$, it can obtain the value $\bm{r}_{t}[j] \in \mathbb{C}$ before post-processing:
\begin{small}
\begin{equation}
    \bm{r}_{t}[j] = \frac{1}{\sqrt{\eta_{t}}}\bm{a}_{t}^{\Htranspose}\bm{y}_{t} = \frac{1}{\sqrt{\eta_{t}}}\bracket{\bm{a}_{t}^{\Htranspose}\sum_{i\in\mathcal{S}_{t}}\tilde{\bm{h}}_{t,i}b_{t,i}\card{\mathcal{D}_{i}}\bm{p}_{t,i}[j] + \bm{a}_{t}^{\Htranspose}\bm{e}_{t}} \;,
    \label{process_received_signal}
\end{equation}
\end{small}
where $\bm{a}_{t} \in \mathbb{C}^{k}$ represents the receiver beamforming vector and $\eta_{t}$ is the scaling factor.
For convenience, we use $\bm{H}_{t} = [\tilde{\bm{h}}_{t,1},\ldots,\tilde{\bm{h}}_{t,|\mathcal{S}_{t}|}]$ to denote the effective channel coefficient matrix,
$\bm{B}_{t} = \text{diag}\bracket{b_{t,1},\ldots,b_{t,|\mathcal{S}_{t}|}}$ to denote the power transmission matrix, 
$\bm{G}_{t} = [\card{\mathcal{D}_{1}}\bm{p}_{t,1}, \ldots, \card{\mathcal{D}_{|\mathcal{S}_{t}|}}\bm{p}_{t,|\mathcal{S}_{t}|}]^{\transpose}$ to denote the signal transmission matrix, and $\bm{E}_{t} = [\bm{e}_{t,1}, \ldots, \bm{e}_{t,d}]$ to denote the noise matrix.
So the total estimated value vector $\bm{r}_{t} = [\bm{r}_{t}[1], \ldots, \bm{r}_{t}[d]]$ can be written as:
\begin{small}
\begin{equation}
    \bm{r} = \frac{1}{\sqrt{\eta_{t}}}\bracket{\bm{a}^{\Htranspose}\bm{H}_{t}\bm{B}_{t}\bm{G}_{t} + \bm{a}_{t}^{\Htranspose}\bm{E}_{t}} \;.
    \label{rewrite_process_received_signal}
\end{equation}
\end{small}
To alleviate the influence of the distortion caused by noise and improve the performance of over-the-air computation, each entry of $\bm{B}_{t}$ follows the uniform-forcing transceiver design~\cite{chen2018uniform}:
\begin{small}
\begin{equation}
    b_{t,i} = \sqrt{\eta_{t}}\frac{\bracket{\bm{a}_{t}^{\Htranspose}\tilde{\bm{h}}_{t,i}}^{\Htranspose}}{\norm{\bm{a}_{t}^{\Htranspose}\tilde{\bm{h}}_{t,i}}^{2}} \;. 
    \label{transmit_vector}
\end{equation}
\end{small}
where the transmission scalar $b_{t,i}$ can be computed after the calculation of receiver beamforming vector in system optimization, and then feed back to each device~\cite{yang2020federated}. Substituting~\eqref{transmit_vector} into~\eqref{rewrite_process_received_signal}, we can get a simplified version of $\bm{r}_{t}$:
\begin{small}
\begin{equation}
    \bm{r}_{t} = \sum_{i \in \mathcal{S}_{t}}\card{\mathcal{D}_{i}}\bm{p}_{t,i}^{\transpose} + \frac{1}{\sqrt{\eta_{t}}}\bm{a}_{t}^{\Htranspose}\bm{E}_{t} \;.
    \label{simplified_recovered_signal}
\end{equation}
\end{small}
Finally, through the post-processing function of $\psi$, the server obtains the global descent direction vector $\hat{\bm{p}}_{t}$:
\begin{small}
\begin{equation}
    \hat{\bm{p}}_{t} = \psi\bracket{\bm{r}_{t}} = \frac{1}{\sum_{i \in \mathcal{S}_{t}}\card{\mathcal{D}_{i}}}\bm{r}_{t}^{\Htranspose} = \tilde{\bm{p}}_{t} + \frac{1}{\bracket{\sum_{i \in \mathcal{S}_{t}}\card{\mathcal{D}_{i}}}\sqrt{\eta_{t}}}\bracket{\bm{a}_{t}^{\Htranspose}\bm{E}_{t}}^{\Htranspose} \;,
    \label{global_update_direction}
\end{equation}
\end{small}
where $\tilde{\bm{p}}_{t,i}$ is the averaged local descent direction vector as defined in~\eqref{averaged_local_descent_direction}.

\begin{algorithm}
\caption{Over-the-Air Second-Order Federated Algorithm}
\label{ACCADE}
\small
\For{each iteration t}
{
    server chooses devices participating in this iteration and stores them as $\mathcal{S}_{t}$.\\
    server broadcasts the current model parameter vector $\bm{w}_{t}$ to all devices.\\
    \For{each participating device $i \;\; \textbf{in parallel}$ }{
        compute local gradient $\bm{g}_{t,i} = \frac{1}{\left|\mathcal{D}_{i}\right|} \sum_{\bm{z}_{i, j} \in \mathcal{D}_{i}} \nabla f\left(\bm{w}_{t} ,\bm{z}_{i, j}\right)+\gamma \bm{w}_{t}$.\\
        compute local Hessian matrix $\bm{H}_{t, i} = \frac{1}{\left|\mathcal{D}_{i}\right|} \sum_{\bm{z}_{i, j} \in \mathcal{D}_{i}}  \nabla^{2} f\left(\bm{w}_{t}, \bm{z}_{i, j}\right)+\gamma \bm{I}_{d}$.\\
        compute local Newton descent direction $\bm{p}_{t,i} = \bm{H}_{t, i}^{-1} \bm{g}_{t,i}$.\\
        encode $\bm{p}_{t, i}$ as $\bm{s}_{t,i}$ according to \eqref{preprocess}. \\
        transmit the signal $\bm{x}_{t,i} = b_{t,i}\bm{s}_{t,i}$ through wireless channel.
    }

    server receives the signal $\bm{y}_{t}$ \eqref{received_signal} and maintains $\hat{\bm{p}_{t}}$ \eqref{global_update_direction}.\\
    server performs an update step $\bm{w}_{t+1} = \bm{w}_{t} - \alpha \hat{\bm{p}_{t}}$.
}
\end{algorithm}
Based on the AirComp-based communication model and second-order optimization algorithm, we propose our over-the-air second-order federated algorithm, as shown in Algorithm \ref{ACCADE}.


%
%

\section{Theoretical Convergence Analysis}
\label{Theoretical Analysis}
In this section, we provide the convergence analysis of our proposed algorithm.
A major challenge of convergence analysis is to tackle the distortion of the descent direction vector caused by channel noise,
device selection, and the use of local Newton step.  
To address this issue, we study the impact of distortion with respect to these influencing factors. 
In particular, we exploit the idea of sketching to analyze the approximation of local gradients and Hessian matrices.
To better elaborate our analysis, some preliminaries are firstly presented.

\subsection{Preliminaries} 
The core of our proposed algorithm is using local Hessian matrices and local gradients, which is calculated through subsets of the total data set, to construct local Newton descent directions and aggregate them. 
This brings the benefits of fewer communication rounds between the server and the devices.
However, since we rely on local information to approximate Newton descent directions, the quality of local Hessian/gradients, in other words, the difference between the local ones and global ones, are of concern.
In order to tackle this issue, we adopt the idea of matrix sketching~\cite{drineas2016randnla, woodruff2014sketching}.
Specifically, for a given input matrix $\bm{M} \in \mathbb{R}^{n \times d}$, we can replace it with $\bm{C} = \bm{L}^{\transpose}\bm{M} \in \mathbb{R}^{s\times d}$, where matrix $\bm{C}$ acts as the sketch of $\bm{M}$ with the sketching matrix $\bm{L} \in \mathbb{R}^{n \times s}$.
In this way, the original problem related to $\bm{M}$ can be solved more efficiently using the smaller alternative matrix $\bm{C}$ without losing too much information.
The construction of the sketch is similar to the calculation of local Hessian/gradients, where we adopt partial information of the global Hessian/gradients to serve as the local Hessian/gradients.
In this paper, we consider the row sampling scheme in matrix sketching.
The sketch $\bm{C}$ is constructed by the uniform sampled and re-scaled subset of rows of $\bm{M}$ with sampling probability $\mathbb{P}\bracket{\bm{c}_{i} = \frac{\bm{m}_{j}}{\sqrt{sp}}} = p,\; p =\frac{1}{n}$, where $\bm{c}_{i}$ and $\bm{m}_{j}$ are the $i$-th row of $\bm{C}$ and $j$-th row of $\bm{M}$, respectively.
Consequently, the sketching matrix $\bm{L}$ has only one non-zero entry in each column, 
and we shall measure the difference between the local Hessian/gradients and global ones with the help of such sketching matrices.

In the following, we consider a linear predictor model $\ell: \mathbb{R} \rightarrow \mathbb{R}$, which is frequently used in machine learning research, e.g., logistic and linear regression, support vector machines, neural networks and graphical models.
The function $f\bracket{\bm{w}, \bm{z}_{i,j}}$ can thus be rewritten as $\ell\bracket{\bm{w}^{\transpose}\bm{u}_{i,j}}$.
Accordingly, we define $\boldsymbol{M}_{t}=\left[\boldsymbol{m}_{1}^{\transpose}, \ldots, \boldsymbol{m}_{n}^{\transpose}\right]^{\transpose} \in \mathbb{R}^{n \times d}$ with $\bm{m}_{t,j} = \sqrt{\ell^{\prime\prime}\bracket{\bm{w}^{\transpose}\bm{u}_{i,j}} / n}\bm{u}_{i,j} \in \mathbb{R}^{d}$, so the global Hessian matrix can be represented as $\boldsymbol{H}_{t}=\boldsymbol{M}_{t}^{\transpose} \boldsymbol{M}_{t}+\gamma \boldsymbol{I}_{d}$.
Moreover, by defining $\bm{N}_{t} = \left[\bm{n}_{1}, \ldots, \bm{n}_{n}\right] \in \mathbb{R}^{d \times n}$ with $\bm{n}_{i}= \nabla f \left(\bm{w}_{t}, \bm{z}_{i}\right)$, the global gradient $\nabla F\left(\bm{w}_{t}\right)$ can be denoted by $\bm{g}_{t}=\frac{1}{n} \bm{N}_{t} \bm{1}+\gamma \bm{w}_{t}$. 
Let $\left\{\bm{L}_{i}\right\}_{i = 1}^{m}$ be the sketching matrices, the local Hessian matrices and local gradients can be reformulated as:
\begin{small}
\begin{equation}
    \bm{H}_{t, i}=\bm{M}_{t}^\mathsf{T} \bm{L}_{i} \bm{L}_{i}^\mathsf{T} \bm{M}_{t}+\gamma \bm{I}_{d} \;,\quad \bm{g}_{t, i}=\frac{1}{n} \bm{N}_{t} \bm{L}_{i} \bm{L}_{i}^\mathsf{T} \bm{1}+\gamma \bm{w} \;.
\end{equation}
\end{small}
In addition, we define an auxiliary quadratic function as follows to facilitate our analysis:
\begin{small}
\begin{equation}
    \phi(\bm{p})=\frac{1}{2} \bm{p}^\mathsf{T} \bm{H}_{t} \bm{p}-\bm{g}_{t}^\mathsf{T} \bm{p}=\frac{1}{2} \bm{p}^\mathsf{T}\left(\bm{M}_{t}^\mathsf{T} \bm{M}_{t}+\gamma \bm{I}_{d}\right) \bm{p}-\bm{g}_{t}^\mathsf{T} \bm{p} \;.
    \label{quadratic function}
\end{equation}
\end{small}
As a quadratic function, the minimum point of $\phi(\bm{p})$ denoted by $\bm{p}^{*}$ can be analytically obtained, which is the same as the exact Newton descent direction vector in \eqref{Newton step}, i.e.,
\begin{small}
\begin{equation}
    \bm{p}^{*}=\arg \min \phi(\bm{p})= \nabla^{2}F^{-1}\left(\bm{w}_{t}\right)\nabla F\left(\bm{w}_{t}\right) = \bm{H}_{t}^{-1} \bm{g}_{t}=\left(\bm{M}_{t}^{\top} \bm{M}_{t}+\gamma \bm{I}_{d}\right)^{-1} \bm{g}_{t} \;.
\end{equation}
\end{small}
Due to the effect of channel noise, device selection, and the use of local Newton step, the actual descent direction rather than the exact Newton step $\bm{p}^{*}$ is given by:
\begin{small}
\begin{equation}
    \begin{aligned}
        \hat{\bm{p}}_{t} =\,& \bm{p}^{*} + \underbrace{\left(\bar{\bm{p}}_{t} - \bm{p}^{*}\right)}_{\text{Local Hessian}} + \underbrace{\left(\bm{p}_{t} - \bar{\bm{p}}_{t}\right)}_{\text{Local Gradient}} + \underbrace{\left(\tilde{\bm{p}}_{t} - \bm{p}_{t}\right)}_{\text{Device Selection}} + \underbrace{\left(\hat{\bm{p}}_{t} - \tilde{\bm{p}}_{t}\right)}_{\text{Channel Noise}} \;,
    \end{aligned}
    \label{decomposition_global_update_direction}
\end{equation}
\end{small}
where $\hat{\bm{p}}_{t}$ and $\tilde{\bm{p}}_{t}$ are defined in~\eqref{global_update_direction} and~\eqref{averaged_local_descent_direction}, respectively, 
$\bm{p}_{t} = \frac{1}{\underset{i \in \mathcal{S}}{\sum}\card{\mathcal{D}_{i}}}\sum_{i \in \mathcal{S}} \card{\mathcal{D}_{i}}\bm{p}_{t,i}$ is the averaged local descent direction vector without device selection, and $\bar{\bm{p}}_{t} = \frac{1}{\underset{i \in \mathcal{S}}{\sum}\card{\mathcal{D}_{i}}}\sum_{i \in \mathcal{S}} \card{\mathcal{D}_{i}}\bar{\bm{p}}_{t,i} = \frac{1}{\underset{i \in \mathcal{S}}{\sum}\card{\mathcal{D}_{i}}}\sum_{i \in \mathcal{S}} \card{\mathcal{D}_{i}}\bm{H}_{t,i}^{-1}\bm{g}_{t}$ is the Newton descent direction with the exact global gradient.
In the following analysis, we will use the quadratic function~\eqref{quadratic function} to illustrate how close $\hat{\bm{p}}_{t}$ and $\bm{p}^{*}$ are.
Besides, the error of model parameter vector in the iterates $\bm{\Delta}_{t} = \bm{w}_{t} - \bm{w}^{*}$ acts as the metric, where $\bm{w}^{*}$ denotes the optimal solution. 

Throughout this paper, we consider the following assumptions, which are widely adopted in FL problems \cite{chen2020joint, liu2020reconfigurable, xiaowen2021optimized}.

\noindent\textbf{Assumption 1.} The global loss function $F$ is $L$-smooth.

\noindent\textbf{Assumption 2.} The global loss function $F$ is strongly convex, which indicates the unique optimal model parameter vector $\bm{w}^{*}$ of the FL task.

\noindent\textbf{Assumption 3.} The local loss function $F_{i}$ is twice-differentiable, smooth and convex.



\subsection{Convergence Analysis}
Since the local gradients and Hessian matrices are adopted to approximate the global descent direction, the gap between the local direction and the global direction is essential for the convergence analysis. Therefore, we first recall two lemmas to reveal their relationships.

\begin{lemma}[{\cite[variant of Lemma 2]{ghosh2020distributed}}]
Let $\lambda , \delta = \sum_{i=1}^{m} \delta_{i}, \{\delta_{i}\} \in(0,1)$ be fixed parameters, $r=\operatorname{rank}\left(\bm{M}_{t}\right)$, and $\bm{U} \in \mathbb{R}^{n \times r}$ be the orthonormal bases of the matrix $\bm{M}_{t}$.
Let $\mu \in\left[1, \frac{n}{d}\right]$ be the coherence of $\bm{M}_{t}$ defined in~[30].
Let $\left\{\bm{L}_{i} \in \mathbb{R}^{n \times \card{\mathcal{D}_{i}}}\right\}_{i=1}^{m}$ be independent uniform sampling sketching matrices with $\card{\mathcal{D}_{i}} \geq \frac{3 \mu d}{\lambda^{2}} \log \frac{d}{\delta_{i}}$.
It holds with the probability exceeding $1-\delta$ that:
\begin{small}
\begin{equation}
    \left\|\bm{U}^\mathsf{T} \bm{L}_{i} \bm{L}_{i}^\mathsf{T} \bm{U}-\bm{I}\right\|_{2} \leq \lambda\;, \quad \forall i \in \mathcal{S}\;.
\end{equation}
\end{small}
\end{lemma}

\begin{lemma}[{\cite[variant of Lemma 3]{ghosh2020distributed}}]
Let $\left\{\bm{L}_{i} \in \mathbb{R}^{n \times \card{\mathcal{D}_{i}}}\right\}_{i=1}^{m}$ be independent uniform sampling sketching matrices, $\delta = \sum_{i=1}^{m} \delta_{i}, \{\delta_{i}\} \in(0,1)$ be fixed parameters, then with the probability exceeding $1-\delta$, we have:
\begin{small}
\begin{equation}
    \left\|\frac{1}{n} \bm{N}_{t}\bm{L}_{i} \bm{L}_{i}^\mathsf{T} \bm{1}-\frac{1}{n} \bm{N}_{t} \bm{1}\right\| \leq\left(1+\sqrt{2 \ln \left(\frac{1}{\delta_{i}}\right)}\right) \sqrt{\frac{1}{\card{\mathcal{D}_{i}}}} \max _{j}\left\|\bm{n}_{j}\right\|\;.
\end{equation}
\end{small}
\end{lemma}

With Lemma 1 and Lemma 2, we further propose Lemma 3 to characterize the gap between $\hat{\bm{p}}_{t}$ and $\bm{p}^{*}$ via the support quadratic function.

\begin{lemma}
Let $\left\{\bm{L}_{i}\right\}_{i=1}^{m} \in \mathbb{R}^{n \times \card{\mathcal{D}_{i}}}$ be independent uniform sampling sketching matrices, $\phi_{t}$ be the quadratic function as defined in~\eqref{quadratic function}, $\lambda , \{\delta_{i}\} \in(0,1)$ be fixed parameters with $\tilde{\delta} = \min\{\delta_{i}\}$ and $\hat{\bm{p}}_{t}$ be the approximate descent direction vector defined in~\eqref{decomposition_global_update_direction}.
It holds that:
$$
 \phi_{t}(\bm{p}^{*}) \leq \phi_{t}\left(\hat{\bm{p}}_{t}\right) \leq \epsilon^{2} +\left(1-\zeta^{2}\right) \phi_{t}(\bm{p}^{*}) \;,
$$
where
\begin{small}
\begin{equation}
    \zeta^{2} = 3\tau^{2}\left(\lambda+\frac{\lambda^{2}}{1-\lambda}\right)^{2} + 24\vartheta^{2}\left(\tau\left(\lambda+\frac{\lambda^{2}}{1-\lambda}\right)+1\right)^{2}
    \label{zeta}
\end{equation} 
\end{small}
with $\tau=\frac{\sigma_{\max }\left(\bm{M}^{\top} \bm{M}\right)}{\sigma_{\max }\left(\bm{M}^{\top} \bm{M}\right)+n\gamma}$, $\vartheta = \max _{t} \left(1 - \frac{\sum_{i \in \mathcal{S}_{t}}\card{\mathcal{D}_{i}}}{n}\right) < 1$ and
\begin{small}
\begin{equation}
    \begin{aligned}
        \epsilon^{2}
        =& \frac{3}{\sigma_{\min}\left(\bm{H}_{t}\right)}\left\|\frac{1}{\bracket{\sum_{i \in \mathcal{S}_{t}}\card{\mathcal{D}_{i}}}\sqrt{\eta_{t}}}\bm{a}_{t}^\mathsf{H}\bm{E}_{t}\right\|^{2} + \left[24\left(1 - \frac{\sum_{i \in \mathcal{S}_{t}}\card{\mathcal{D}_{i}}}{n}\right)^{2}\frac{1}{\min_{i\in\mathcal{S}_{t}}\card{\mathcal{D}_{i}}}+\frac{m}{n}\right] \\
        &\qquad\qquad\qquad\cdot \bigg[\frac{1}{1-\lambda} \frac{1}{\sqrt{\sigma_{\min }\left(\bm{H}_{t}\right)}}\left(1+\sqrt{2 \ln \left(\frac{1}{\tilde{\delta}}\right)}\right)  \max _{j} \left\| \bm{n}_{j} \right\|\bigg]^{2}.
    \end{aligned}
\label{epsilon}
\end{equation}
\end{small}
\end{lemma}

The proof of Lemma~3 can be found in Appendix A. 
To illustrate that  $\hat{\bm{p}}_{t}$ is a good descending direction, we introduce Lemma 4 supported by the property of the quadratic function introduced in Lemma 3.
\begin{lemma}[{\cite[Lemma~6]{ghosh2020distributed}}] 
Let $\zeta \in(0,1)$, $\epsilon$ be any fixed parameter, if $\hat{\bm{p}_{t}}$ satisfies $\phi\left(\hat{\bm{p}}_{t}\right) \leq \epsilon^{2}+(1-\left.\zeta^{2}\right) \min _{\bm{p}} \phi(\bm{p})$, then under Assumption 1, the error of model parameter vector $\bm{\Delta}_{t}=\bm{w}_{t}-\bm{w}^{*}$ in iterations satisfies
\begin{small}
\begin{equation}
    \bm{\Delta}_{t+1}^\mathsf{T} \bm{H}_{t} \bm{\Delta}_{t+1} \leq L\left\|\bm{\Delta}_{t+1}\right\|\left\|\bm{\Delta}_{t}\right\|^{2}+\frac{\zeta^{2}}{1-\zeta^{2}} \bm{\Delta}_{t}^\mathsf{T} \bm{H}_{t} \bm{\Delta}_{t}+2 \epsilon^{2} \;,
\end{equation}
\end{small}
\end{lemma}
Based on Lemma 3 and Lemma 4, we can derive the main result:
\begin{theorem}
\label{theorem 1}
Suppose the size of local dataset at each device $\card{\mathcal{D}_{i}} \geq \frac{3 \mu d}{\lambda^{2}} \log \frac{d}{\delta_{i}}$ for some $\lambda, \delta_{i} \in(0,1)$, then under Assumption 1 with the probability exceeding $1-\delta$ we have 
\begin{small}
\begin{align*}
    \mathbb{E}\left(\left\|\bm{\Delta}_{t+1}\right\|\right) \leq \max \left\{\sqrt{\kappa_{t}\left(\frac{\zeta^{2}}{1-\zeta^{2}}\right)}\left\|\bm{\Delta}_{t}\right\|, \frac{L}{\sigma_{\min }\left(\bm{H}_{t}\right)}\left\|\bm{\Delta}_{t}\right\|^{2}\right\} +\epsilon^{\prime}\;,
\end{align*}
\end{small}
where the expectation takes with respect to the channel noise $\bm{e}_{t}$, $\zeta$ is defined as \eqref{zeta},
$\kappa_{t}=\frac{\sigma_{\max }\left(\bm{H}_{t}\right)}{\sigma_{\min }\left(\bm{H}_{t}\right)}$ denotes the condition number of $\bm{H}_{t}$, 
and 
\[
\begin{small}
\begin{aligned}
    \epsilon^{\prime} =& \frac{2\sqrt{3}}{\sigma_{\min}\left(\bm{H}_{t}\right)}\frac{d\sigma}{\sum_{i\in\mathcal{S}_{t}}\card{\mathcal{D}_{i}}}\frac{\norm{\bm{a}_{t}}}{\sqrt{\eta_{t}}} + \sqrt{24\left(1 - \frac{\sum_{i\in\mathcal{S}_{t}}\card{\mathcal{D}_{i}}}{n}\right)^{2}\frac{1}{\min_{i\in\mathcal{S}_{t}}\card{\mathcal{D}_{i}}}+\frac{m}{n}}\\
    &\qquad\qquad\qquad\qquad\cdot\frac{1}{1-\lambda} \frac{2}{\sigma_{\min }\left(\bm{H}_{t}\right)}\left(1+\sqrt{2 \ln \left(\frac{1}{\tilde{\delta}}\right)}\right)\max _{j} \left\| \bm{n}_{j} \right\|\;.
\end{aligned}
\end{small}
\]
\end{theorem}
%
The proof can be found in Appendix B. From Theorem \ref{theorem 1}, we have the following observations.
\subsubsection{\bf The proposed algorithm keeps a linear-quadratic convergence rate}
From the analysis results, it can be seen that the term $\left\|\bm{\Delta}_{t}\right\| = \left\|\bm{w}_{t} - \bm{w}^{*}\right\|$ keeps the property in this form:
$\mathbb{E}\left(\left\|\bm{\Delta}_{t+1}\right\|\right) \leq \max \left\{\omega_{1}\left\|\bm{\Delta}_{t}\right\|, \omega_{2}\left\|\bm{\Delta}_{t}\right\|^{2}\right\} + \epsilon^{\prime}$. 
When $\left\|\bm{\Delta}_{t}\right\| > \frac{\omega_{1}}{\omega_{2}}$, this property can be simplified as $ \mathbb{E}\left(\left\|\bm{\Delta}_{t+1}\right\|\right) \leq \omega_{2}\left\|\bm{\Delta}_{t}\right\|^{2} + \epsilon^{\prime} $.
It is obvious that the proposed algorithm keeps the same quadratic convergence rate as the canonical Newton's method. 
At the beginning of the algorithm it can converge to the neighbor of the optimal point quickly. 
When $\left\|\bm{\Delta}_{t}\right\| < \frac{\omega_{1}}{\omega_{2}}$, this property turns into $ \mathbb{E}\left(\left\|\bm{\Delta}_{t+1}\right\|\right) \leq \omega_{1}\left\|\bm{\Delta}_{t}\right\| + \epsilon^{\prime} $, 
which means when $\left\|\bm{\Delta}_{t}\right\|$ is small enough during the process of the algorithm, it degenerates into the linear convergence rate. 
In conclusion, the proposed algorithm keeps a linear-quadratic convergence rate and performs better than first-order algorithms.

\subsubsection{\bf The proposed algorithm is accompanied by an accumulative error term} Notice that there is an error term $\epsilon^{\prime}$ in each iteration, which comes from the approximation, device selection and channel noise. 
Consider the noise-free case without device selection, which means $\sigma = 0$ and $\left|\mathcal{S}_{t}\right| = m$, this error term degenerates to:
\begin{small}
\begin{equation}
    \begin{aligned}
    \epsilon^{\prime}_{1} = 
    \frac{1}{1-\lambda} \frac{2}{\sigma_{\min }\left(\bm{H}_{t}\right)}\left(1+\sqrt{2 \ln \left(\frac{1}{\tilde{\delta}}\right)}\right)\sqrt{\frac{m}{n}}\max _{j} \left\| \bm{n}_{j} \right\|\;,
\end{aligned}
\label{epsilon1}
\end{equation}
\end{small}
which is exactly the same as the error term introduced in~\cite{ghosh2020distributed}.
With the algorithm executed iteratively, the gap between the expected global loss function value and the optimal one is upper bounded by this accumulative error term.
Therefore, the active device set $\mathcal{S}_{t}$, the receiver beamforming vectors $\left\{\bm{a}_{t}\right\}$ and the scaling factors $\left\{\eta_{t}\right\}$ need to be tuned in each iteration so as to reduce the error gap.

\section{System Optimization}
\label{System Optimization}
In this section, we first formulate a system optimization problem to minimize the error term in the convergence analysis results.
Then, we propose our approach for joint optimization of device selection and receiver beamforming vector.

\subsection{Problem Formulation}
In light of convergence analysis, to obtain a precise model parameter vector, minimizing the error gap demonstrated in Theorem 1 is a key issue. 
It is observed that the coefficients of the two iterative terms $\bm{\Delta}_{t}$ and $\bm{\Delta}_{t+1}$ in Theorem 1 are independent of variables $\mathcal{S}_{t}$, $\bm{a}_{t}$ and $\eta_{t}$. 
Therefore, in order to achieve the minimization of the total error gap, we only need to minimize the error term $\epsilon^{\prime}$ in each iteration as follows 
\begin{small}
\begin{equation}
\label{power constraint}
\begin{aligned}
    \min_{\mathcal{S}_{t}, \bm{a}_{t}, \eta_{t}} 
    \quad&\frac{2\sqrt{3}}{\sigma_{\min}\left(\bm{H}_{t}\right)}\frac{d\sigma}{\sum_{i\in\mathcal{S}_{t}}\card{\mathcal{D}_{i}}}\frac{\norm{\bm{a}_{t}}}{\sqrt{\eta_{t}}} + \sqrt{24\left(1 - \frac{\sum_{i\in\mathcal{S}_{t}}\card{\mathcal{D}_{i}}}{n}\right)^{2}\frac{1}{\min_{i\in\mathcal{S}_{t}}\card{\mathcal{D}_{i}}}+\frac{m}{n}}\\
    &\qquad\qquad\qquad\qquad\cdot\frac{1}{1-\lambda} \frac{2}{\sigma_{\min }\left(\bm{H}_{t}\right)}\left(1+\sqrt{2 \ln \left(\frac{1}{\tilde{\delta}}\right)}\right)\max _{j} \left\| \bm{n}_{j} \right\| \\
    \text{s.t.}\;\quad  
    & \frac{\eta_{t}}{\left\|\bm{a}_{t}^\mathsf{H}\tilde{\bm{h}}_{t,i}\right\|^{2}} \leq dP_{0} \quad \forall i \in \mathcal{S}_{t} \;
\end{aligned}
\end{equation}
\end{small}
The power constraint in \eqref{power constraint} can be rewritten in the form of the restriction of scaling factor: $\eta_{t} \leq dP_{0}\left\|\bm{a}_{t}^\mathsf{H}\tilde{\bm{h}}_{t,i}\right\|^{2}, \ \forall i \in \mathcal{S}_{t}$. 
We take the negative correlation between the scaling factor $\eta_{t}$ and the objective function value $\epsilon^{\prime}$ into consideration.
$\eta_{t}$ can be set as $\eta_{t} = dP_{0}\min_{i \in \mathcal{S}_{t}}\left\|\bm{a}_{t}^\mathsf{H}\tilde{\bm{h}}_{t,i}\right\|^{2}$~\cite{chen2018uniform}, and the problem can be simplified as $\mathscr{P}$:
\begin{small}
\begin{equation}
\begin{aligned}
    \mathscr{P}: \min_{\mathcal{S}_{t}, \bm{a}_{t}}\quad
     & \frac{\sqrt{3d}\sigma}{\sqrt{P_{0}}\sum_{i\in\mathcal{S}_{t}}\card{\mathcal{D}_{i}}}\;\; \max_{i \in \mathcal{S}_{t}}\;\; \left(\frac{\left\|\bm{a}_{t}\right\|}{\left\|\bm{a}_{t}^\mathsf{H}\tilde{\bm{h}}_{t,i}\right\|}\right) + \sqrt{24\left(1 - \frac{\sum_{i\in\mathcal{S}_{t}}\card{\mathcal{D}_{i}}}{n}\right)^{2}\frac{1}{\min_{i\in\mathcal{S}_{t}}\card{\mathcal{D}_{i}}}+\frac{m}{n}}\\
    &\qquad\qquad\qquad\qquad\cdot\frac{1}{1-\lambda} \frac{2}{\sigma_{\min }\left(\bm{H}_{t}\right)}\left(1+\sqrt{2 \ln \left(\frac{1}{\tilde{\delta}}\right)}\right)\max _{j} \left\| \bm{n}_{j} \right\|\;.
    \end{aligned}
    \label{system_optimization_problem}
\end{equation}
\end{small}
We have the following key observations for solving \eqref{system_optimization_problem}:
\begin{itemize}
    \item [\textbullet] Intuitively, to achieve the minimization of the objective value of $\mathscr{P}$ , the number of selected devices is supposed to be maximized, then $\mathscr{P}$ will degenerate into the form of traditional beamforming optimization. However, the term $\max_{i \in \mathcal{S}_{t}}\; \left(\frac{\left\|\bm{a}_{t}\right\|}{\left\|\bm{a}_{t}^\mathsf{H}\tilde{\bm{h}}_{t,i}\right\|}\right)$
    is related to device selection, which further
    results in the incorrectness to directly maximize $|\mathcal{S}_{t}|$.
    \item [\textbullet] By searching over all the possible participating device sets, the optimal $\mathcal{S}_{t}$ can be determined. Still, the number of devices $m$ can be very large, leading to an exponential growth of the optimization procedure in the number of devices $m$.
    \item [\textbullet] After the search of participating devices, the remaining problem is a typical beamforming optimization problem, but it is still non-convex and intractable.
\end{itemize}

In conclusion, since a combinatorial search of participating devices and minimization of the non-convex objective function are involved, it is evident that $\mathscr{P}$ is a mixed-integer non-convex problem.
In order to tackle the complexity of computation and the difficulty of non-convexity, we propose an efficient method to iteratively search the optimal set of selected devices $\mathcal{S}_{t}$ while jointly optimizing the receiver beamforming vector $\bm{a}_{t}$ for each given $\mathcal{S}_{t}$.

\subsection{Receiver Beamforming Optimization}
For a given set of selected devices $\mathcal{S}_{t}$, $\mathscr{P}$ can be simplified as $\mathscr{P}_{1}: \min_{\bm{a}_{t}} \max_{i \in \mathcal{S}_{t}} \frac{\left\|\bm{a}_{t}\right\|}{\left\|\bm{a}_{t}^\mathsf{H}\tilde{\bm{h}}_{t,i}\right\|}$,
which is equivalent to:
$ \min_{\bm{a}_{t}} \max_{i \in \mathcal{S}_{t}} \frac{\left\|\bm{a}_{t}\right\|^{2}}{\left\|\bm{a}_{t}^\mathsf{H}\tilde{\bm{h}}_{t,i}\right\|^{2}}\;.$
This can be further reformulated as $\mathscr{P}_{1}^{\prime}$ according to the analysis in~\cite{chen2018uniform}:
\begin{small}
\begin{align*}
    \mathscr{P}_{1}^{\prime}:  \min_{\bm{a}_{t}} \quad \left\|\bm{a}_{t}\right\|^{2} \qquad \text{s.t.} \quad \left\|\bm{a}_{t}^\mathsf{H}\tilde{\bm{h}}_{t,i}\right\|^{2} \geq 1 \quad \forall i \in \mathcal{S}_{t}\;.
\end{align*}
\end{small}
It can be seen that $\mathscr{P}_{1}^{\prime}$ is actually a quadratically constrained quadratic programming problem, which is difficult to solve.
We first use the matrix lifting technique to pre-process $\mathscr{P}_{1}^{\prime}$ and turn it into a low-rank optimization form.
Specifically, let $\bm{A} = \bm{a}_{t}\bm{a}_{t}^\mathsf{H}$ with $\text{rank}\left(\bm{A}\right)=1$ and $\bm{Q}_{i}=\tilde{\bm{h}}_{t,i}\tilde{\bm{h}}_{t,i}^\mathsf{H}$, $\mathscr{P}_{1}^{\prime}$ can be recast as:
\begin{small}
\begin{align*}
    \min_{\bm{A}} \quad  \operatorname{Tr}\left(\bm{A}\right)\qquad 
    \text{s.t.} \quad  \bm{A} \succeq \bm{0} ,\;\;\operatorname{rank}(\bm{A})=1,\;\; \operatorname{Tr}\left(\bm{A}\bm{Q}_{i}\right) \ge 1 \quad \forall i \in \mathcal{S}_{t}\;.
\end{align*}
\end{small}
The key to solving this low-rank optimization problem is to deal with the troublesome rank-one constraint. 
A common method to solve such a problem is semidefinite relaxation (SDR)~\cite{luo2007approximation, sidiropoulos2006transmit}, which drops the rank-one constraint to obtain a relaxed problem in the form of semidefinite programming. 
By this means, SDR can arrive at an approximate solution efficiently through solving the relaxed problem. 
However, as the size of the problem grows, the rank-one constraint is usually unsatisfied. 
In this situation, the approximate solution needs to be scaled through randomization methods, leading to an alternative solution with low accuracy~\cite{chen2018uniform}, which will further affect the learning performance of FL.
To guarantee the rank-one constraint, we can replace it with its equivalent form~\cite{yang2020federated, hua2019device}:
$
\operatorname{Tr}\left(\bm{A}\right) - \left\|\bm{A}\right\|_{2} = 0 \quad \text{with} \quad \operatorname{Tr}\left(\bm{A}\right) > 0\;.
$
Then, the original problem turns into a difference-of-convex-function (DC) program.
By solving this DC program, a more precise solution can be obtained since all constraints are satisfied.
Therefore, we develop a DC Algorithm (DCA) based on the principles in~\cite{tao1997convex, khamaru2018convergence} to solve this
problem.
Specifically, we can get the following problem by taking the new constraint as a penalty term:
\begin{small}
\begin{align*}
    \min_{\bm{A}} \quad  \operatorname{Tr}\left(\bm{A}\right)+\theta \left(\operatorname{Tr}\left(\bm{A}\right) - \left\|\bm{A}\right\|_{2}\right) \quad
    \text{s.t.} \quad  \bm{A} \succeq \bm{0}, \; \operatorname{Tr}(\bm{A})>0,\; \operatorname{Tr}\left(\bm{A}\bm{Q}_{i}\right) \ge 1 \;\; \forall i \in \mathcal{S}_{t} \; ,
\end{align*}
\end{small}
where $\theta$ is the penalty factor. Although this is still a non-convex problem owing to the concave term $- \left\|\bm{A}\right\|_{2}$, 
we can take the linearization of $\left\|\bm{A}\right\|_{2}$ and convert it into a convex subproblem:
\begin{small}
\begin{align*}
    \mathscr{P}_{DCA}:\min_{\bm{A}} \ & (1+\theta) \operatorname{Tr}(\bm{A})-\theta\left\langle\partial\left\|\bm{A}_{j}\right\|_{2}, \bm{A}\right\rangle \ \;
    \text{s.t.} \ \bm{A} \succeq \bm{0},\; \operatorname{Tr}(\bm{A})>0,\; \operatorname{Tr}\left(\bm{A}\bm{Q}_{i}\right) \ge 1 \ \forall i \in \mathcal{S}_{t},
\end{align*}
\end{small}
where $\langle\cdot,\cdot \rangle$ denotes the inner product of two matrices and $\partial\left\|\bm{A}_{j}\right\|_{2}$ represents the subgradient of $\left\|\bm{A}_{j}\right\|_{2}$ at $\bm{A}_{j}$.
Therefore, the result can be obtained by iteratively solving $\mathscr{P}_{DCA}$ until $\operatorname{Tr}\left(\bm{A}\right) - \left\|\bm{A}\right\|_{2}$ is sufficiently small. 
The overall procedure of DCA is as summarized in Algorithm \ref{DCA}.
\begin{algorithm}
    \caption{DC Algorithm for Receiver Beamforming Optimization (DCA)}
    \label{DCA}
    \small
    \textbf{input:} effective channel coefficients $\left\{\tilde{\bm{h}}_{t,i}\right\}$, penalty factor $\theta$, threshold $\xi $
    
    turn $\mathscr{P}$ into the DCA form $\mathscr{P}_{DCA}$.

    choose $\bm{A}_{0} \succeq 0$, set $j = 1$. 
    
    \While{$\left|\operatorname{Tr}\left(\bm{A}_{j-1}\right) - \left\|\bm{A}_{j-1}\right\|_{2}\right| \ge \xi$}{
        compute the subgradient $\partial\left\|\bm{A}_{j-1}\right\|_{2}$.

        substitute $\partial\left\|\bm{A}_{j-1}\right\|_{2}$ into $\mathscr{P}_{DCA}$, solve the subproblem and set the result as $\bm{A}_{j}$.

        $j \leftarrow j+1 \;.$ 
    }
  \end{algorithm}

\begin{algorithm}
    \caption{System optimization approach GS+DCA}
    \label{GS-DCA}
    \small
    \textbf{input:} effective channel coefficients $\left\{\tilde{\bm{h}}_{t,i}\right\}$, $T^{(0)}$, $\rho$, $K$

    \textbf{output:} $\mathcal{S}^{(k+1)}$ and its corresponding $\bm{a}^{(K+1)}$.

    \textbf{initialization:} $\mathcal{S}^{(0)} = \mathcal{S}$
    
    \For{iteration $k = 0, 1, 2, ... , K$}{
        generate the neighboring solution set $\mathcal{F}^{(k)}$.

        \For{each $\tilde{\mathcal{S}} \in \mathcal{F}^{(k)}$}{
            substitute $\tilde{\mathcal{S}}$ into $\mathscr{P}_{1}$, then solve the problem using DCA to get the corresponding optimal $\tilde{\bm{a}}$.
        }

        sample $\tilde{\mathcal{S}}^{(k)}$ according to the probability $\mathbb{P}\left(\tilde{\mathcal{S}}^{(k)}\right)= \frac{ \exp \left(-J\left(\tilde{\mathcal{S}}^{(k)}, \tilde{\bm{a}}^{(k)}\right)/T^{(k)}\right)}{\sum_{\tilde{\mathcal{S}} \in \mathcal{F}^{(k)}} \exp \left(-J\left(\tilde{\mathcal{S}}, \tilde{\bm{a}}\right)/T^{(k)}\right)}.$

        $\mathcal{S}^{(k+1)} \gets \tilde{\mathcal{S}}^{(k)},\;T^{(k+1)} \gets \rho T^{(k)}$.
        
    }
    
\end{algorithm}
\subsection{Device Selection Optimization}
As mentioned above, the device selection is a combinatorial optimization problem, which is impossible to perform a traversal in the whole solution space. Thus, we adopt the well-known Gibbs Sampling (GS) \cite{geman1984stochastic} method to optimize the selection of device set iteratively.
The main idea of GS is that in each iteration, a device set is sampled from the neighbors of the current device set according to an appropriate distribution.
In this way, the set of selected devices can gradually approach the global optimal solution. 

To be specific, we treat different sets of selected devices as states, and the goal is to find the state which can minimize the objective value in $\mathscr{P}$. 
For the sake of such state, at iteration $k$ of GS's process, with the set of selected devices $\mathcal{S}^{(k-1)}$ given in the last iteration, we first generate the neighboring solution set of $\mathcal{S}^{(k-1)}$.
The neighboring solution set, denoted by $\mathcal{F}^{(k)}$, contains the device sets that differ from the $\mathcal{S}^{(k-1)}$ in only one entry. 
For example, by assuming $\mathcal{S} = \{0, 1, 2\}$ and $\mathcal{S}^{(k-1)} = \{1, 2\}$, then we have $\mathcal{F}^{(k)} = \{\{0, 1, 2\}, \{2\}, \{1\}\}$.

After the identification of the neighboring solution set, the candidate states are also determined according to the sets in $\mathcal{F}^{(k)}$, and we need to choose a state to approach the optimal set. 
Based on the distribution introduced in \cite{bremaud2013markov}, we sample a device set in $\mathcal{F}^{(k)}$ with the probability
\begin{small}
\begin{equation}
    \mathbb{P}\left(\tilde{\mathcal{S}}^{(k)}\right)= \frac{ \exp \left(-J\left(\tilde{\mathcal{S}}^{(k)}, \tilde{\bm{a}}^{(k)}\right)/T^{(k)}\right)}{\sum_{\tilde{\mathcal{S}} \in \mathcal{F}^{(k)}} \exp \left(-J\left(\tilde{\mathcal{S}}, \tilde{\bm{a}}\right)/T^{(k)}\right)}\;,
    \label{Gibbs_distribution}
\end{equation}
\end{small}
where $J(\bm{x}, \bm{y})$ denotes the objective function value of $\mathscr{P}$. Here, the receiver beamforming vector is calculated through DCA with the given set of selected devices.

In the distribution \eqref{Gibbs_distribution}, there is a special parameter $T^{(k)}$ serving as the temperature. 
The algorithm starts from a relatively high temperature $T^{(0)}$ in order to move around the solution space freely, rather than being stuck in a local minimum point. 
As the algorithm proceeds, the algorithm slowly decreases the temperature by the factor $\rho$ to focus on the states that minimize the objective function.
Besides, to reduce the computational complexity, we have adopted a similar warm start technique as in~\cite{liu2020reconfigurable}.
The optimal beamforming vector in the previous iteration is used to serve as the initial point to accelerate the process of beamforming optimization.
The overall process of system optimization is outlined in Algorithm~\ref{GS-DCA}.

\section{Simulation Results}
\label{Simulation Results}
In this section, we evaluate the performance of the proposed schemes to demonstrate the advantage of our proposed second-order federated optimization algorithm and the effectiveness of our system optimization approach.
Code for our experiments are available at: https://github.com/Golden-Slumber/AirFL-2nd.
We first consider logistic regression with the loss function of the $i$-th device 
$ F_{i}\left(\bm{w}\right)=\frac{1}{|\mathcal{D}_{i}|} \sum_{\bm{z}_{i, j}=\left(\bm{u}_{i, j}, v_{i, j}\right) \in \mathcal{D}_{i}} \log \left(1+\exp \left(-v_{i, j} \bm{u}_{i, j}^\mathsf{T} \bm{w}\right)\right)+\frac{\gamma}{2}\|\bm{w}\|_{2}^{2}$, 
where the regularization parameter is set to be $\gamma = 10^{-8}$. 
As for datasets, we adopt four different standard datasets from the LIBSVM library: Covtype, a9a, w8a, and phishing.
In this paper, we consider a distributed wireless scenario, where these data samples are uniformly distributed in $m=20$ devices, the server is equipped with $k=5$ antennas. 
The channel coefficients are given by the small-scale fading coefficients $\{\bm{h}_{t,i}^{\prime}\}$ multiplied by the path loss gain $PL_{i}$, i.e., $\bm{h}_{t,i} = PL_{i}\bm{h}_{t,i}^{\prime}$.
Here, the small-scale fading coefficients follow the i.i.d complex normal distribution $\mathcal{CN}\bracket{0,\bm{I}}$.
The path loss gain is given by $PL_{i} = \sqrt{G_{0}}\bracket{d_{0}/d_{i}}^{\nu/2}$, where $G_{0} = 10^{-3.35}$ is the average channel power gain with the distance to the server $d_{0} = 1$ m, $d_{i} \in [100, 120]$  stands for the distance between the $i$-th device and the server, and $\nu  = 3.76$ represents the path loss exponent factor.
For the step size $\alpha$, we use backtracking line search to find $\alpha$ satisfying the Armijo–Goldstein condition~\cite[Chapter~3]{nocedal2006numerical}. 
For the system optimization, we set $\lambda = 0.1$, $\tilde{\delta} = 0.01$, penalty factor $\theta = 1$, threshold $\xi = 10^{-10}$, initial temperature $T_{0} = 100$, $\rho=0.9$, and $K=30$.
Besides, we use Baseline 0 to denote the centralized training setting in all experiments.

Furthermore, we also consider an image classification problem on a non-i.i.d dataset constructed from the Fashion-MNIST dataset at the end of this section. 
To address it, we train a softmax classifier with cross-entropy loss and $\ell_{2}$ regularization term. 
To be specific, the loss function of the $i$-th device is given as 
$F_{i}\left(\bm{W}\right)=\frac{1}{|\mathcal{D}_{i}|} \sum_{\left(\bm{u}, v\right) \in \mathcal{D}_{i}} \sum_{c=1}^{C} \bm{1}\{v = c\} \log \frac{\exp (\bm{u}^\mathsf{T} \bm{w}_{c})}{\sum_{j=1}^{C}\exp(\bm{u}^\mathsf{T} \bm{w}_{j})}+\frac{\gamma}{2}\sum_{c=1}^{C}\|\bm{w}_{c}\|_{2}^{2}$, 
where $\bm{W} = [\bm{w}_{1}, \dots, \bm{w}_{C}]$ is the concatenation of parameter vectors related to different classes, and $C = 10$ represents the total number of classes.


\subsection{Comparison with First-Order Algorithms}
We compared our proposed algorithm with two existing AirComp-based first-order algorithms in this experiment, where SNR is set to 80 dB:
\begin{enumerate}
    \item Baseline $1$: AirComp-based Federated Averaging (FedAvg) algorithm with DC-based optimization framework \cite{yang2020federated}, where the threshold of MSE is set to $5$ dB.
    
    \item Baseline $2$: AirComp-based Fedsplit algorithm \cite{xia2020fast}, where the threshold for device selection is set to $0.5$. 

\end{enumerate}
\begin{figure*}[htbp!]
    \centering
    \subfloat[Covtype]{
        \begin{minipage}[c][1\width]{0.23\textwidth}
            \centering
            \includegraphics[width=1\linewidth]{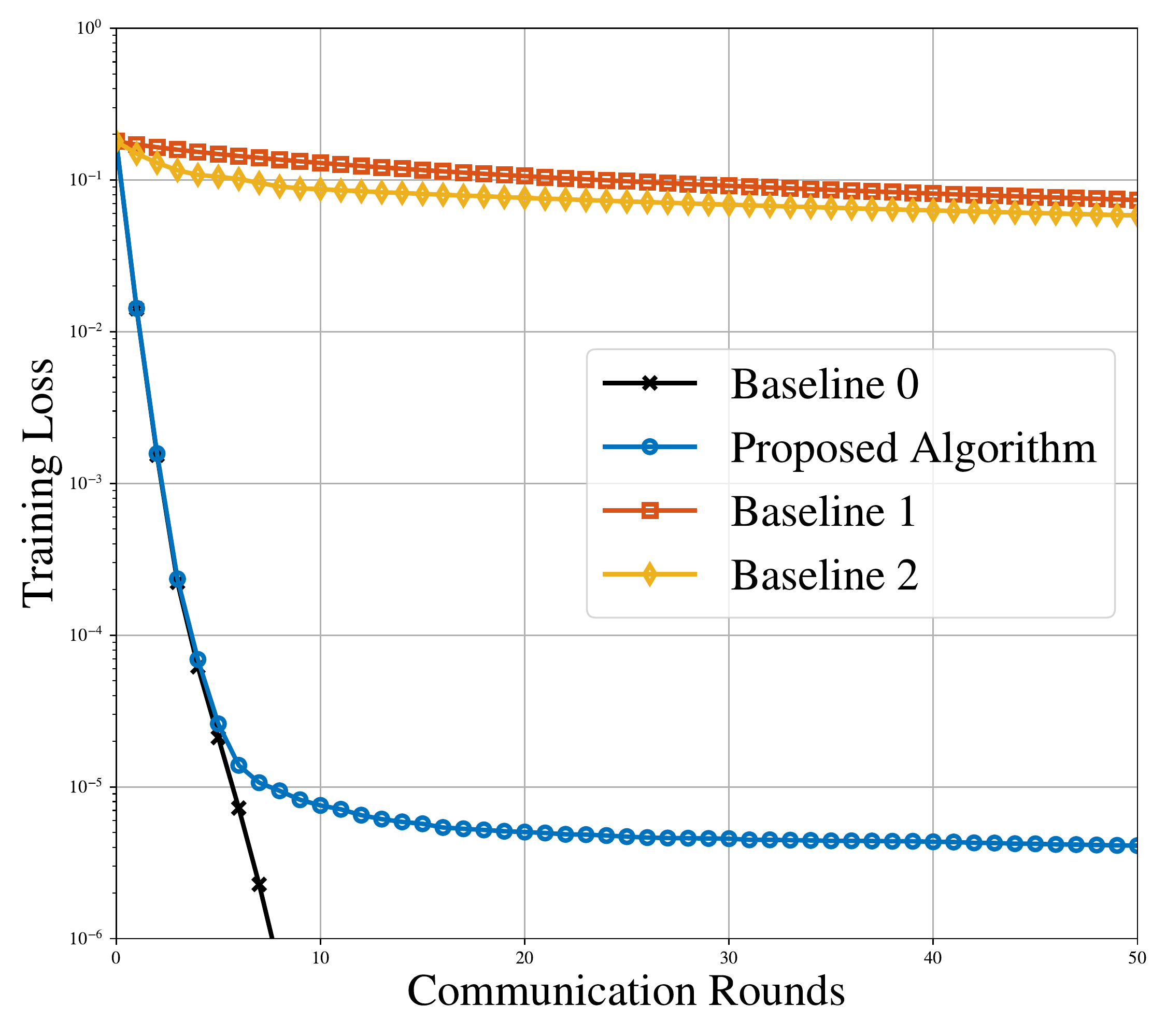}
        \end{minipage}
    }
    \hfill
    \subfloat[a9a]{
        \begin{minipage}[c][1\width]{0.23\textwidth}
            \centering
            \includegraphics[width=1\linewidth]{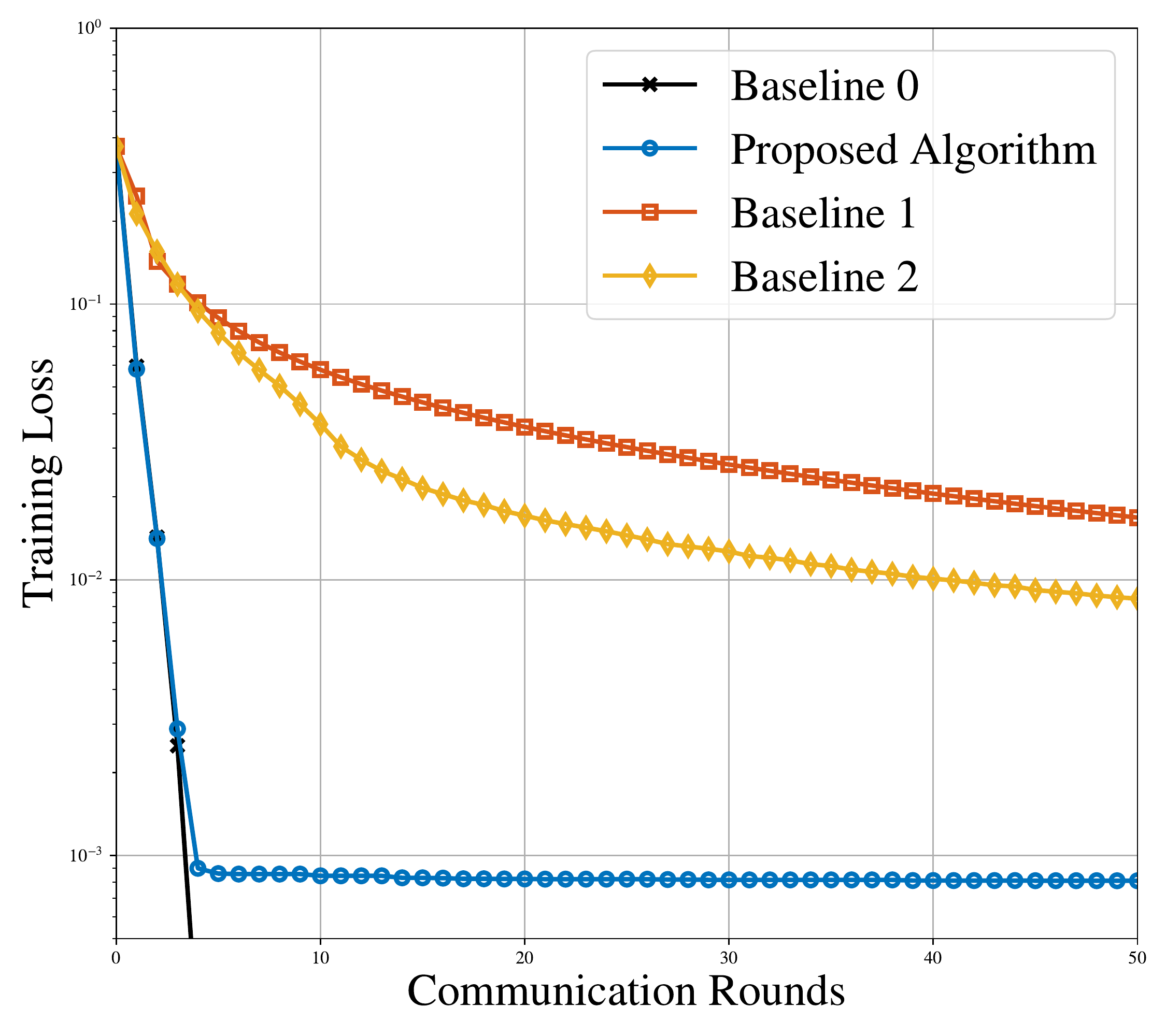}
        \end{minipage}
    }
    \hfill
    \subfloat[w8a]{
        \begin{minipage}[c][1\width]{0.23\textwidth}
            \centering
            \includegraphics[width=1\linewidth]{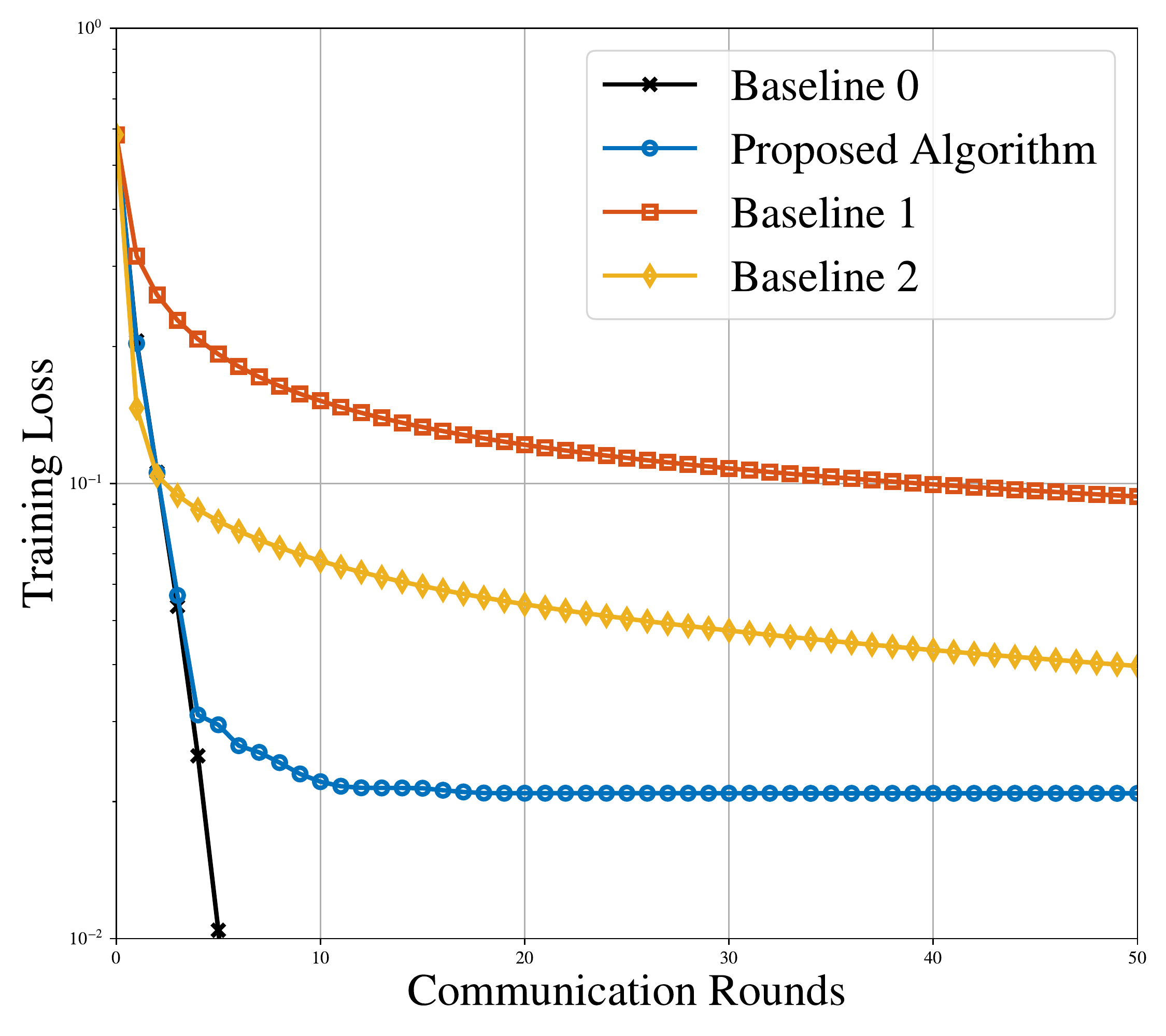}
        \end{minipage}
    }
    \hfill
    \subfloat[phishing]{
        \begin{minipage}[c][1\width]{0.23\textwidth}
            \centering
            \includegraphics[width=1\linewidth]{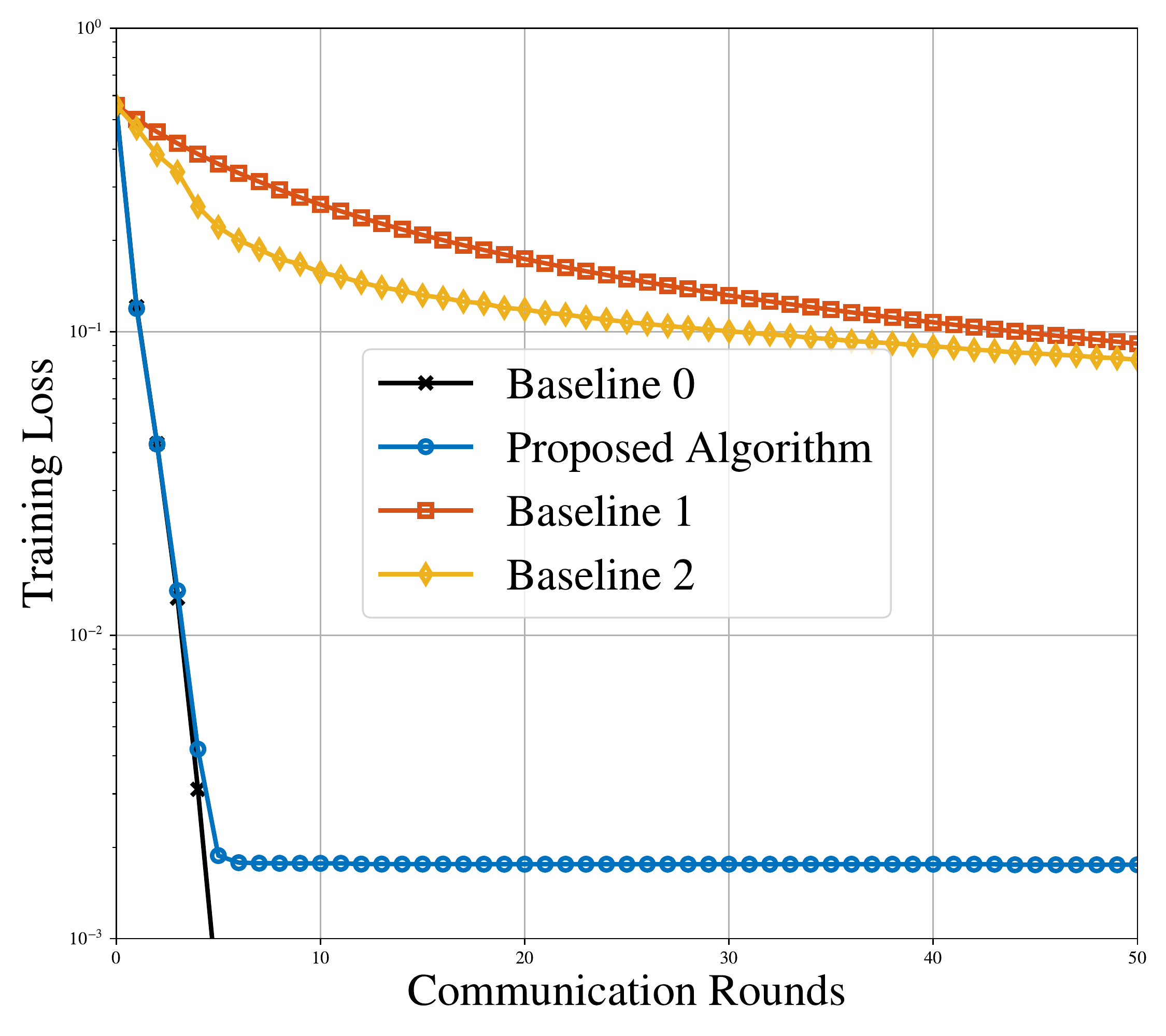}
        \end{minipage}
    }
    \caption{Training loss of the proposed algorithm and two first-order algorithms.}
    \label{with first-order loss}
    \vspace{-0.5cm}
\end{figure*}

\begin{figure*}[htbp!]
    \centering
    \subfloat[Covtype]{
        \begin{minipage}[c][1\width]{0.23\textwidth}
            \centering
            \includegraphics[width=1\linewidth]{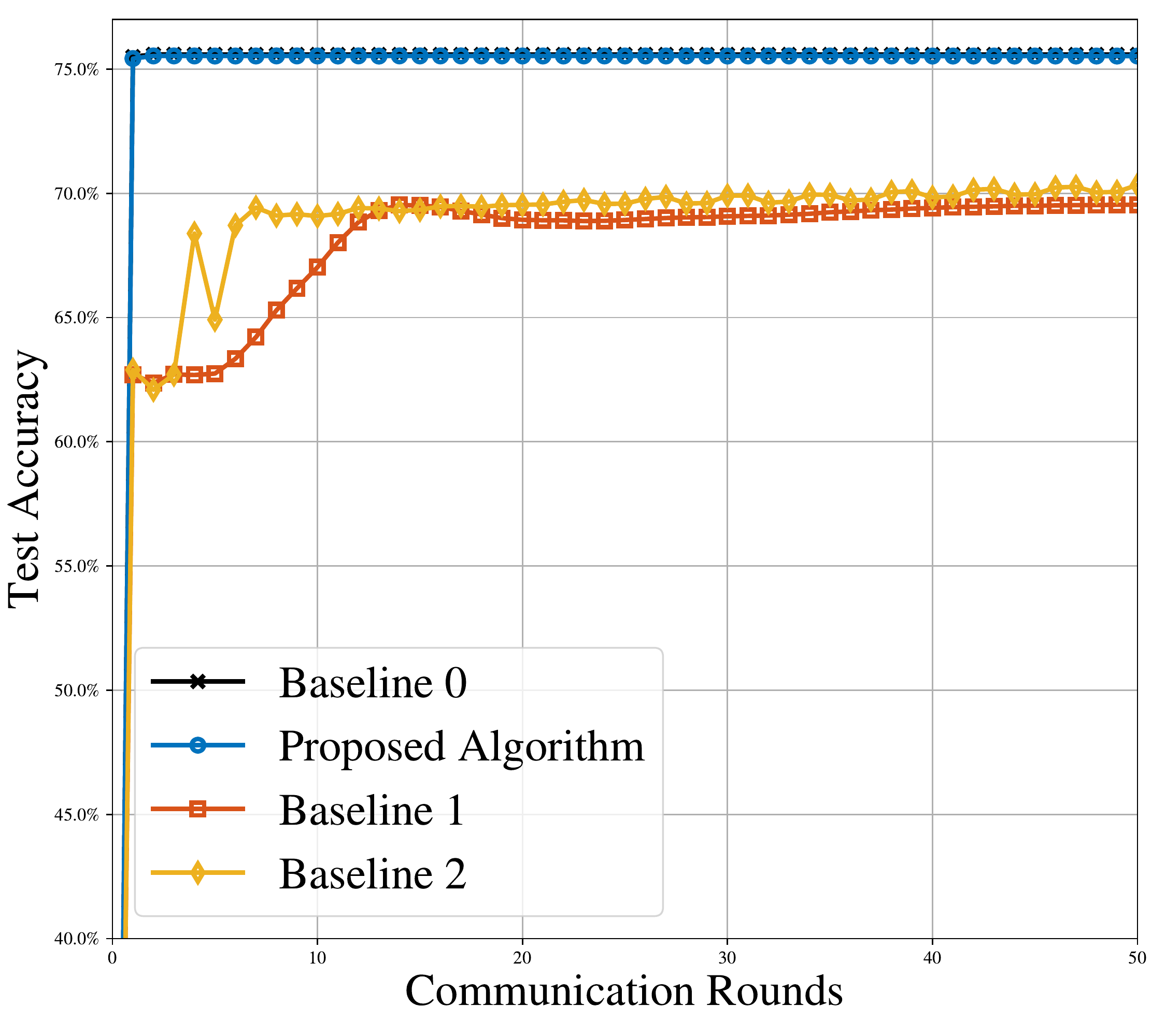}
        \end{minipage}
    }
    \hfill
    \subfloat[a9a]{
        \begin{minipage}[c][1\width]{0.23\textwidth}
            \centering
            \includegraphics[width=1\linewidth]{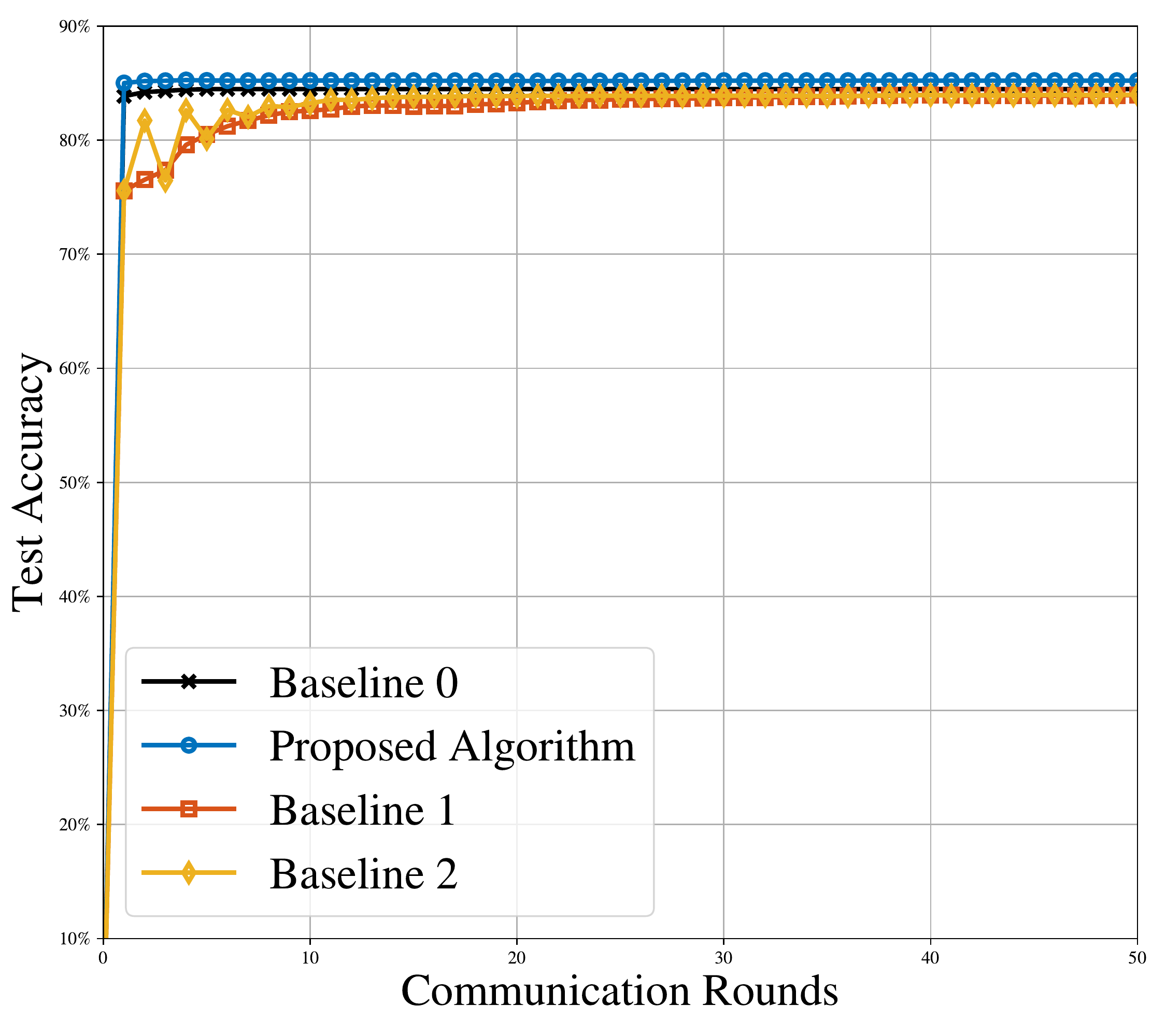}
        \end{minipage}
    }
    \hfill
    \subfloat[w8a]{
        \begin{minipage}[c][1\width]{0.23\textwidth}
            \centering
            \includegraphics[width=1\linewidth]{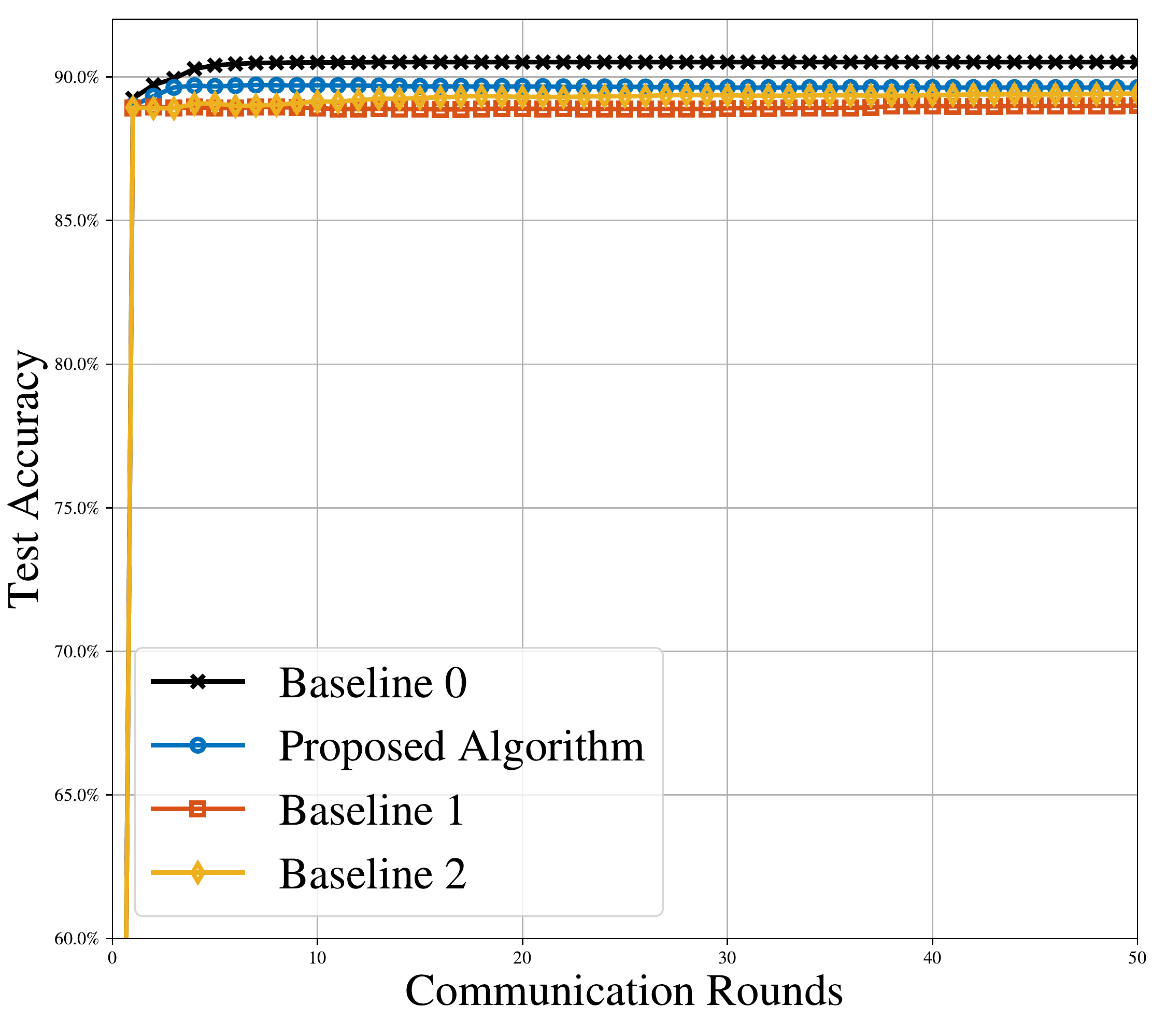}
        \end{minipage}
    }
    \hfill
    \subfloat[phishing]{
        \begin{minipage}[c][1\width]{0.23\textwidth}
            \centering
            \includegraphics[width=1\linewidth]{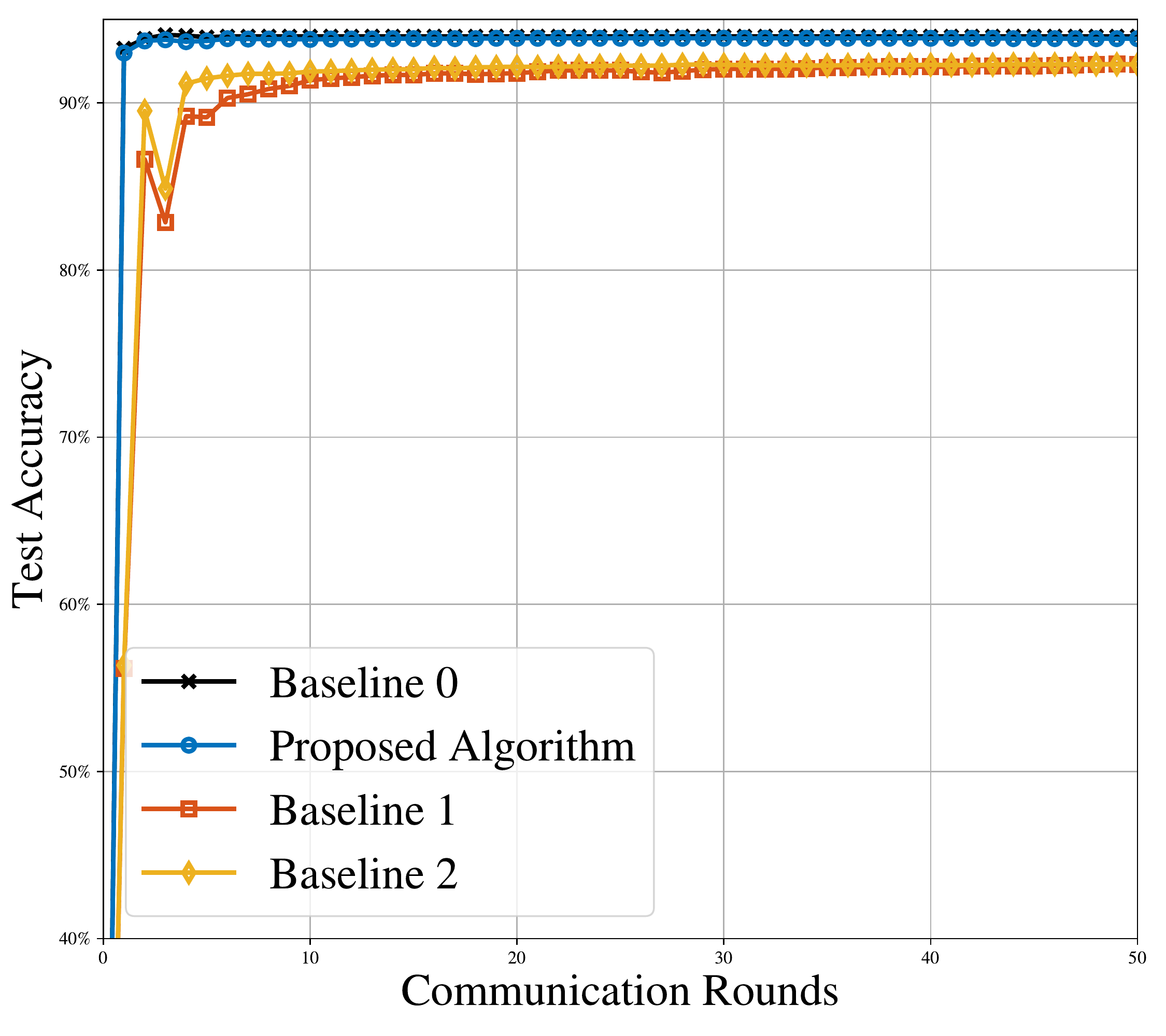}
        \end{minipage}
    }
    
    \caption{Test accuracy of the proposed algorithm and two first-order algorithms.}
    \label{with first-order accuracy}
    \vspace{-0.5cm}
\end{figure*}
Fig. \ref{with first-order loss} and Fig. \ref{with first-order accuracy} show the performance of these algorithms in training loss and test accuracy. 
Regarding the optimality gap, benefiting from the linear-quadratic convergence rate, the proposed algorithm reaches a small optimality gap in the first few dozen communication rounds, while that of the first-order methods remains at a relatively higher level. 
As for the test accuracy, our proposed algorithm can quickly reach and stabilize at a high accuracy level, while the first-order methods have relatively low and fluctuating accuracy. 
Overall, our proposed algorithm keeps a quadratic convergence rate at the beginning of FL process, resulting in fewer communication rounds to complete the learning task than first-order algorithms.
This further leads to less wireless channel impact and better learning performance, as illustrated in the simulation results.
\begin{figure*}[htbp!]
    \centering
    \subfloat[Covtype]{
        \begin{minipage}[c][0.85\width]{0.23\textwidth}
            \centering
            \includegraphics[width=1\linewidth]{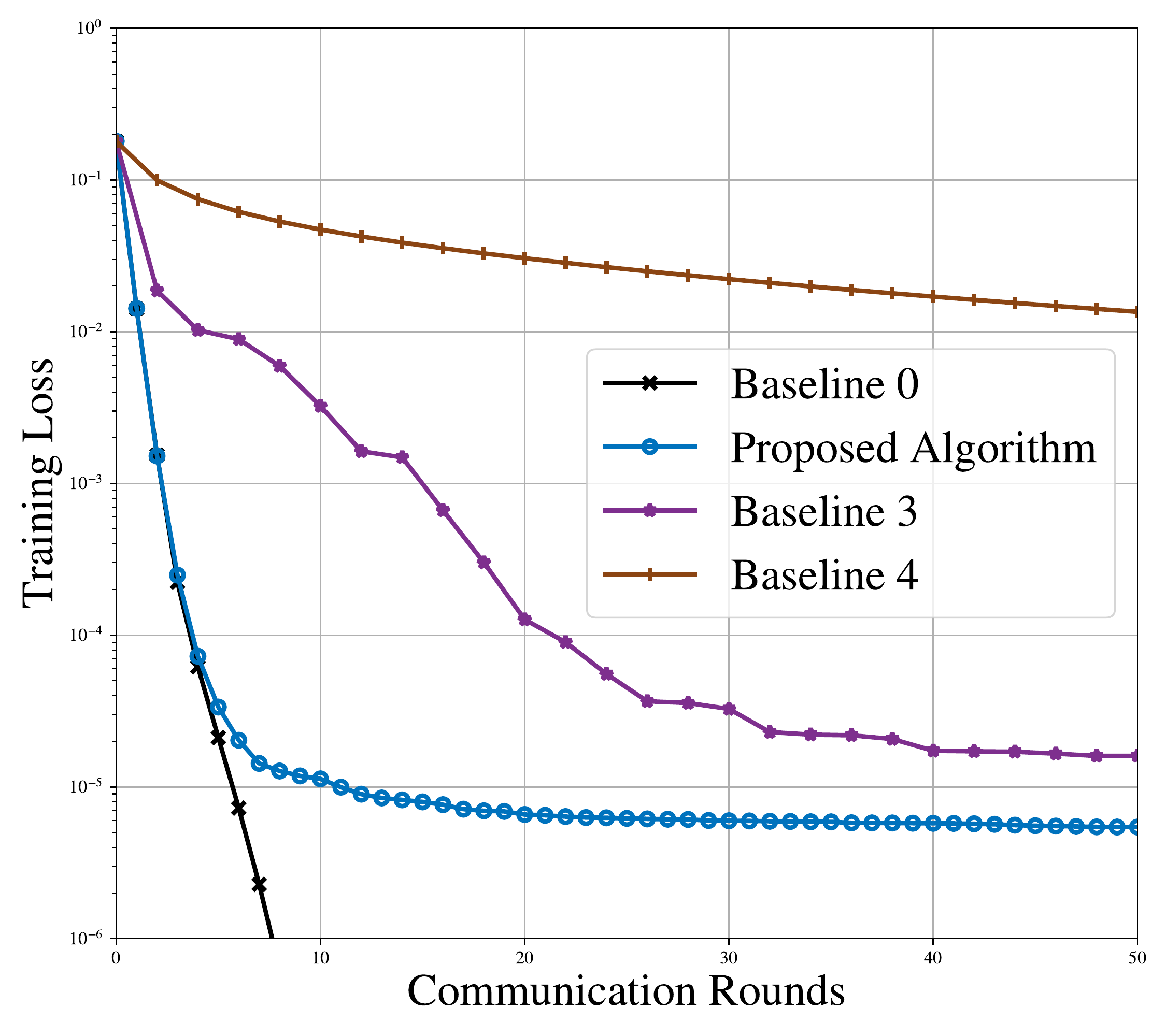}
        \end{minipage}
    }
    \hfill
    \subfloat[a9a]{
        \begin{minipage}[c][0.85\width]{0.23\textwidth}
            \centering
            \includegraphics[width=1\linewidth]{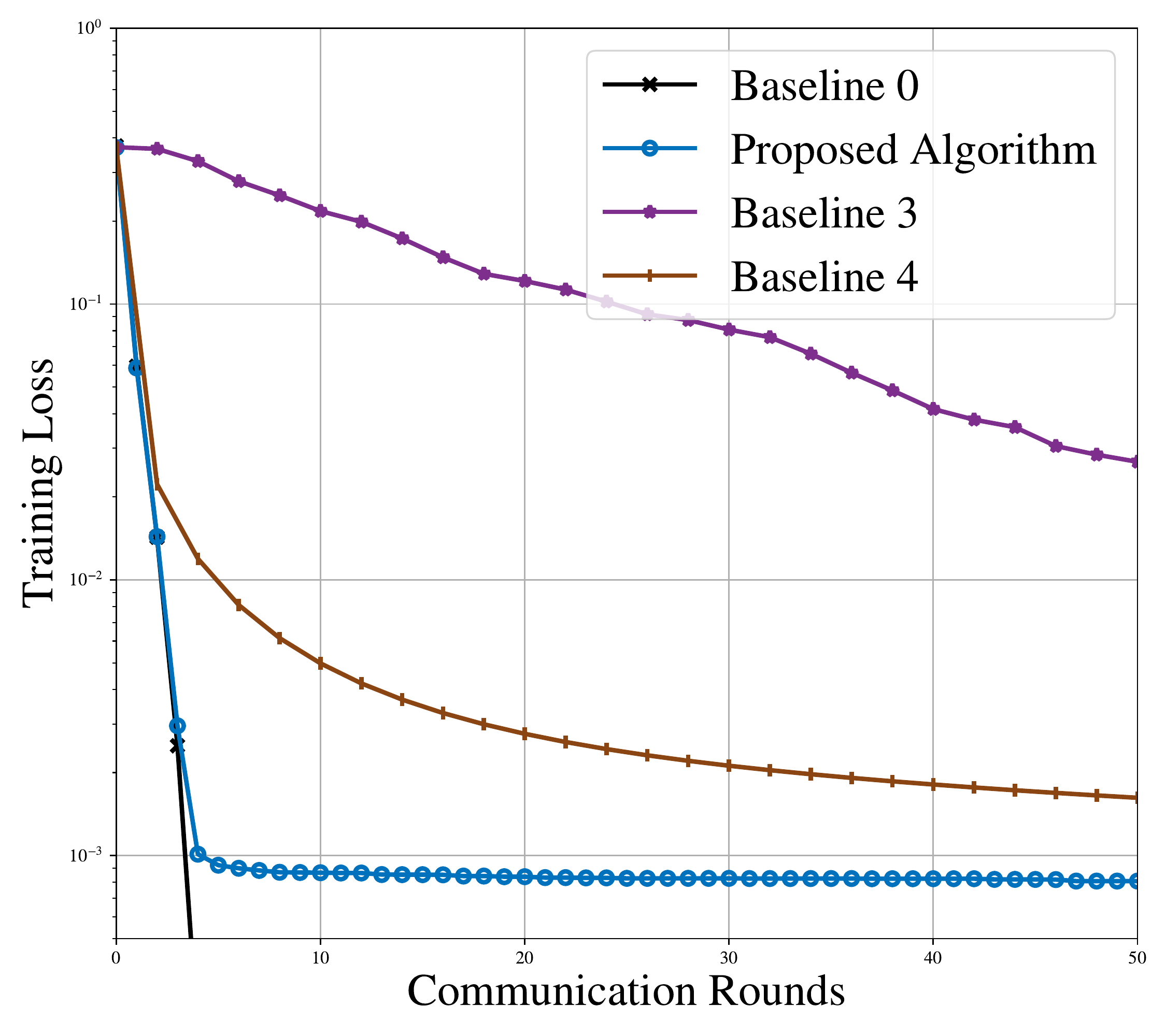}
        \end{minipage}
    }
    \hfill
    \subfloat[w8a]{
        \begin{minipage}[c][0.85\width]{0.23\textwidth}
            \centering
            \includegraphics[width=1\linewidth]{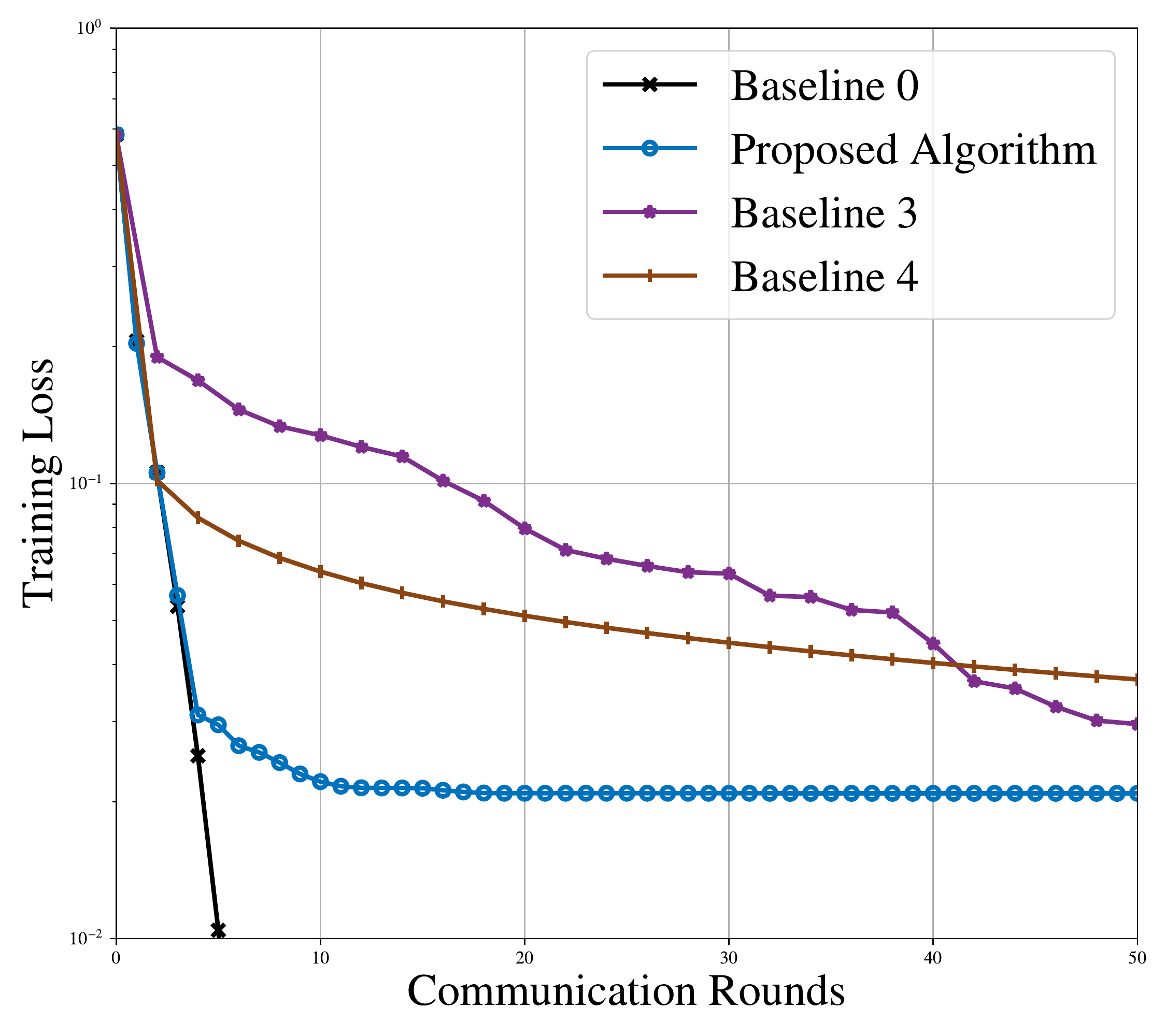}
        \end{minipage}
    }
    \hfill
    \subfloat[phishing]{
        \begin{minipage}[c][0.85\width]{0.23\textwidth}
            \centering
            \includegraphics[width=1\linewidth]{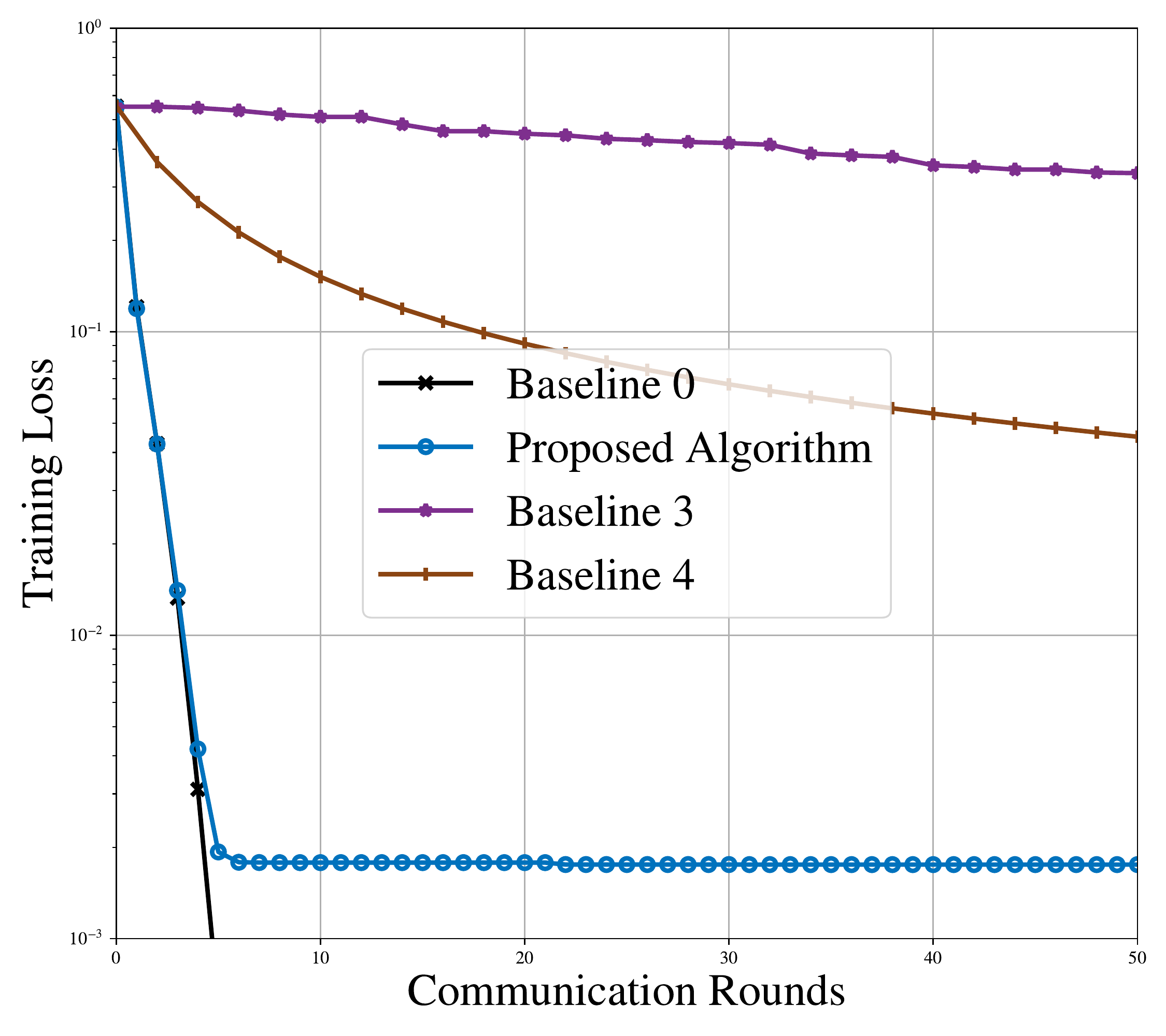}
        \end{minipage}
    }

    \caption{Training loss of the proposed algorithm and two second-order algorithms.}
    \label{with second-order loss}
    \vspace{-0.5cm}
\end{figure*}

\begin{figure*}[htbp!]
    \centering
    \subfloat[Covtype]{
        \begin{minipage}[c][0.85\width]{0.23\textwidth}
            \centering
            \includegraphics[width=1\linewidth]{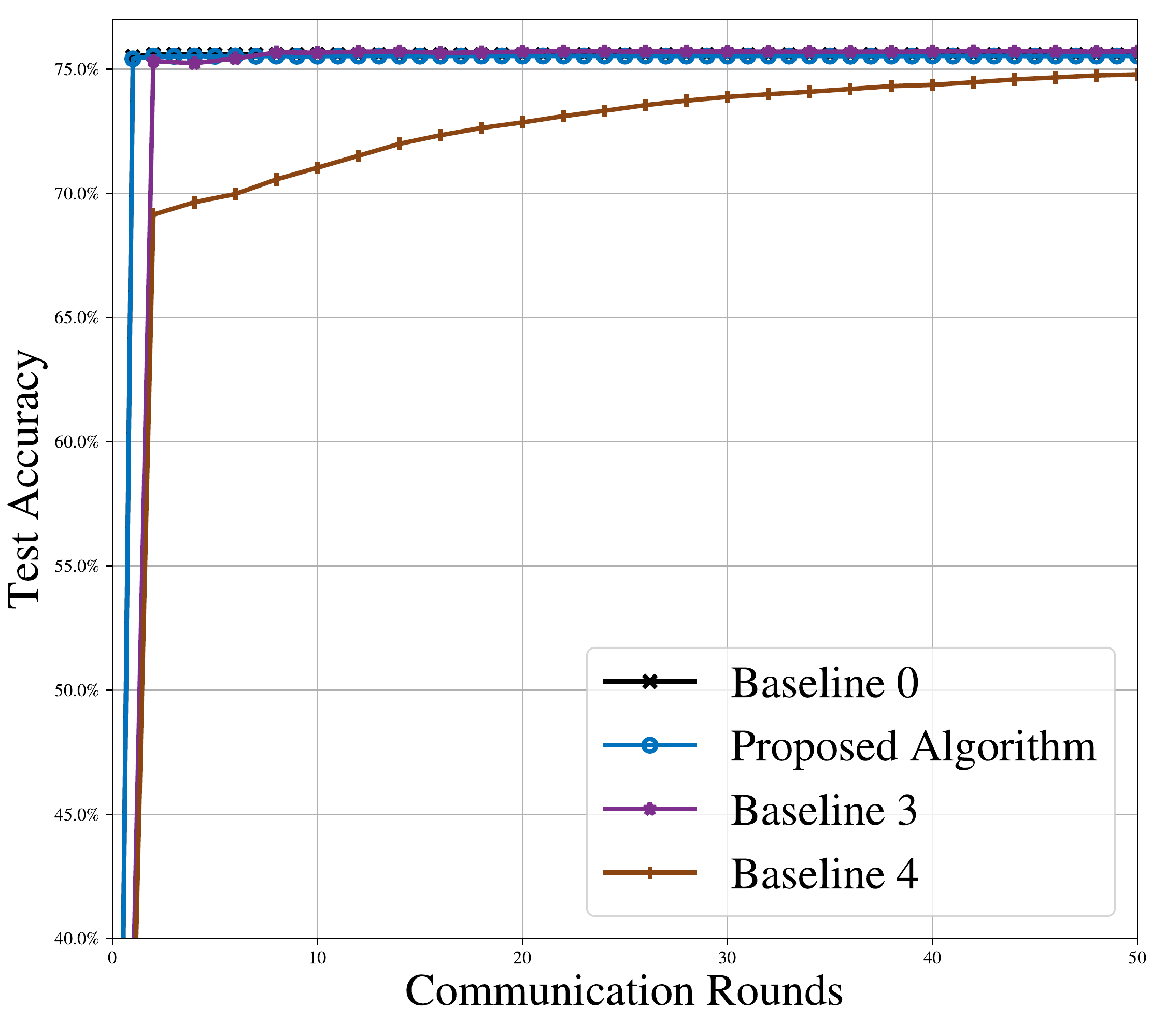}
        \end{minipage}
    }
    \hfill
    \subfloat[a9a]{
        \begin{minipage}[c][0.85\width]{0.23\textwidth}
            \centering
            \includegraphics[width=1\linewidth]{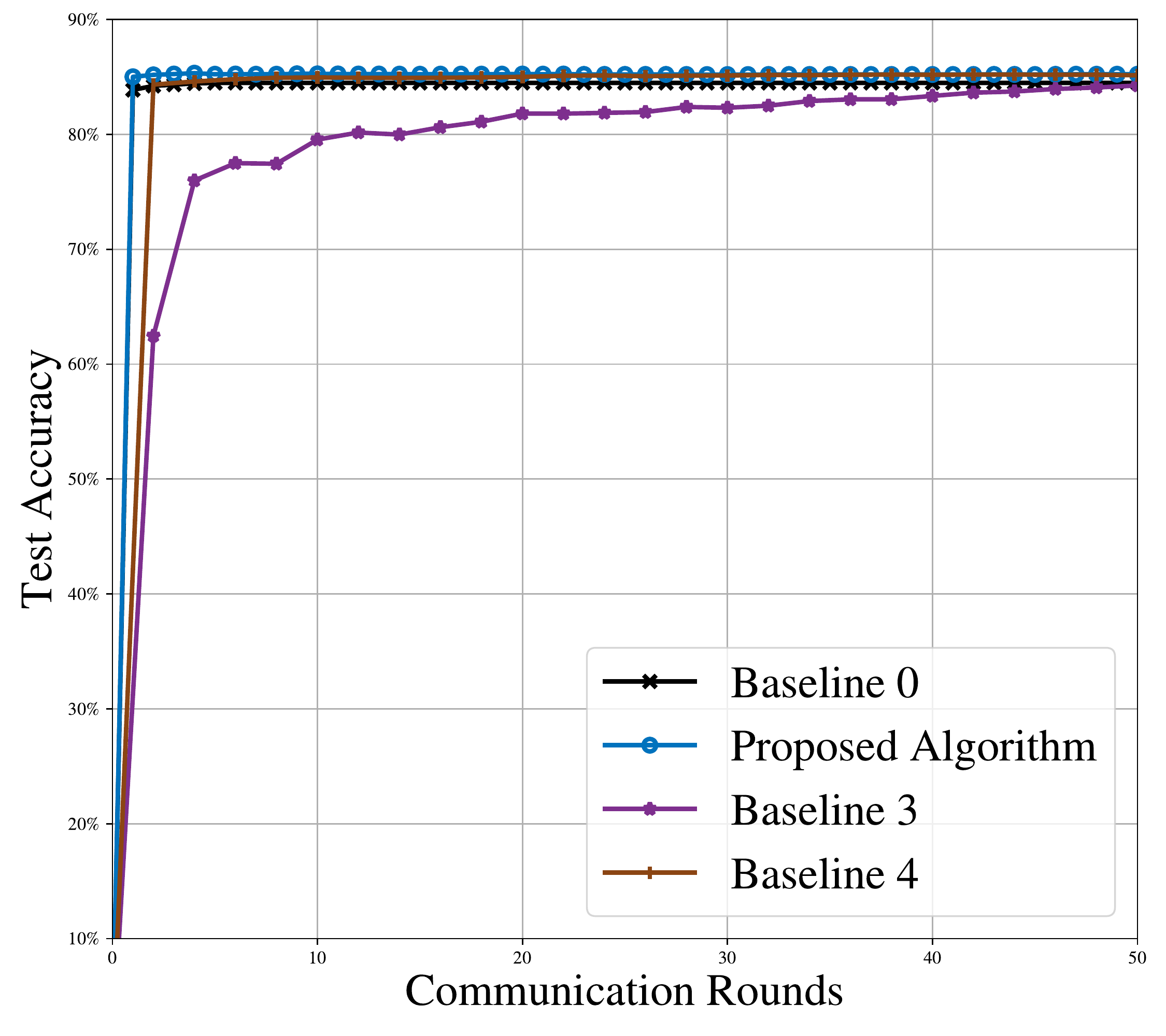}
        \end{minipage}
    }
    \hfill
    \subfloat[w8a]{
        \begin{minipage}[c][0.85\width]{0.23\textwidth}
            \centering
            \includegraphics[width=1\linewidth]{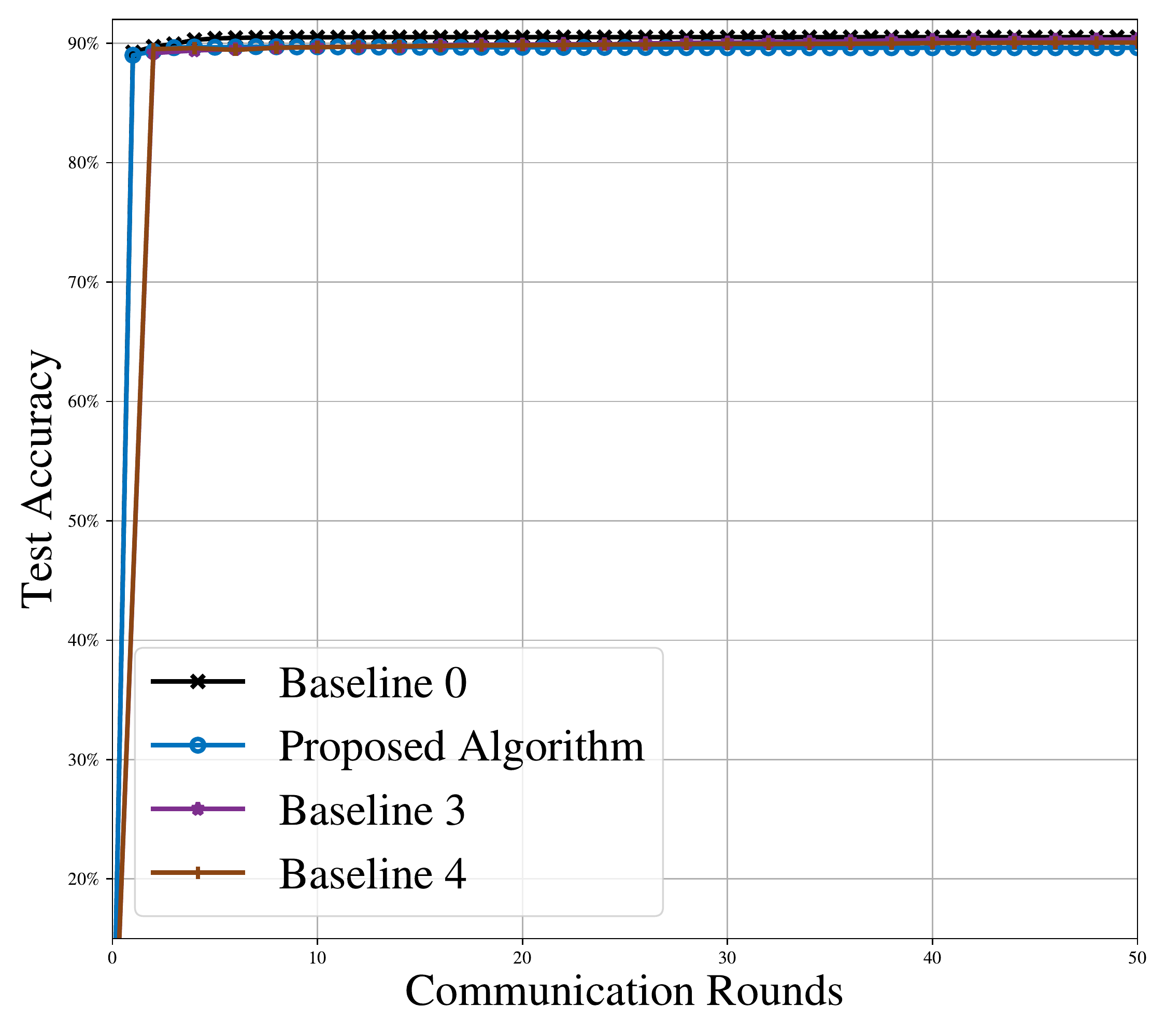}
        \end{minipage}
    }
    \hfill
    \subfloat[phishing]{
        \begin{minipage}[c][0.85\width]{0.23\textwidth}
            \centering
            \includegraphics[width=1\linewidth]{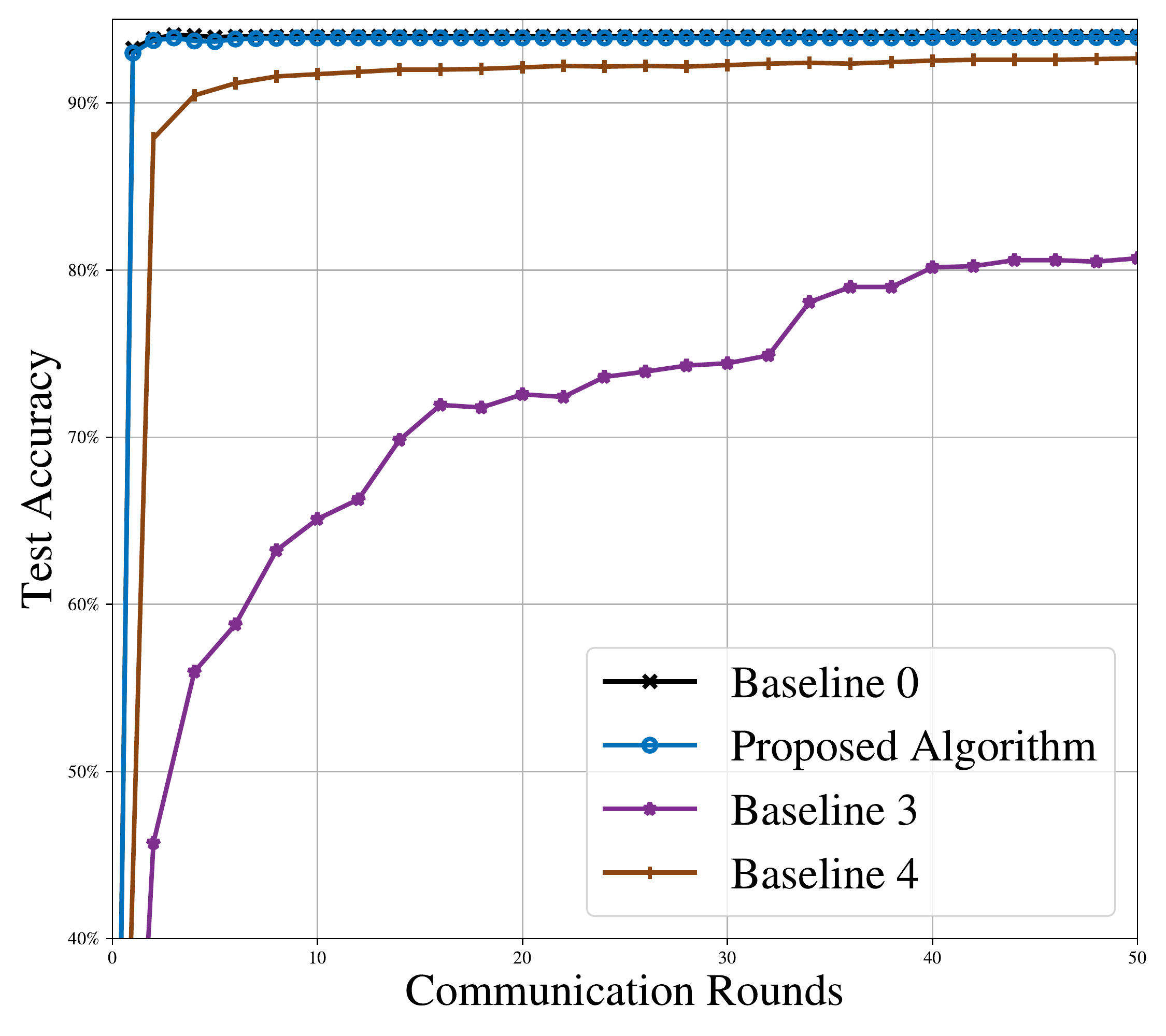}
        \end{minipage}
    }

    \caption{Test accuracy of the proposed algorithm and two second-order algorithms.}
    \label{with second-order accuracy}
    \vspace{-0.5cm}
\end{figure*}

\subsection{Comparison with Second-Order Algorithms}
In this experiment, we compared our proposed algorithm with the following two state-of-the-art second-order algorithms under over-the-air computation:
\begin{enumerate}
    \item Baseline 3: GIANT \cite{wang2018giant} with over-the-air computation. GIANT requires an extra aggregation of local gradients, leading to two communication rounds in each iteration. 
    The communication model of this gradients aggregation is implemented in the same way of $\bm{p}_{t}$, as illustrated in Section II-C. Here, we set $|\mathcal{S}_{t}| = m$, and the receiver beamforming vector is optimized through DCA.
    \item Baseline 4: DANE \cite{shamir2014communication} with over-the-air computation. Similar to GIANT, It also requires an aggregation of local gradients, so its implementation is the same as GIANT. 
\end{enumerate}

Fig. \ref{with second-order loss} and Fig. \ref{with second-order accuracy} plot the training loss and the test accuracy, respectively, where SNR is set to 70 dB.
It is observed that our proposed algorithm converges faster and remains stable at a relatively high level of accuracy, while the compared methods, AirComp-based GIANT and AirComp-based DANE, have a slower convergence rate.
This is because both the procedures of GIANT and DANE involve aggregating local gradients to calculate the global gradient in each iteration.
This extra transmission of local gradients through a wireless environment aggravates the impact of channel noise, leading to a relatively poor convergence rate.
Therefore, we can see that our proposed algorithm outperforms AirComp-based GIANT and AirComp-based DANE.

\subsection{Effectiveness of Proposed System Optimization Approach}
In this experiment, we evaluated the performance using GS+DCA to accomplish system optimization with four settings:
\begin{enumerate}
    \item perfect aggregation, where the model is aggregated without wireless channel impact.
    \item GS+SDR, where the receiver beamforming optimization is performed through SDR.
    \item DCA only, where we only perform beamforming optimization through DCA.
    \item SDR only, where we only perform beamforming optimization through SDR.
\end{enumerate}
\begin{figure*}[htbp]
    \centering
    \subfloat{
        \begin{minipage}[c][0.9\width]{0.3\textwidth}
            \centering
            \includegraphics[width=1\linewidth]{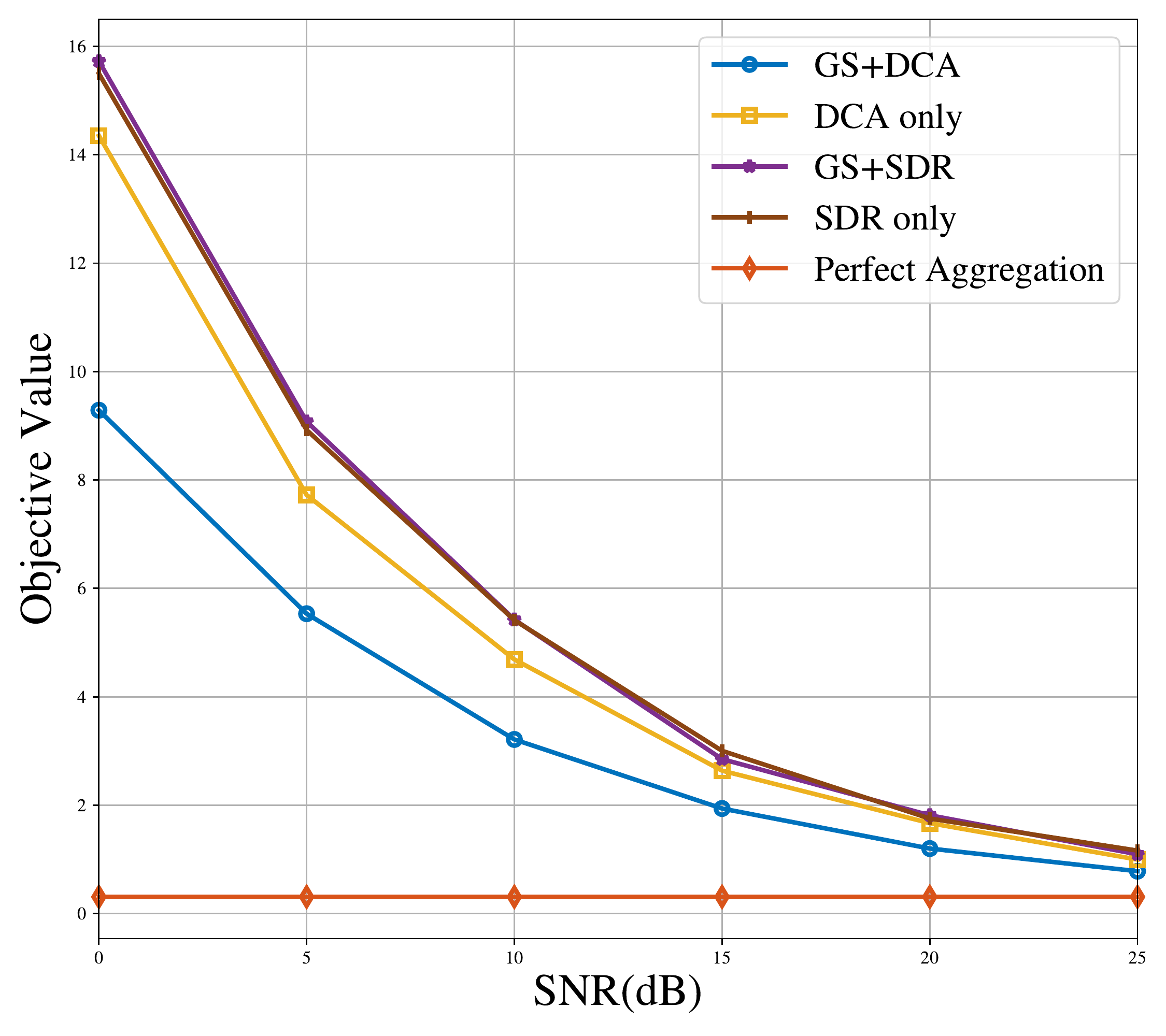}
            \label{with SNR}
        \end{minipage}
    }
    \hspace{8em}
    \subfloat{
        \begin{minipage}[c][0.9\width]{0.3\textwidth}
            \centering
            \includegraphics[width=1\linewidth]{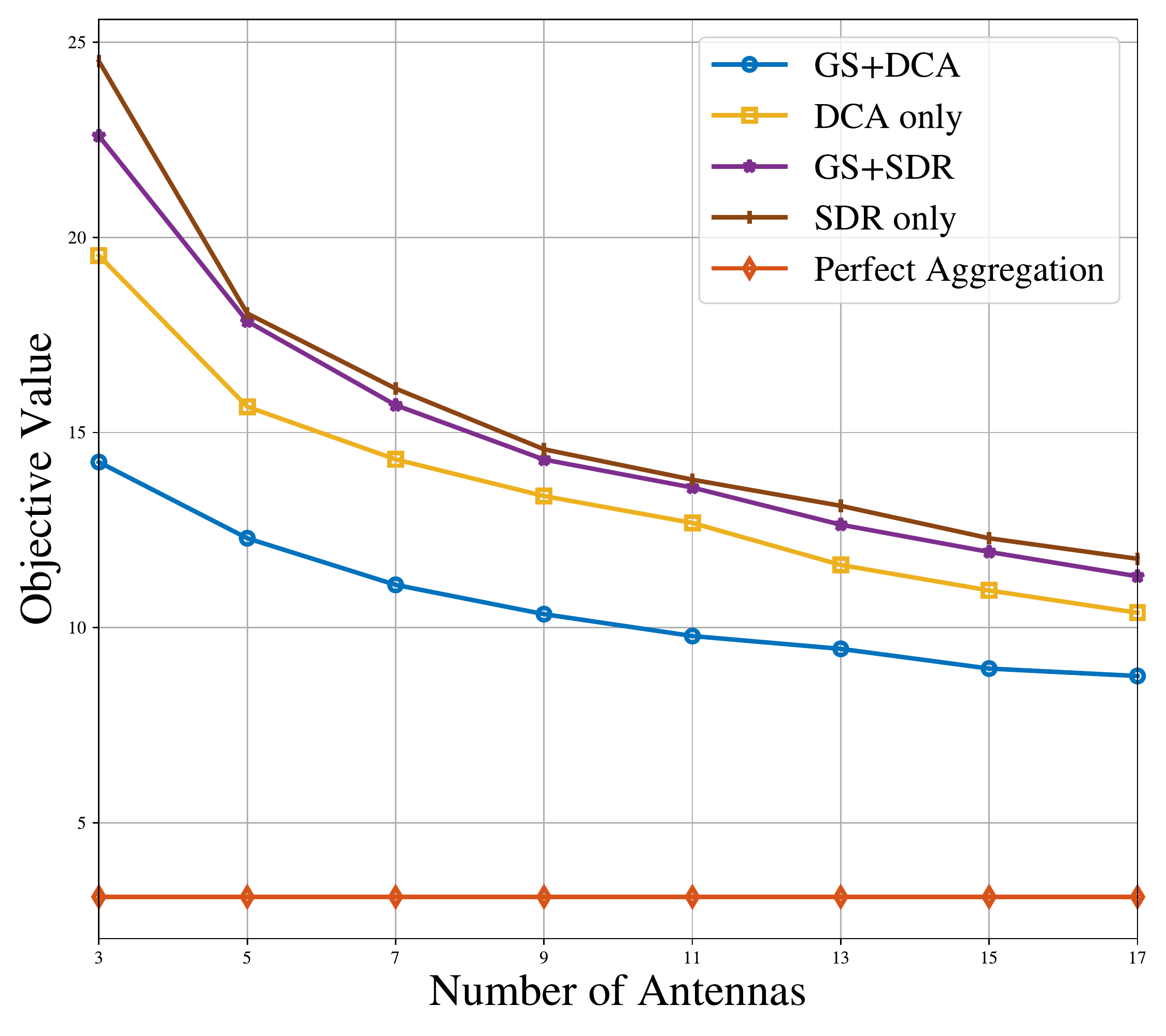}
            \label{with antenna}
        \end{minipage}
    }
    \caption{Objective value of system optimization problem $\mathscr{P}$ versus SNR and number of antennas.}
    \label{with optimization}
    \vspace{-0.5cm}
\end{figure*}
To verify the effectiveness of the device selection, we consider the distance heterogeneity and data size heterogeneity in this experiment.
Specifically, as for distance heterogeneity, we set the distance of $10\%$ devices to be $d_{i} \in [200, 220]$ while the rest to be $d_{i} \in [50, 60]$.
As for data size heterogeneity, we set the data size of $10\%$ devices to be $\card{\mathcal{D}_{i}} \in [0.008\frac{n}{m}, 0.01\frac{n}{m}]$ while the rest to be 
$\card{\mathcal{D}_{i}} \in [1.01\frac{n}{m}, 1.11\frac{n}{m}]$.

We first numerically evaluate the objective value of the system optimization problem $\mathscr{P}$ under different settings in Fig. \ref{with optimization} by averaging 100 channel realizations.
The objective value of perfect aggregation does not depend on SNR and the number of antennas since the error during the FL process in this situation only comes from the approximation as \eqref{epsilon1} indicates. 
The objective values of all settings decrease as SNR and the number of antennas increase, due to the mitigation of noise effect and the increase of diversity gain \cite{chen2018uniform}, respectively.
However, the objective value of GS+DCA is smaller than that of other settings.
On the one hand, SDR fails to give a precise solution for the receiver beamforming vector as the size of the problem grows.
This further leads to the ineffectiveness of device selection in GS+SDR and worse performance compared with the settings using DCA to perform beamforming optimization.
On the other hand, device selection in GS+DCA mitigates the straggler issue caused by distance heterogeneity and data size heterogeneity, resulting in a better performance compared with DCA only.

\begin{figure*}[htbp!]
    \centering
    \subfloat[Covtype]{
        \begin{minipage}[c][0.85\width]{0.23\textwidth}
            \centering
            \includegraphics[width=1\linewidth]{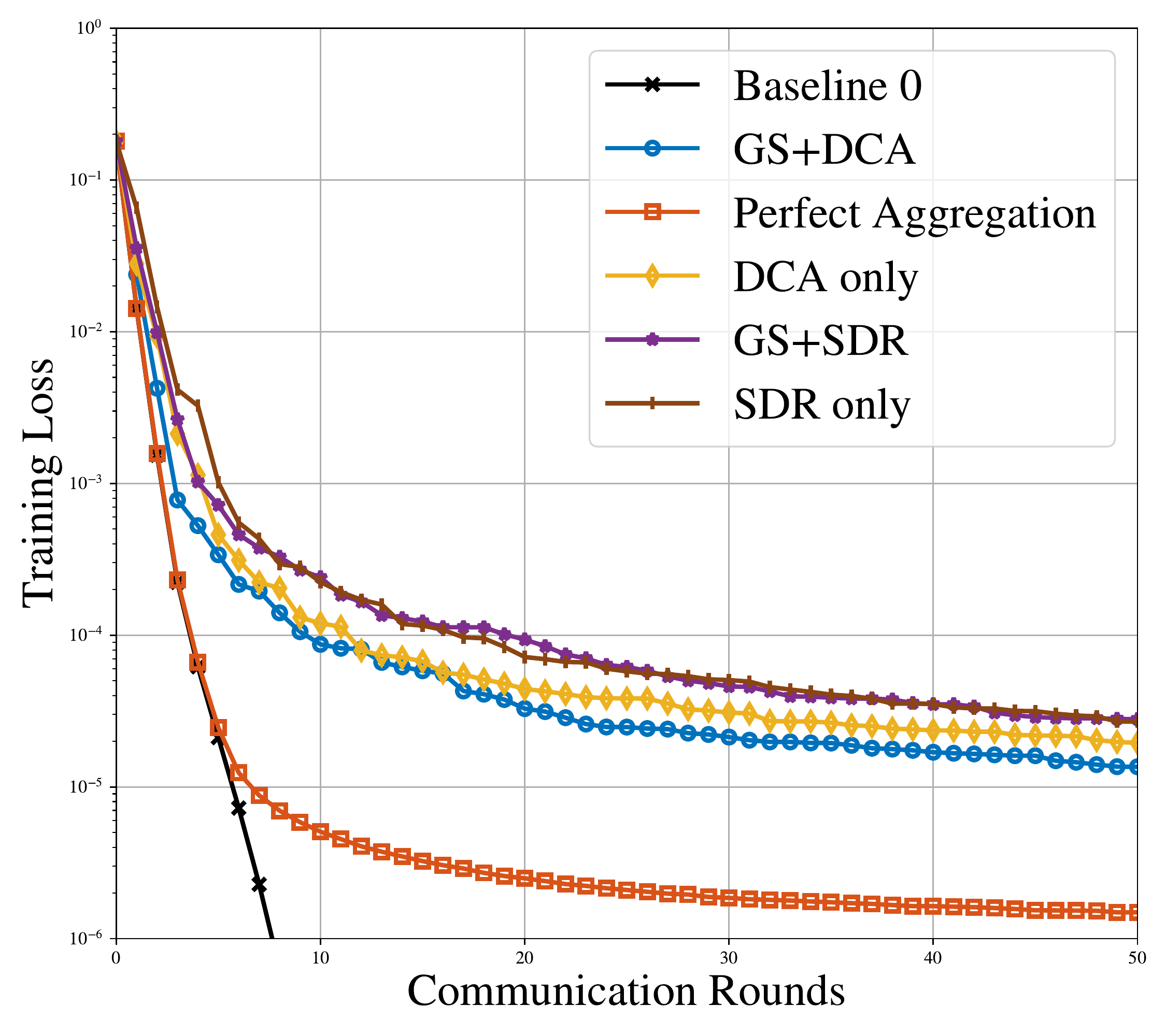}
        \end{minipage}
    }
    \hfill
    \subfloat[a9a]{
        \begin{minipage}[c][0.85\width]{0.23\textwidth}
            \centering
            \includegraphics[width=1\linewidth]{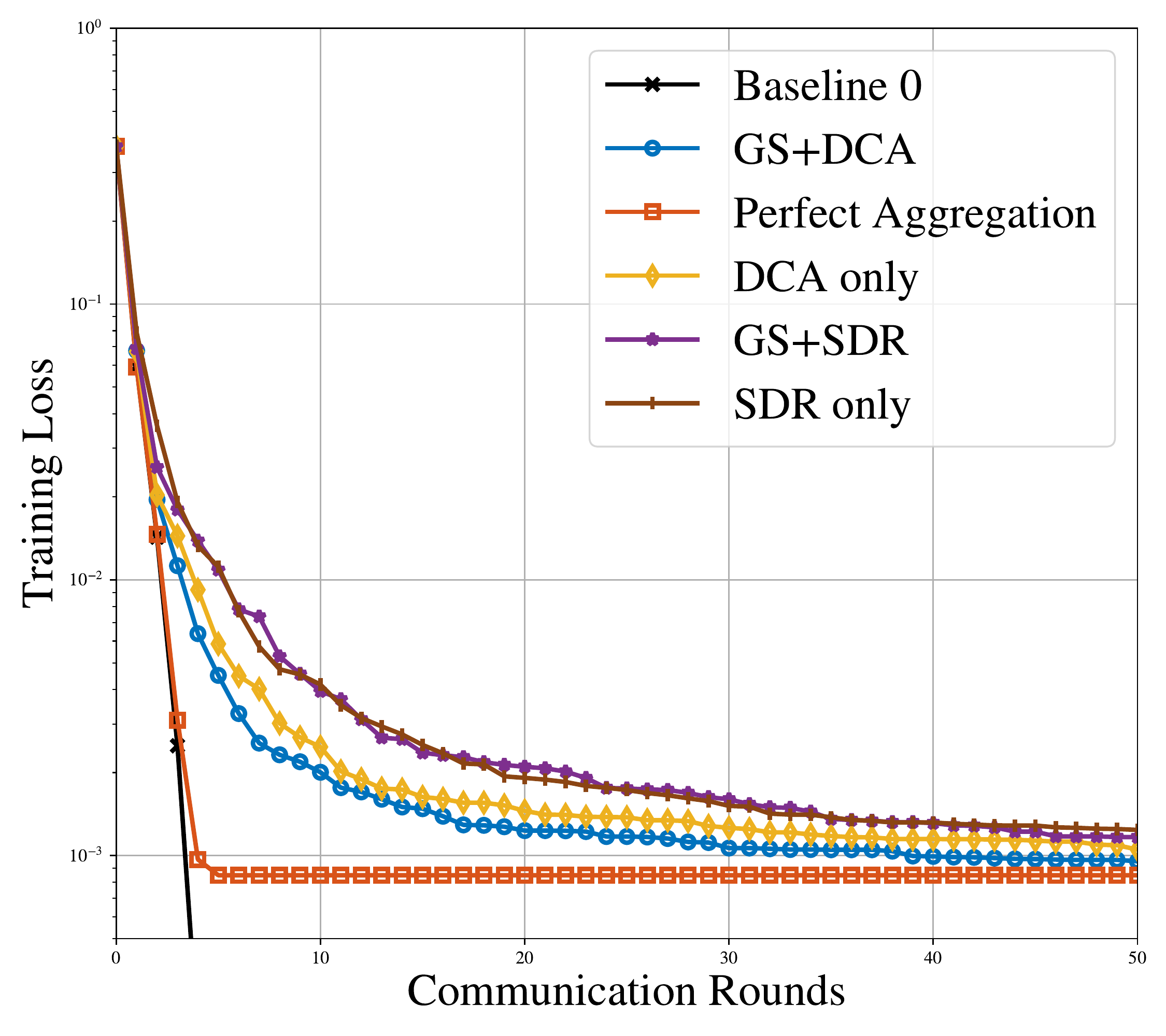}
        \end{minipage}
    }
    \hfill
    \subfloat[w8a]{
        \begin{minipage}[c][0.85\width]{0.23\textwidth}
            \centering
            \includegraphics[width=1\linewidth]{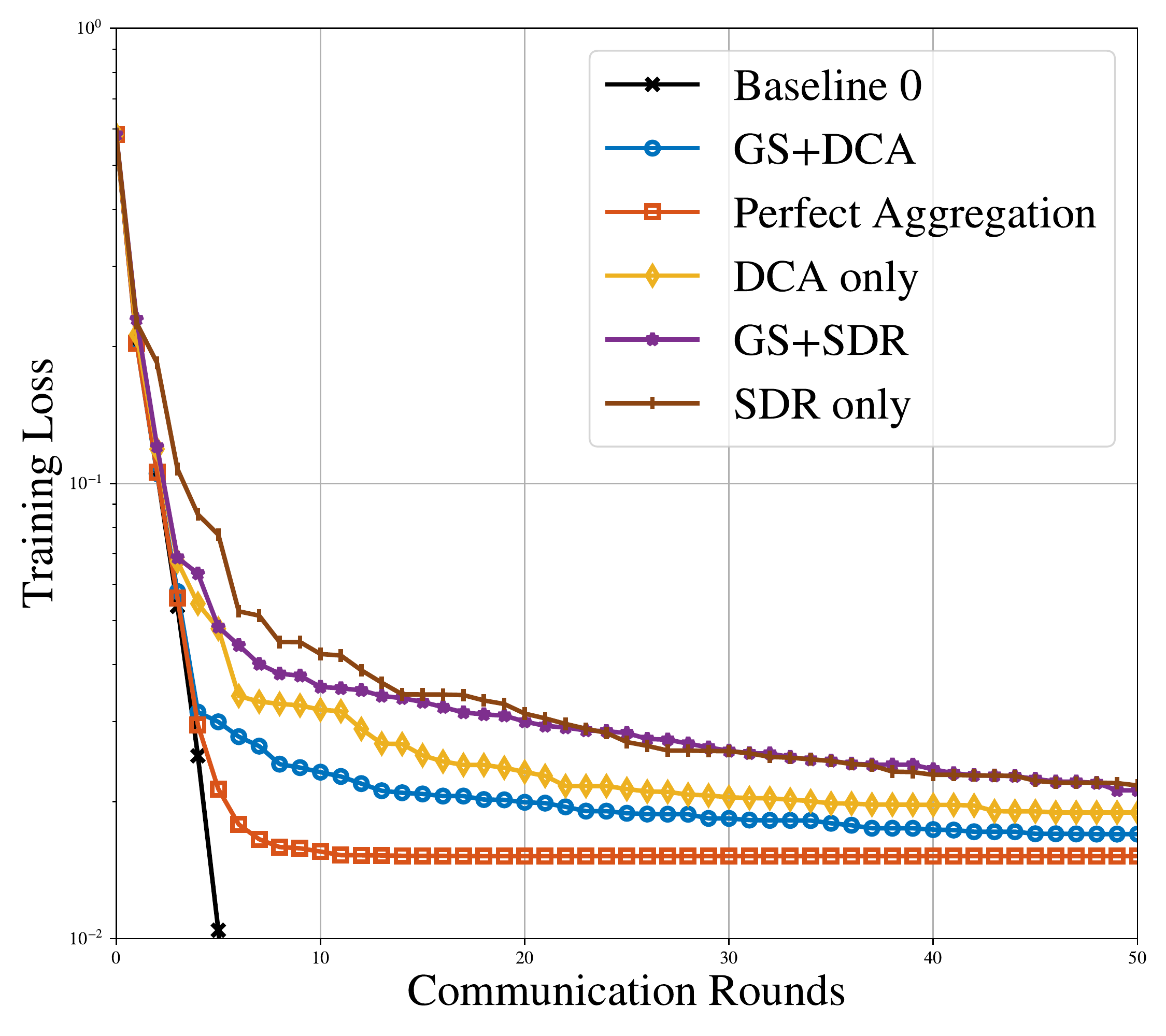}
        \end{minipage}
    }
    \hfill
    \subfloat[phishing]{
        \begin{minipage}[c][0.85\width]{0.23\textwidth}
            \centering
            \includegraphics[width=1\linewidth]{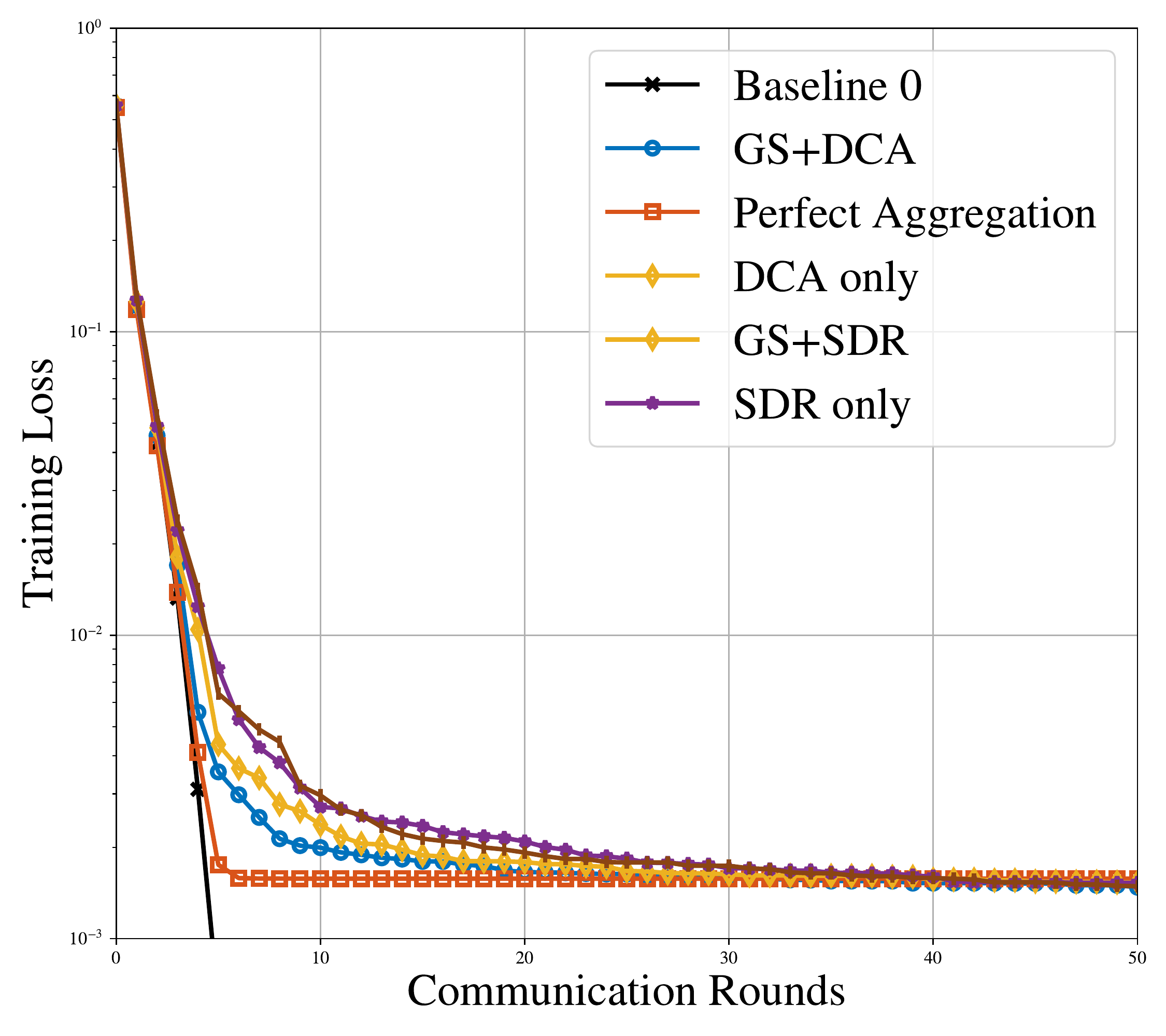}
        \end{minipage}
    }
    \caption{Training loss of the proposed algorithm in different system optimization settings.}
    \label{with optimization loss}
    \vspace{-0.5cm}
\end{figure*}

\begin{figure*}[htbp!]
    \centering
    \subfloat[Covtype]{
        \begin{minipage}[c][0.85\width]{0.23\textwidth}
            \centering
            \includegraphics[width=1\linewidth]{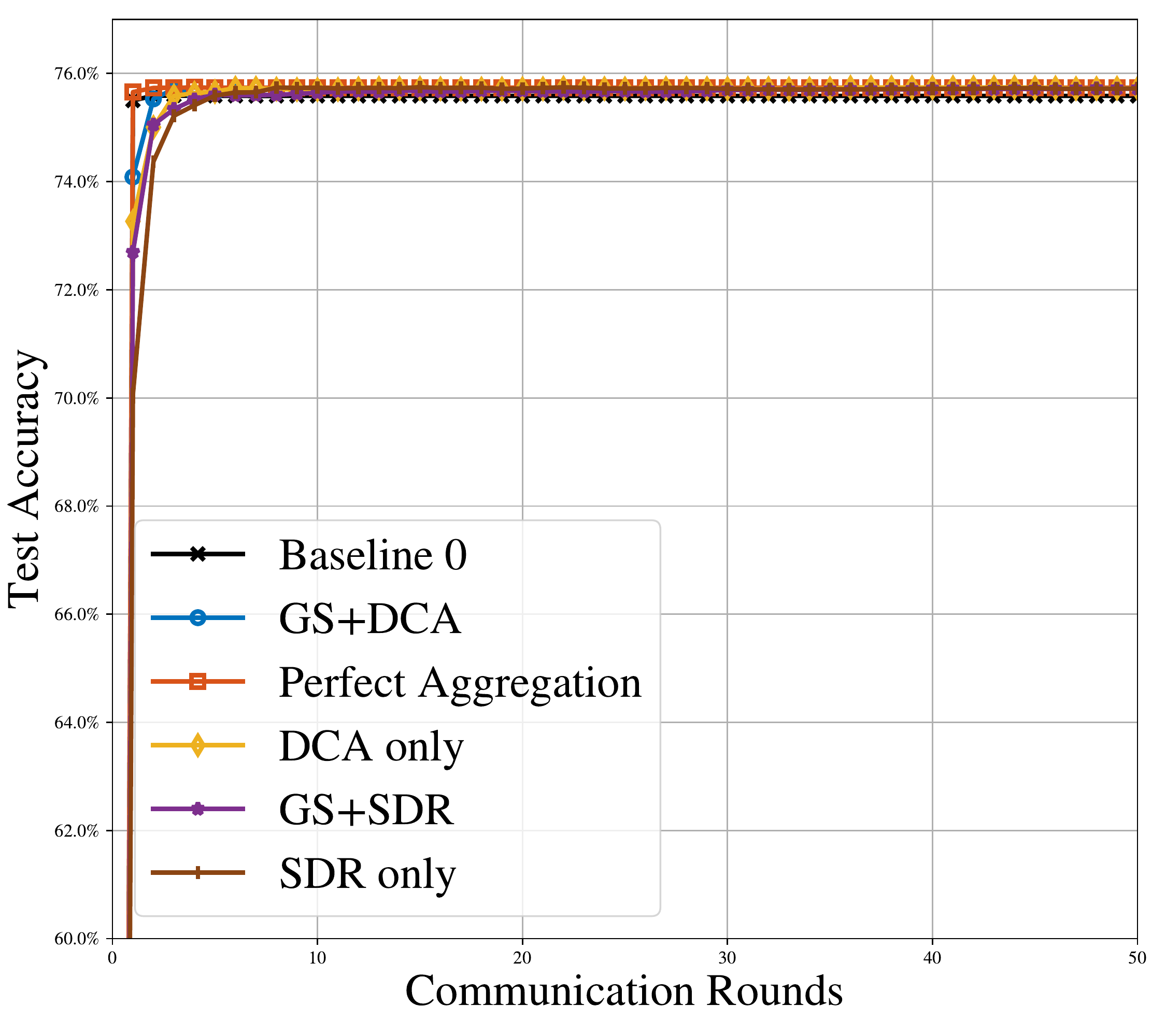}
        \end{minipage}
    }
    \hfill
    \subfloat[a9a]{
        \begin{minipage}[c][0.85\width]{0.23\textwidth}
            \centering
            \includegraphics[width=1\linewidth]{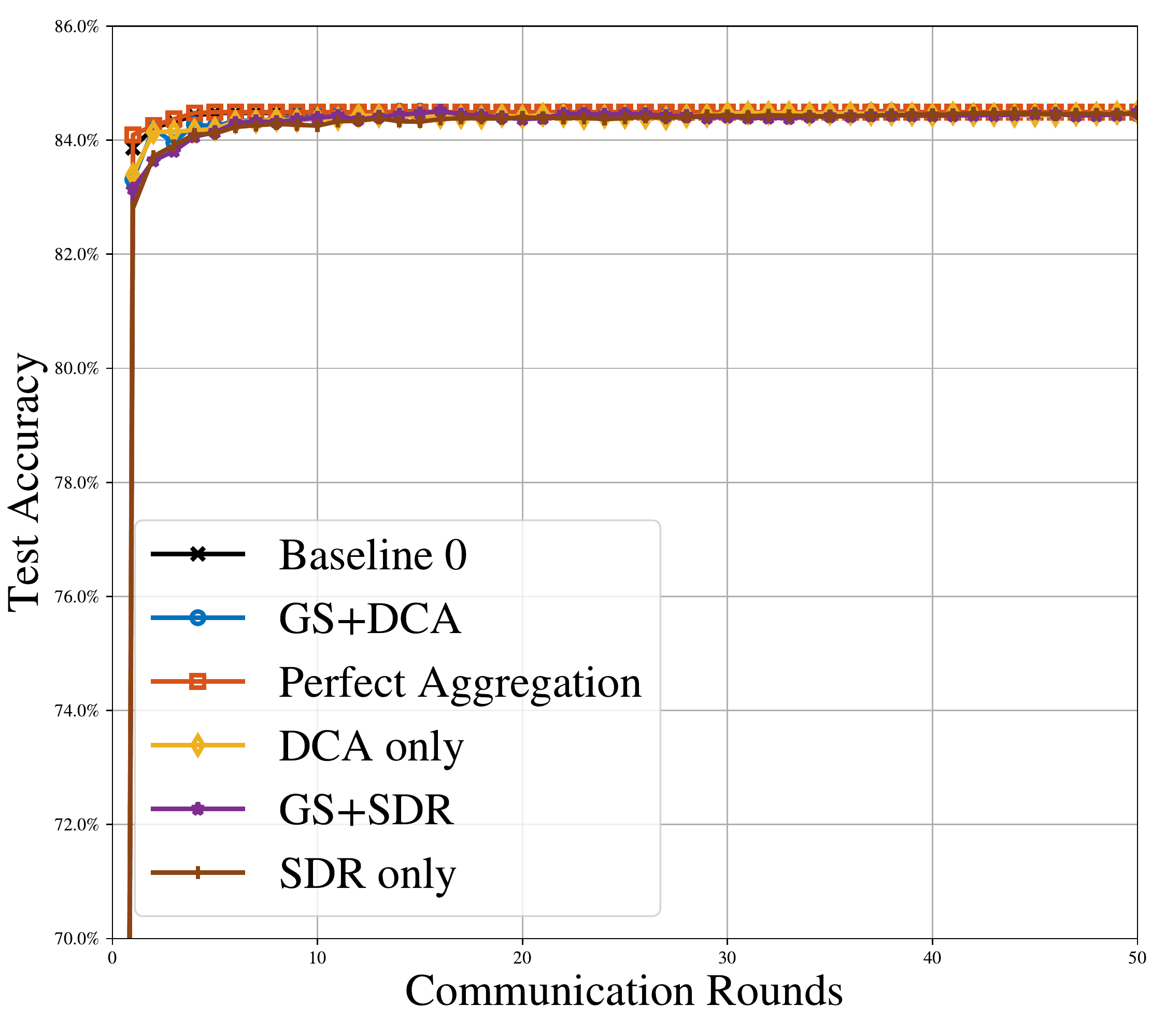}
        \end{minipage}
    }
    \hfill
    \subfloat[w8a]{
        \begin{minipage}[c][0.85\width]{0.23\textwidth}
            \centering
            \includegraphics[width=1\linewidth]{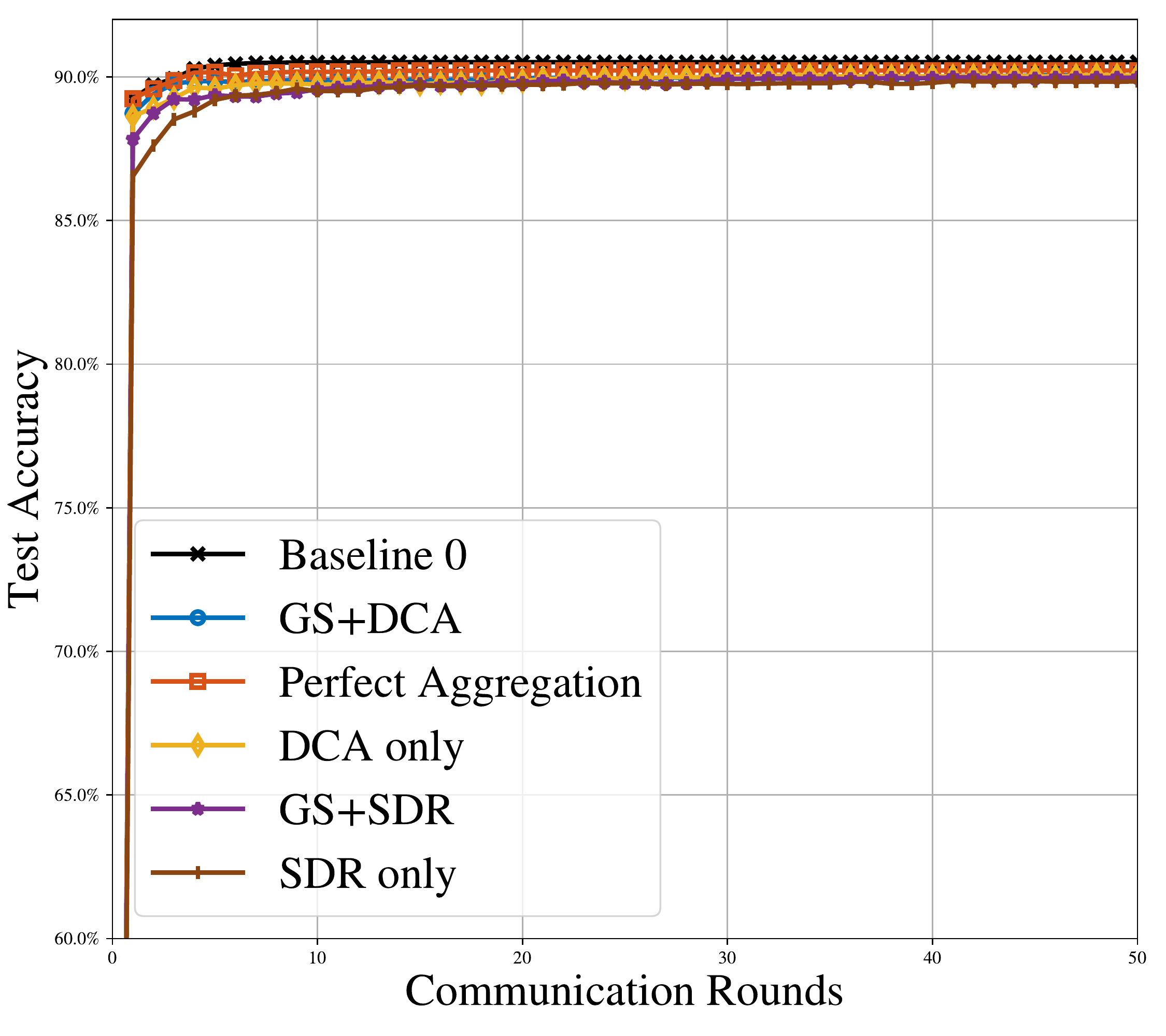}
        \end{minipage}
    }
    \hfill
    \subfloat[phishing]{
        \begin{minipage}[c][0.85\width]{0.23\textwidth}
            \centering
            \includegraphics[width=1\linewidth]{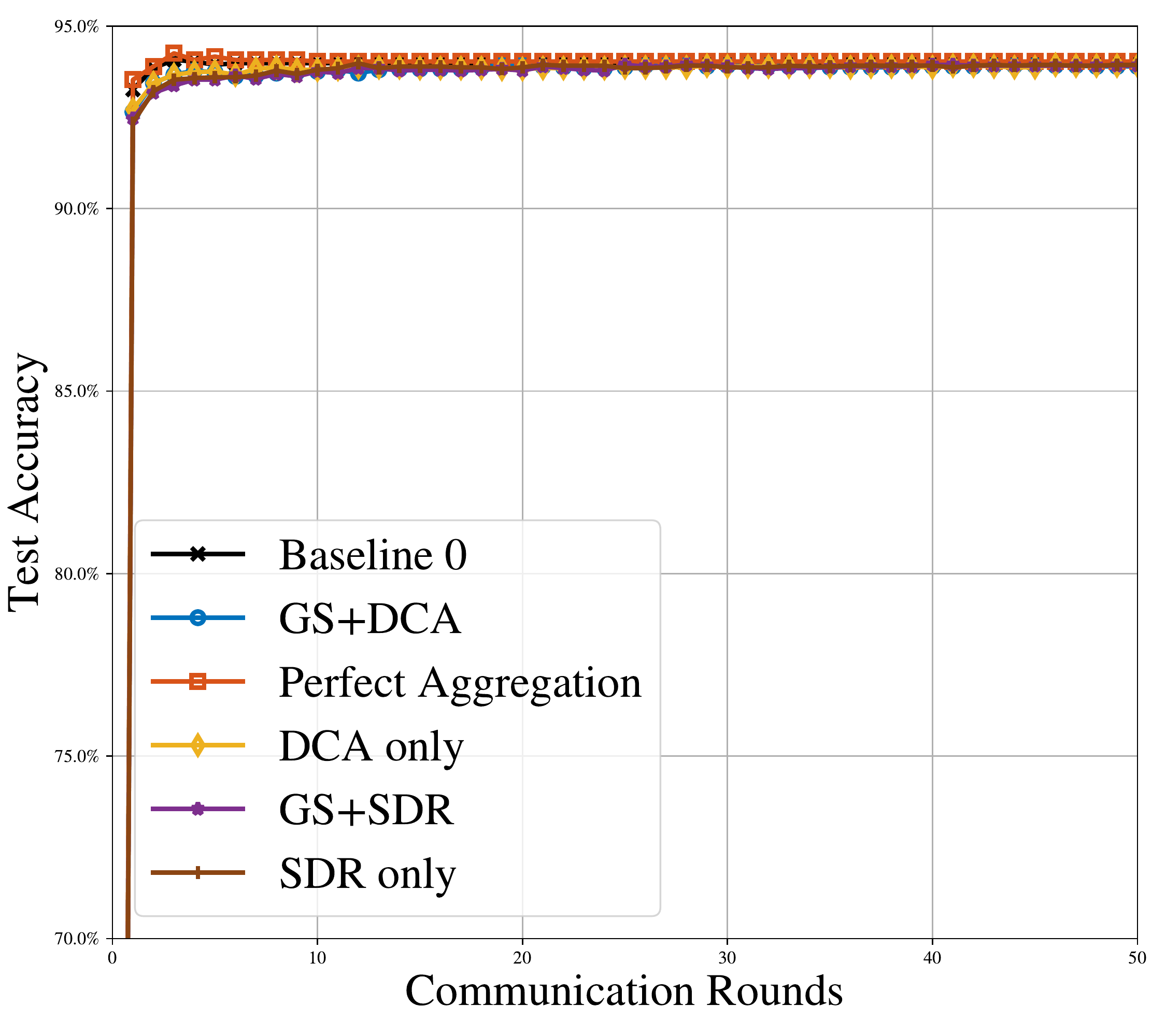}
        \end{minipage}
    }
    \caption{Test accuracy of the proposed algorithm in different system optimization settings.}
    \label{with optimization accuracy}
    \vspace{-0.5cm}
\end{figure*}

Fig. \ref{with optimization loss} plots the training loss for our proposed algorithm in different system optimization settings, where SNR is set to 35 dB. 
The results show that with device selection and a more precise solution given by DCA, the error term can be minimized in each iteration and a smaller optimality gap close to that of perfect aggregation can be obtained.
As revealed in Fig. \ref{with optimization accuracy}, this smaller optimality gap further leads to higher test accuracy, demonstrating that our proposed system optimization approach effectively improves learning performance.

\begin{figure*}[htbp]
    \centering
    \subfloat{
        \begin{minipage}[c][0.8\width]{0.3\textwidth}
            \centering
            \includegraphics[width=1\linewidth]{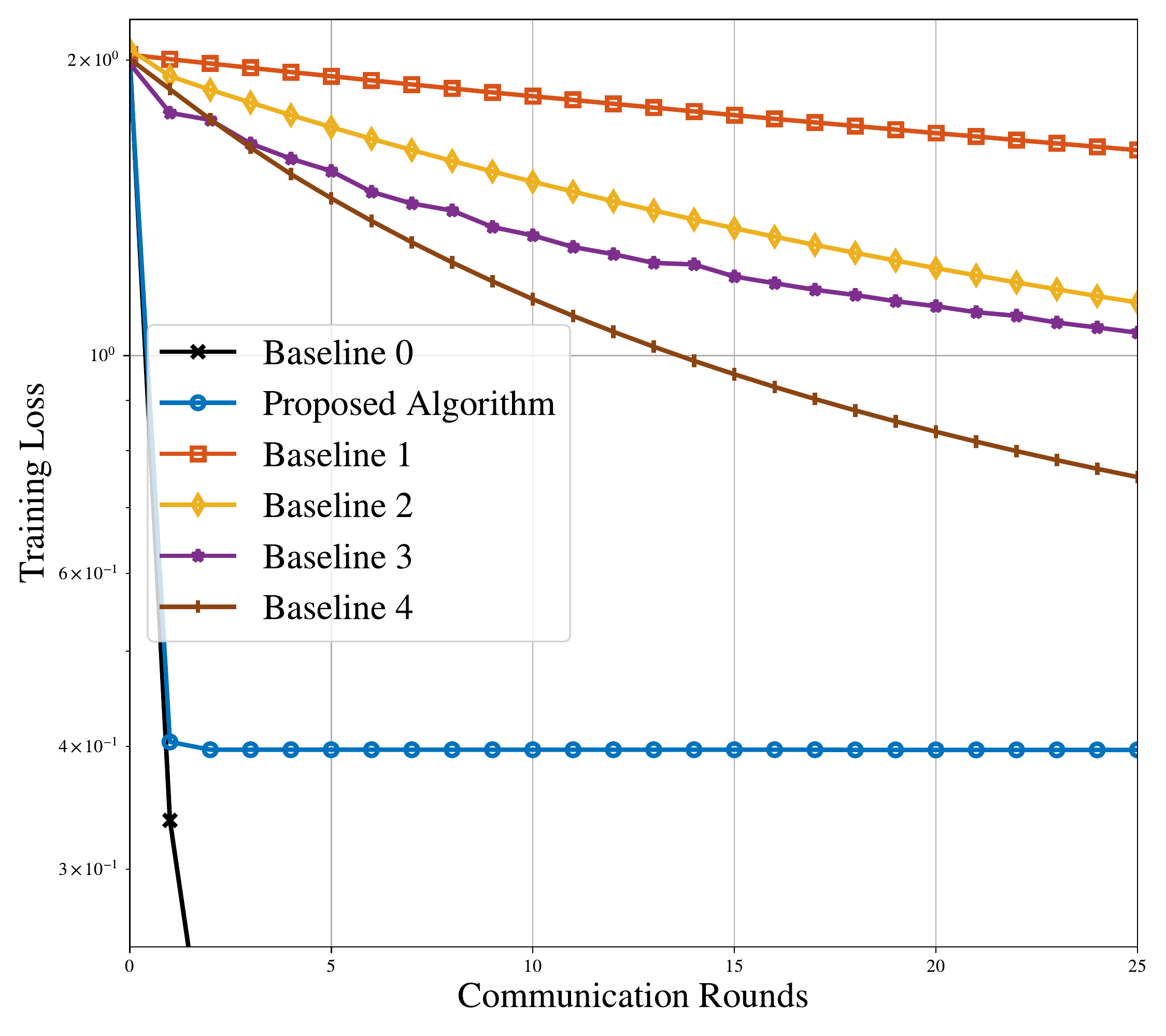}
            \label{with algorithms}
        \end{minipage}
    }
    \hspace{8em}
    \subfloat{
        \begin{minipage}[c][0.8\width]{0.3\textwidth}
            \centering
            \includegraphics[width=1\linewidth]{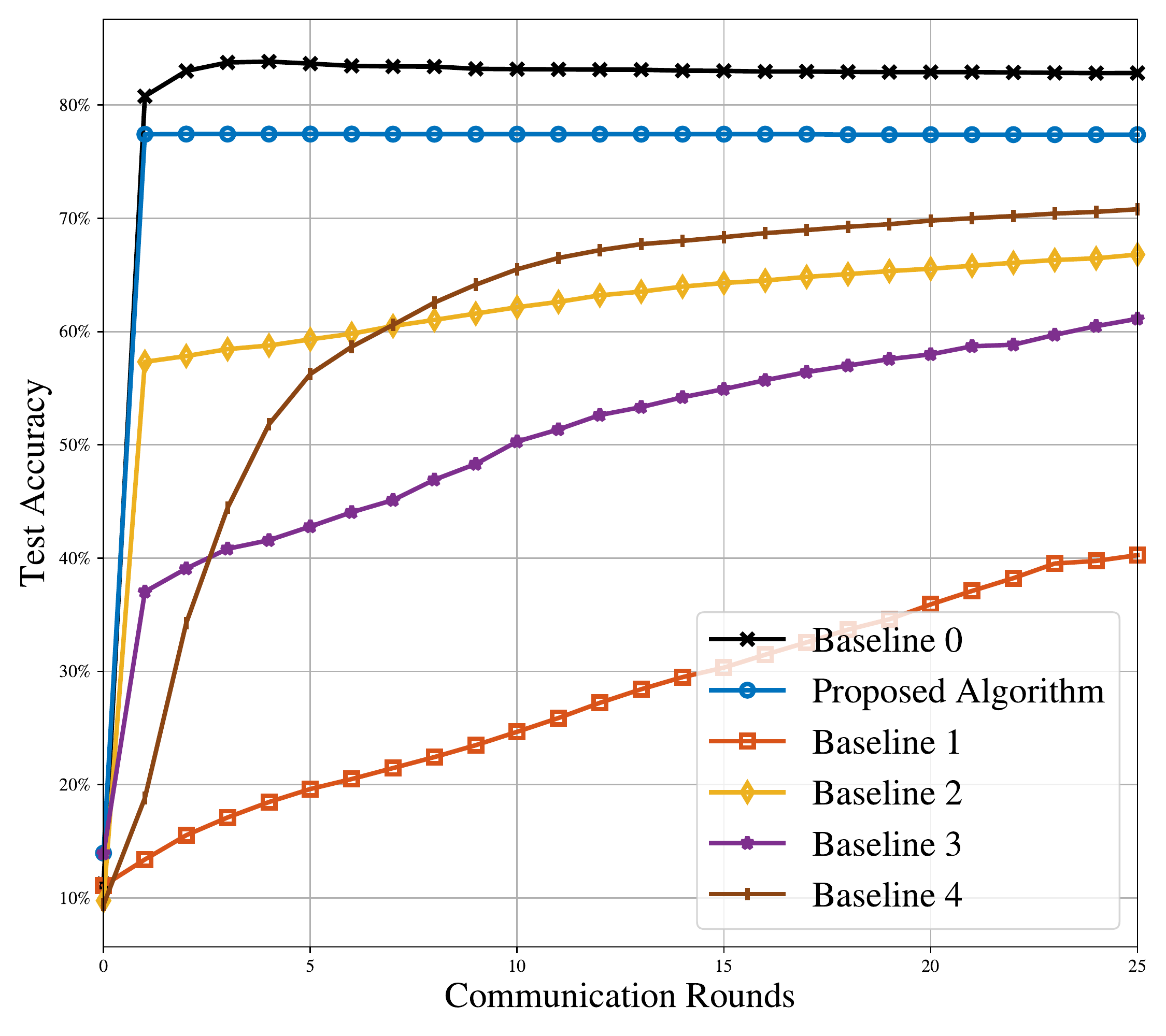}
            \label{with optimization methods}
        \end{minipage}
    }
    \caption{Simulation results on the Fashion-MNIST dataset.}
    \label{with mnist}
    \vspace{-0.8cm}
\end{figure*}

\subsection{Fashion-MNIST Data Set}
We consider an image classification problem on a non-i.i.d dataset constructed from the Fashion-MNIST dataset in this experiment, where $m=10$ and SNR is set to $90$ dB.
The related parameters are set to be the same as the previous experiments, and we use the percentage of correctly classified test images to evaluate the learning performance.

Fig. \ref{with mnist} presents the training loss and test accuracy versus communication rounds of our proposed algorithm and four baseline algorithms. 
It reveals that our proposed algorithm significantly outperforms the baseline algorithms. 
On the one hand, it keeps a better convergence rate than first-order algorithms, leading to fewer communication rounds between the devices and the server. 
On the other hand, compared with other second-order algorithms under over-the-air computation, the aggregation operation only occurs once per iteration in our proposed algorithm. 
Therefore, our proposed algorithm is more communication-efficient than baseline algorithms in terms of both the total iteration rounds and the communication within each iteration, which further benefit learning performance, as illustrated in Fig. \ref{with mnist}. 


\section{Conclusion}
\label{Conclusion}
In this paper, we developed a communication-efficient FL system by over-the-air second-order federated optimization algorithm. 
The communication rounds and communication latency at each round can be simultaneously reduced. 
This is achieved by leveraging the second-order information of the learning loss function for achieving fast convergence rates and exploiting the signal superposition property of a multiple access channel for fast model aggregation. 
The characterized convergence behavior reveals a linear-quadratic convergence rate for the proposed algorithm.  
As the proposed algorithm is accompanied by an accumulative error term in each iteration, a system optimization problem was formulated to minimize the total error gap while achieving a precise model. We then presented Gibbs Sampling and DC programming methods to jointly optimize device selection and receiver beamforming. The experimental results illustrated that our proposed algorithm and network optimization approach can achieve high communication efficiency for FL systems.

\begin{appendices}
\section{Proof of Lemma 3}
In order to bound $\hat{\bm{p}_{t}}$ through $\bm{p}^{*}$, the difference between the values of their quadratic functions is essential. 
According to \eqref{decomposition_global_update_direction}, here we decompose this difference as
\begin{small}
\begin{align*}
    \phi_{t}\left(\hat{\bm{p}}_{t}\right)-\phi_{t}\left(\bm{p}^{*}\right) 
    =& \frac{1}{2}\left\|\bm{H}_{t}^{\frac{1}{2}}\left(\hat{\bm{p}}_{t}-\bm{p}^{*}\right)\right\|^{2} 
    = \frac{1}{2}\left\|\bm{H}_{t}^{\frac{1}{2}}\left[\left(\bar{\bm{p}}_{t}-\bm{p}^{*}\right)+\left(\bm{p}_{t}-\bar{\bm{p}}_{t}\right)+\left(\tilde{\bm{p}}_{t}-\bm{p}_{t}\right)+\left(\hat{\bm{p}}_{t}-\tilde{\bm{p}}_{t}\right) \right]\right\|^{2} \\
    \leq& \underbrace{\left\|\bm{H}_{t}^{\frac{1}{2}}\left(\bm{p}_{t}-\bar{\bm{p}}_{t}\right)\right\|^{2}}_{\text{Term 1}} +\underbrace{3\left\|\bm{H}_{t}^{\frac{1}{2}}\left(\bar{\bm{p}}_{t}-\bm{p}^{*}\right)\right\|^{2}}_{\text{Term 2}} + \underbrace{3\left\|\bm{H}_{t}^{\frac{1}{2}}\left(\tilde{\bm{p}}_{t} - \bm{p}_{t}\right)\right\|^{2}}_{\text{Term 3}} + \underbrace{3\left\|\bm{H}_{t}^{\frac{1}{2}}\left(\hat{\bm{p}}_{t}-\tilde{\bm{p}}_{t}\right)\right\|^{2}}_{\text{Term 4}}\;,
\end{align*}
\end{small}
As for Term 1, by Lemma 1, we have $(1-\lambda) \bm{M}_{t}^\mathsf{T} \bm{M}_{t} \preceq \bm{M}_{t}^\mathsf{T} \bm{L}_{i} \bm{L}_{i}^\mathsf{T} \bm{M}_{t} \preceq (1+\lambda) \bm{M}_{t}^\mathsf{T} \bm{M}_{t}$.
Through this we can get $(1-\lambda) \bm{H}_{t} \preceq \bm{H}_{i, t} \preceq(1+\lambda) \bm{H}_{t}$. Thus, there exists matrix $\xi_{i}$
satisfying $\bm{H}_{t}^{\frac{1}{2}} \bm{H}_{t,i}^{-1} \bm{H}_{t}^{\frac{1}{2}}=\bm{I}+\xi_{i} $ and $-\frac{\lambda}{1+\lambda} \preceq \xi_{i} \preceq \frac{\lambda}{1-\lambda}$,
which leads to a useful property: $\left\|\bm{H}_{t}^{\frac{1}{2}} \bm{H}_{t,i}^{-1} \bm{H}_{t}^{\frac{1}{2}}\right\| \leq 1+\frac{\lambda}{1-\lambda}=\frac{1}{1-\lambda}$. 
With this property and Lemma 2, we can get the following inequality:
\begin{small}
\begin{align*}
    \norm{\bm{H}_{t}^{\frac{1}{2}}\bracket{\bm{p}_{t} - \bar{\bm{p}}_{t}}} 
    &\leq \frac{1}{n} \sum_{i\in \mathcal{S}} \card{\mathcal{D}_{i}} \norm{\bm{H}_{t}^{\frac{1}{2}}\bm{H}_{t,i}^{-1}\bm{H}_{t}^{\frac{1}{2}}}\norm{\bm{H}_{t}^{-\frac{1}{2}}\bracket{\bm{g}_{t,i}-\bm{g}_{t}}} \\
    &\leq \frac{1}{1-\lambda} \frac{1}{\sigma_{min}\bracket{\bm{H}_{t}}}\frac{1}{n} \sum_{i\in \mathcal{S}} \card{\mathcal{D}_{i}} \bracket{1 + \sqrt{2 \ln{\frac{1}{\delta_{i}}}}}\sqrt{\frac{1}{\card{\mathcal{D}_{i}}}}\max_{j}\norm{\bm{n}_{j}} \\
    &\leq \frac{1}{1-\lambda} \frac{1}{\sigma_{min}\bracket{\bm{H}_{t}}}\frac{1}{n}\bracket{1 + \sqrt{2 \ln{\frac{1}{\tilde{\delta}}}}}\max_{j}\norm{\bm{n}_{j}}\sqrt{\sum_{i\in \mathcal{S}}m\card{\mathcal{D}_{i}}}\\
    &= \frac{1}{1-\lambda} \frac{1}{\sigma_{min}\bracket{\bm{H}_{t}}}\bracket{1 + \sqrt{2 \ln{\frac{1}{\tilde{\delta}}}}}\sqrt{\frac{m}{n}}\max_{j}\norm{\bm{n}_{j}} \;.
\end{align*}
\end{small}
For convenience, we denote:
$\mathcal{G} = \frac{1}{1-\lambda} \frac{1}{\sigma_{min}\bracket{\bm{H}_{t}}}\bracket{1 + \sqrt{2 \ln{\frac{1}{\tilde{\delta}}}}}\max_{j}\norm{\bm{n}_{j}}\;,$
and Term 1 is bounded by $\text{Term 1} \leq \frac{m}{n}\mathcal{G}^2$.
As for Term 2, based on the analysis in \cite[Lemma 6]{wang2018giant}, we have 
\begin{small}
\begin{align*}
    \left\|\bm{H}_{t}^{\frac{1}{2}}\left(\bar{\bm{p}}_{t}-\bm{p}^{*}\right)\right\| 
    &\leq \norm{\frac{1}{n}\sum_{i\in\mathcal{S}}\card{\mathcal{D}_{i}}\bm{H}_{t}^{\frac{1}{2}}\left(\bar{\bm{p}}_{t,i}-\bm{p}^{*}\right)} 
     \leq \frac{1}{n}\sum_{i\in\mathcal{S}}\card{\mathcal{D}_{i}}\norm{\bm{H}_{t}^{\frac{1}{2}}\left(\bar{\bm{p}}_{t,i}-\bm{p}^{*}\right)} 
    \leq 
    \zeta_{1} \norm{\bm{H}_{t}^{\frac{1}{2}}\bm{p}^{*}} \;,
\end{align*}
\end{small}
with $\zeta_{1}=\tau\left(\lambda+\frac{\lambda^{2}}{1-\lambda}\right)$  and $\tau=\frac{\sigma_{\max }\left(\bm{M}^{\top} \bm{M}\right)}{\sigma_{\max }\left(\bm{M}^{\top} \bm{M}\right)+n\gamma}$.
Then Term 2 is bound by:
\begin{small}
\begin{align*}
    \text{Term 2} = 3 \left\|\bm{H}_{t}^{\frac{1}{2}}\left(\bar{\bm{p}}_{t}-\bm{p}^{*}\right)\right\|^{2} \leq 3\zeta_{1}^{2} \norm{\bm{H}_{t}^{\frac{1}{2}}\bm{p}^{*}}^{2} = - 3\zeta_{1}^{2} \phi\bracket{\bm{p}^{*}}.
\end{align*}
\end{small}
As for Term 3, it can be reformulated as follows:
\begin{small}
\begin{align*}
    \text{Term 3}  
    =3\left\|\bm{H}_{t}^{\frac{1}{2}}\left(\tilde{\bm{p}}_{t} - \bm{p}_{t}\right)\right\|^{2} = 3\left\|\bm{H}_{t}^{\frac{1}{2}}\left(\frac{1}{\sum_{i\in\mathcal{S}_{t}}\card{\mathcal{D}_{i}}}\sum_{i\in\mathcal{S}_{t}}\card{\mathcal{D}_{i}}\bm{p}_{t,i}-\frac{1}{n}\sum_{i\in\mathcal{S}}\card{\mathcal{D}_{i}}\bm{p}_{t,i}\right)\right\|^{2}\;.
\end{align*}
\end{small}
According to the analysis in \cite[Section 3.1]{friedlander2012hybrid}, it follows:
\begin{small}
\begin{align*}
    \text{Term 3}
    \leq& 12\bracket{1 - \frac{\sum_{i\in\mathcal{S}_{t}}\card{\mathcal{D}_{i}}}{n}}^{2}\bracket{\norm{\bm{H}_{t}^{\frac{1}{2}}\bm{p}_{t,i} - \bm{H}_{t}^{\frac{1}{2}}\bar{\bm{p}}_{t,i}} + \norm{\bm{H}_{t}^{\frac{1}{2}}\bar{\bm{p}}_{t,i} - \bm{H}_{t}^{\frac{1}{2}}\bm{p}^{*}} + \norm{\bm{H}_{t}^{\frac{1}{2}}\bm{p}^{*}}}^{2} \\
    \stackrel{(a)}{\leq}& 12 \bracket{1 - \frac{\sum_{i\in\mathcal{S}_{t}}\card{\mathcal{D}_{i}}}{n}}^{2}\bracket{\sqrt{\frac{1}{\min_{i\in\mathcal{S}_{t}}\card{\mathcal{D}_{i}}}}\mathcal{G} +
    \norm{\bm{H}_{t}^{\frac{1}{2}}\bar{\bm{p}}_{t,i} - \bm{H}_{t}^{\frac{1}{2}}\bm{p}^{*}} + \norm{\bm{H}_{t}^{\frac{1}{2}}\bm{p}^{*}}}^{2} \\
    \stackrel{(b)}{\leq}& 12 \bracket{1 - \frac{\sum_{i\in\mathcal{S}_{t}}\card{\mathcal{D}_{i}}}{n}}^{2}\bracket{\sqrt{\frac{1}{\min_{i\in\mathcal{S}_{t}}\card{\mathcal{D}_{i}}}}\mathcal{G} +
    \bracket{\zeta_{1}+ 1}\norm{\bm{H}_{t}^{\frac{1}{2}}\bm{p}^{*}}}^{2} \\
    \leq& 24\bracket{1 - \frac{\sum_{i\in\mathcal{S}_{t}}\card{\mathcal{D}_{i}}}{n}}^{2}\frac{1}{\min_{i\in\mathcal{S}_{t}}\card{\mathcal{D}_{i}}}\mathcal{G}^{2} - 24\vartheta^{2}\bracket{\zeta_{1}+ 1}^{2}\phi\bracket{\bm{p}^{*}}
\end{align*}
\end{small}
where $\zeta_{1}=\tau\left(\lambda+\frac{\lambda^{2}}{1-\lambda}\right)$, $\tau=\frac{\sigma_{\max }\left(\bm{M}^{\top} \bm{M}\right)}{\sigma_{\max }\left(\bm{M}^{\top} \bm{M}\right)+n\gamma}$, $\vartheta = \max _{t}\bracket{1 - \frac{\sum_{i\in\mathcal{S}_{t}}\card{\mathcal{D}_{i}}}{n}} < 1$, $\left(a\right)$ and $\left(b\right)$ are obtained in the way similar to the analysis of Term 1 and Term 2.
As for Term 4, we have:
\begin{small}
\begin{align*}
    \text{Term 4} 
    = 3\norm{\bm{H}_{t}^{\frac{1}{2}}\frac{1}{\bracket{\sum_{i\in\mathcal{S}_{t}}\card{\mathcal{D}_{i}}}\sqrt{\eta_{t}}}\bm{a}_{t}^\mathsf{H}\bm{E}_{t}}^{2} 
    \leq \frac{3}{\sigma_{\min}\left(\bm{H}_{t}\right)}\left\|\frac{1}{\bracket{\sum_{i\in\mathcal{S}_{t}}\card{\mathcal{D}_{i}}}\sqrt{\eta_{t}}}\bm{a}_{t}^\mathsf{H}\bm{E}_{t}\right\|^{2}\;.
\end{align*}
\end{small}
We can get the final result by combining the bound of Term 1, 2, 3 and 4 together:
\begin{small}
\begin{align*}
\phi\left(\hat{\bm{p}}_{t}\right)-\phi\left(\bm{p}^{*}\right) \leq \epsilon^{2}-\zeta^{2} \phi\left(\bm{p}^{*}\right) 
\Rightarrow \phi\left(\bm{p}^{*}\right) \leq \phi\left(\hat{\bm{p}}_{t}\right) \leq \epsilon^{2}+\left(1-\zeta^{2}\right) \phi\left(\bm{p}^{*}\right)\;,
\end{align*}
\end{small}
where $\epsilon$ and $\zeta$ are defined as \eqref{zeta} and \eqref{epsilon}.

\section{Proof of Theorem 1}
Based on Lemma 3 and Lemma 4, we have:
\begin{small}
\begin{align*}
    \bm{\Delta}_{t+1}^{\top} \bm{H}_{t} \bm{\Delta}_{t+1}
    \leq& L\left\|\bm{\Delta}_{t+1}\right\|\left\|\bm{\Delta}_{t}\right\|^{2}+\frac{\zeta^{2}}{1-\zeta^{2}} \bm{\Delta}_{t}^{\top} \bm{H}_{t} \bm{\Delta}_{t}+2 \epsilon^{2} 
    \leq L\left\|\bm{\Delta}_{t+1}\right\|\left\|\bm{\Delta}_{t}\right\|^{2}+\left(\frac{\zeta^{2}}{1-\zeta^{2}} \sigma_{\max }\left(\mathbf{H}_{t}\right)\right)\left\|\bm{\Delta}_{t}\right\|^{2}+2 \epsilon^{2}\;.
\end{align*}
\end{small}
According to the analysis in \cite[Appendix A]{ghosh2020distributed}, this leads to:
\begin{small}
\begin{equation}
\begin{aligned}
    \left\|\bm{\Delta}_{t+1}\right\|
    \leq& \max \left\{\sqrt{\frac{\sigma_{\max }\left(\bm{H}_{t}\right)}{\sigma_{\min }\left(\bm{H}_{t}\right)}\left(\frac{\zeta^{2}}{1-\zeta^{2}}\right)}\left\|\bm{\Delta}_{t}\right\|, \frac{L}{\sigma_{\min }\left(\bm{H}_{t}\right)}\left\|\bm{\Delta}_{t}\right\|^{2}\right\} +\frac{2 \epsilon}{\sqrt{\sigma_{\min }\left(\mathbf{H}_{t}\right)}}\;.
\end{aligned}
\label{inequality}
\end{equation}
\end{small}
As for the error term $\epsilon$, we have:
\begin{small}
\begin{align*}
    \epsilon =& \Bigg\{ \frac{3}{\sigma_{\min}\left(\bm{H}_{t}\right)}\left\|\frac{1}{\bracket{\sum_{i \in \mathcal{S}_{t}}\card{\mathcal{D}_{i}}}\sqrt{\eta_{t}}}\bm{a}_{t}^\mathsf{H}\bm{E}_{t}\right\|^{2} + \left[24\left(1 - \frac{\sum_{i \in \mathcal{S}_{t}}\card{\mathcal{D}_{i}}}{n}\right)^{2}\frac{1}{\min_{i\in\mathcal{S}_{t}}\card{\mathcal{D}_{i}}}+\frac{m}{n}\right] \mathcal{G}^{2} \Bigg\}^{\frac{1}{2}} \\
    \leq& \sqrt{\frac{3}{\sigma_{\min}\left(\bm{H}_{t}\right)}}\frac{d}{\bracket{\sum_{i \in \mathcal{S}_{t}}\card{\mathcal{D}_{i}}}\sqrt{\eta_{t}}}\left\|\bm{a}_{t}\right\|\left\|\bm{e}_{t}\right\| + \sqrt{24\left(1 - \frac{\sum_{i \in \mathcal{S}_{t}}\card{\mathcal{D}_{i}}}{n}\right)^{2}\frac{1}{\min_{i\in\mathcal{S}_{t}}\card{\mathcal{D}_{i}}}+\frac{m}{n}} \cdot \mathcal{G} \;.
\end{align*}
\end{small}
To handle the random variable $\bm{e}_{t}$ in $\epsilon$, 
we take expectations over $\bm{e}_{t}$ on both sides of~\eqref{inequality} :
\begin{small}
\begin{align*}
     \mathbb{E}\left(\left\|\bm{\Delta}_{t+1}\right\|\right)
     \leq& \max \left\{\sqrt{\frac{\sigma_{\max }\left(\bm{H}_{t}\right)}{\sigma_{\min }\left(\bm{H}_{t}\right)}\left(\frac{\zeta^{2}}{1-\zeta^{2}}\right)}\left\|\bm{\Delta}_{t}\right\|, \frac{L}{\sigma_{\min }\left(\bm{H}_{t}\right)}\left\|\bm{\Delta}_{t}\right\|^{2}\right\}+\frac{2\sqrt{3}}{\sigma_{\min}\left(\bm{H}_{t}\right)}\frac{ d\left\|\bm{a}_{t}\right\|\mathbb{E}\left(\left\|\bm{e}_{t}\right\|\right)}{\bracket{\sum_{i \in \mathcal{S}_{t}}\card{\mathcal{D}_{i}}}\sqrt{\eta_{t}}} \\
     &+\sqrt{24\left(1 - \frac{\sum_{i\in\mathcal{S}_{t}}\card{\mathcal{D}_{i}}}{n}\right)^{2}\frac{1}{\min_{i\in\mathcal{S}_{t}}\card{\mathcal{D}_{i}}}+\frac{m}{n}}\cdot\frac{1}{1-\lambda} \frac{2}{\sigma_{\min }\left(\bm{H}_{t}\right)}\left(1+\sqrt{2 \ln \left(\frac{1}{\tilde{\delta}}\right)}\right)\max _{j} \left\| \bm{n}_{j} \right\| \\
    \stackrel{(c)}{\leq}& \max \left\{\sqrt{\frac{\sigma_{\max }\left(\bm{H}_{t}\right)}{\sigma_{\min }\left(\bm{H}_{t}\right)}\left(\frac{\zeta^{2}}{1-\zeta^{2}}\right)}\left\|\bm{\Delta}_{t}\right\|, \frac{L}{\sigma_{\min }\left(\bm{H}_{t}\right)}\left\|\bm{\Delta}_{t}\right\|^{2}\right\}+\frac{2\sqrt{3}}{\sigma_{\min}\left(\bm{H}_{t}\right)}\frac{ d\sigma\left\|\bm{a}_{t}\right\|}{\bracket{\sum_{i \in \mathcal{S}_{t}}\card{\mathcal{D}_{i}}}\sqrt{\eta_{t}}} \\
    &+ \sqrt{24\left(1 - \frac{\sum_{i\in\mathcal{S}_{t}}\card{\mathcal{D}_{i}}}{n}\right)^{2}\frac{1}{\min_{i\in\mathcal{S}_{t}}\card{\mathcal{D}_{i}}}+\frac{m}{n}}\cdot\frac{1}{1-\lambda} \frac{2}{\sigma_{\min }\left(\bm{H}_{t}\right)}\left(1+\sqrt{2 \ln \left(\frac{1}{\tilde{\delta}}\right)}\right)\max _{j} \left\| \bm{n}_{j} \right\| \;.
\end{align*}
\end{small}
\end{appendices}

\bibliographystyle{ieeetr}
\bibliography{ref}

\end{document}